\documentclass[10pt,journal,compsoc]{IEEEtran}

\usepackage{meta/lang}
\usepackage{color}
\usepackage{etoolbox}
\usepackage{listings}
\usepackage[T1]{fontenc}

\usepackage[scaled=0.8]{beramono}

\newcommand{\lstfontfamily}{\ttfamily}

\definecolor{darkviolet}{rgb}{0.5,0,0.4}
\definecolor{darkgreen}{rgb}{0,0.4,0.2} 
\definecolor{darkblue}{rgb}{0.1,0.1,0.9}
\definecolor{darkgrey}{rgb}{0.5,0.5,0.5}
\definecolor{lightblue}{rgb}{0.4,0.4,1}

\definecolor{stringColor}{rgb}{0.16,0.00,1.00}
\definecolor{annotationColor}{rgb}{0.39,0.39,0.39}
\definecolor{keywordColor}{rgb}{0.50,0.00,0.33}
\definecolor{commentColor}{rgb}{0.25,0.50,0.37}
\definecolor{javadocColor}{rgb}{0.25,0.37,0.75}
\definecolor{jTagColor}{rgb}{0.50,0.62,0.75}
\definecolor{eTagColor}{rgb}{0.50,0.62,0.75}
\definecolor{lineNumberColor}{rgb}{0.47,0.47,0.47}

\def\jTags{@author, @deprecated, @exception, @param, @return, @see, @serial, @serialData, @serialField, @since, @throws, @version}

\def\jAnnotations{
    classoffset=1,
    morekeywords={@Override, @Deperecated, @SuppressWarnings, @Retention, @Documented, @Target, @Inherited},
    keywordstyle=\color{annotationColor},
    classoffset=0
}

\def\eTags{FIXME, TODO, XXX}

\newrobustcmd{\markupJavadocs}[1]{%
\edef\mytok{\the\lst@token}%
\renewcommand*{\do}[1]{%
\ifdefstring{\mytok}{##1}%
{\color{jTagColor}\bfseries\listbreak}%
{}%
}%
{\color{javadocColor}%
\expandafter\docsvlist\expandafter{\jTags}%
\renewcommand*{\do}[1]{%
\ifdefstring{\mytok}{##1}%
{\color{eTagColor}\bfseries\listbreak}%
{}%
}%
\expandafter\docsvlist\expandafter{\eTags}%
#1}%
}%

\newrobustcmd{\markupComments}[1]{%
\edef\mytok{\the\lst@token}%
\renewcommand*{\do}[1]{%
\ifdefstring{\mytok}{##1}%
{\color{eTagColor}\bfseries\listbreak}%
{}%
}%
{\color{commentColor}%
\expandafter\docsvlist\expandafter{\eTags}#1}%
}%

\lstdefinestyle{eclipse}{
  basicstyle={\lstfontfamily},
  emphstyle=\bfseries,
  keywordstyle=\color{keywordColor}\bfseries,
  commentstyle=\markupComments,
  stringstyle=\color{stringColor},
  numberstyle=\color{lineNumberColor}\lstfontfamily,
  morecomment=[s][\markupJavadocs]{/**}{*/}, 
  showstringspaces=false,
  numbers=left,
}

\lstdefinestyle{black}{
  basicstyle=\small\lstfontfamily,
  numbers=left,
  columns=fullflexible,
  breaklines=true,
  mathescape=true,
  escapechar=\#,
  tabsize=4,
  frame=lines,
  showstringspaces=false
}

\lstdefinestyle{seminar}{
  basicstyle=\small\ttfamily,
  numbers=left,
  breaklines=true,
  mathescape=true,
  escapechar=\#,
  tabsize=4,
  showstringspaces=false
}

\expandafter\lstset\expandafter{\jAnnotations}

\usepackage{hyperref}
\usepackage{mdframed}
\usepackage{lipsum}
\usepackage{amsmath}
\usepackage{amssymb}
\usepackage{amsfonts}
\usepackage[font=small]{caption}
\usepackage{ifpdf}
\usepackage[final]{graphicx}
\usepackage{todonotes}
\usepackage{booktabs} 
\usepackage{url}
\usepackage{amssymb}
\usepackage{pifont}
\usepackage{color}
\usepackage{tcolorbox}
\usepackage{balance}
\usepackage{makecell}
\usepackage{verbatim}
\usepackage{hyperref}
\usepackage{longtable}
\usepackage{rotating}

\usepackage{siunitx}
\usepackage{pifont}
\usepackage{csquotes}
\usepackage{url}

\usepackage{xcolor}
\usepackage{fancybox}
\cornersize{.1} 
\newcommand{\mytcbox}[2]{
\begin{tcolorbox}[title=#1,boxrule=1pt,boxsep=1pt,left=2pt,right=2pt,top=2pt,bottom=2pt]
#2
\end{tcolorbox}
}
\newcommand{\mynewcontent}[2]{\ifnum#1<10{#2}\else{\textcolor{red}{#2}}\fi}

\newcommand{\significant}[1]{\textbf{#1}}

\newcommand{\priostrategy}[1]{\textit{#1}}

\newcommand{\nrcols}[2]{\multicolumn{#1}{c}{#2}}
\newcommand{\tabprojectlinespace}{\addlinespace[0.5ex]}
\providecommand{\inlinecode}[1]{\textcolor{black}{\texttt{#1}}}

\definecolor{greylight}{RGB}{240,240,240}
\definecolor{greymedium}{RGB}{189,189,189}
\definecolor{greydark}{RGB}{99,99,99}

\usepackage{tcolorbox}
\tcbset{colback=greylight, colframe=greydark, arc=1pt, boxrule=0.75pt}

\ifCLASSOPTIONcompsoc
  \usepackage[nocompress]{cite}
\else
  \usepackage{cite}
\fi
\ifCLASSINFOpdf
\else
\fi
\begin{document}
%
\title{Coverage Goal Selector for Combining Multiple Criteria in Search-Based Unit Test Generation} 
%
%
%
%

\author{Zhichao~Zhou,
        Yuming~Zhou,
        Chunrong~Fang, 
        Zhenyu~Chen,
        Xiapu~Luo,
        Jingzhu~He,
        and~Yutian~Tang
\thanks{Z. Zhou and J. He are with School of Information Science and Technology, ShanghaiTech University, China}
\thanks{Y. Zhou, C. Fang, and Z. Chen are with State Key Laboratory for Novel Software Technology, Nanjing University, China}
\thanks{X. Luo is with the Hong Kong Polytechnic University, Hong Kong SAR, China}
\thanks{Y. Tang is with University of Glasgow, United Kingdom}
\thanks{Y. Tang (yutian.tang@glasgow.ac.uk) is the corresponding author.}}

%
%

\markboth{Journal of \LaTeX\ Class Files,~Vol.~14, No.~8, August~2015}%
{Shell \MakeLowercase{\textit{et al.}}: Bare Advanced Demo of IEEEtran.cls for IEEE Computer Society Journals}
%



\IEEEtitleabstractindextext{%
\begin{abstract}
Unit testing is critical to the software development process, ensuring the correctness of basic programming units in a program (e.g., a method). Search-based software testing (SBST) is an automated approach to generating test cases. SBST generates test cases with genetic algorithms by specifying the coverage criterion (e.g., branch coverage). However, a good test suite must have different properties, which cannot be captured using an individual coverage criterion. Therefore, the state-of-the-art approach combines multiple criteria to generate test cases. \mynewcontent{2}{Since} combining multiple coverage criteria brings multiple objectives for optimization, it hurts the test suites' coverage for certain criteria compared with using the single criterion. To cope with this problem, we propose a novel approach named \textbf{smart selection}. Based on the coverage correlations among criteria and \mynewcontent{2}{the subsumption relationships among coverage goals}, smart selection selects a subset of coverage goals to reduce the number of optimization objectives and avoid missing any properties of all criteria. We conduct experiments to evaluate smart selection on $400$ Java classes with three state-of-the-art genetic algorithms \mynewcontent{2}{under the $2$-minute budget}. On average, smart selection outperforms combining all goals on $65.1\%$ of the classes having significant differences between the two approaches. \mynewcontent{2}{Secondly, we conduct experiments to verify our assumptions about coverage criteria relationships. \mynewcontent{6}{Furthermore, we assess the coverage performance of smart selection under varying budgets of $5$, $8$, and $10$ minutes and explore its effect on bug detection, confirming the advantage of smart selection over combining all goals.}}
\end{abstract}

}

\maketitle

\IEEEdisplaynontitleabstractindextext

%
\IEEEpeerreviewmaketitle

\section{Introduction}\label{sec:intro}
\mynewcontent{2}{Unit testing is a common way to ensure software quality by testing individual units or components of a software system in isolation from the rest of the system.} Manually \mynewcontent{2}{writing} unit tests can be a tedious and error-prone process. Hence, developers and researchers put much effort into automatically generating test cases for programming units \mynewcontent{2}{in recent years}.

Search-based software testing (SBST) is considered a promising approach to generating test cases. It generates test cases with genetic algorithms (e.g., Whole Suite Generation (WS)~\cite{FraserWhole}, MOSA~\cite{PanichellaMOSA}, DynaMOSA~\cite{PanichellaDynaMOSA}) \mynewcontent{2}{based on} the coverage criterion (e.g., branch coverage). The execution of a genetic algorithm \mynewcontent{2}{depends} on fitness functions, which quantify the degree to which a solution (i.e., one or more test cases) achieves its goals (i.e., satisfying a certain coverage criterion). \mynewcontent{2}{Each coverage criterion has a corresponding group of fitness functions, and each fitness function describes whether or how far a test case covers a specific coverage goal (e.g., a branch)}.

\begin{figure}[htbp]
    \centering
    \includegraphics[width=0.48\textwidth]{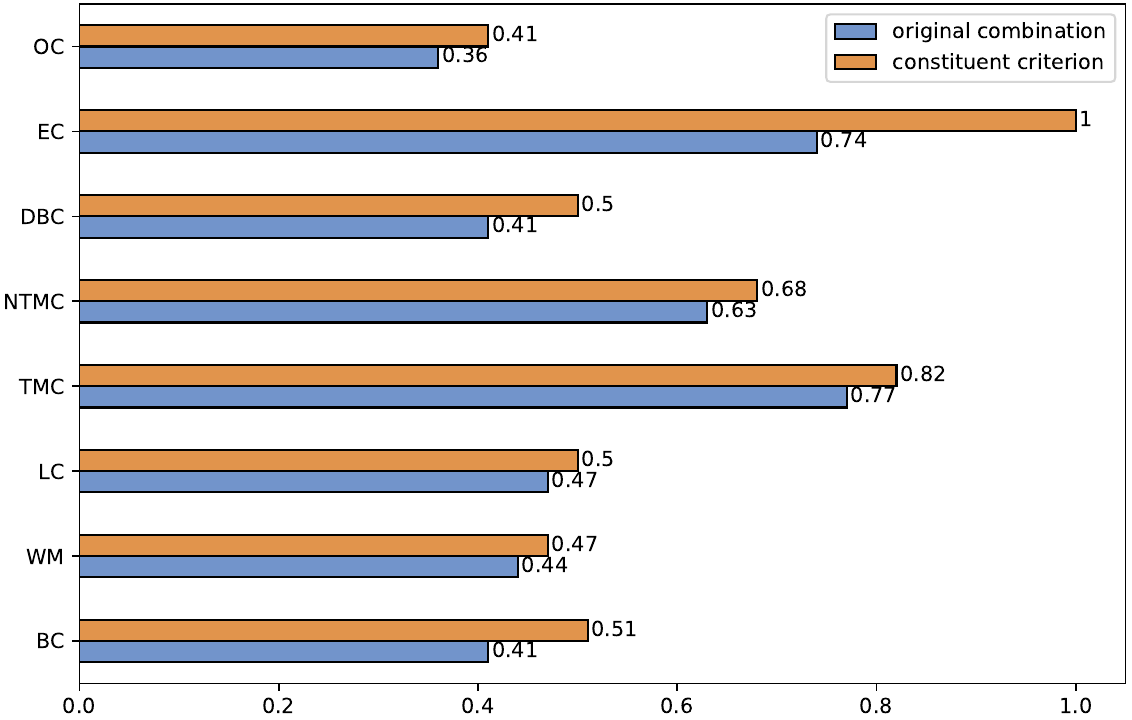}
    \caption{Partial Data of Coverage Gaps}
    \label{fig:coverage-gaps-1}
\end{figure}

\noindent\textbf{The Problem and Motivation.} As claimed in~\cite{Rojas2015CombiningMC}, \mynewcontent{2}{a good test suite must possess different properties that cannot be easily captured by any individual coverage criterion}. Therefore, to generate a good test suite, multiple coverage criteria should be considered in SBST. Hence, the state-of-the-art approach~\cite{Rojas2015CombiningMC} combines multiple coverage criteria to guide genetic algorithms. \mynewcontent{2}{This method} involves eight coverage criteria (see Sec. \ref{subsec:criterion}),  \mynewcontent{2}{which we call} the \textbf{original combination} in this paper. However, \mynewcontent{2}{combining multiple criteria leads to more objectives for optimization, which could impact the effectiveness of the genetic algorithms~\cite{KnowlesSingle, PanichellaMOSA, BrockhoffMulti}}. For example, it can increase the probability of being trapped in local optima. As a result, \mynewcontent{2}{the coverage of the generated test suite} decreases for certain criteria compared with using a single criterion. Fig. \ref{fig:coverage-gaps-1} shows (see Sec. \ref{subsec:rq1}) the average coverage gaps between the original combination and each constituent criterion when applying WS~\cite{FraserWhole} into $85$ \mynewcontent{2}{large} Java classes (i.e., with at least $200$ branches). The average gap of the eight criteria is $8.2\%$, Branch coverage (BC) decreases \mynewcontent{2}{by} $10\%$, and Exception coverage (EC) decreases \mynewcontent{2}{by} $26\%$. Note that \mynewcontent{2}{since the total exceptions in a class cannot be determined}~\cite{Rojas2015CombiningMC}, we normalize the exception coverage values of two approaches ($22.08$ vs. $29.74$) by dividing them by the larger one.






%

\noindent\textbf{Targets.} \mynewcontent{2}{To address this problem, a competent approach should achieve the following targets:}
(1) \textbf{T1: GA Effectiveness.} It should select a subset \mynewcontent{2}{of coverage goals from multiple coverage criteria.} This subset \mynewcontent{2}{should enhance} the effectiveness of guiding genetic algorithms (GAs); and (2) \textbf{T2: Property Consistency.} This subset should \mynewcontent{2}{prevent the omission of} any properties captured by these coverage criteria.  

\noindent\textbf{Our Solution.}
To \mynewcontent{2}{achieve} these targets, we propose a novel approach named \textbf{smart selection} (see Sec. \ref{sec:smart}). In this paper, \mynewcontent{2}{we have considered the eight coverage criteria mentioned above.} However, instead of directly combining them, in smart selection, we first group them into four groups based on coverage correlations (see Sec. \ref{subsec:cc}). Next, we select one representative criterion that is more effective in guiding the genetic algorithms from each group (T1) (see Sec. \ref{subsec:rc}). These selected coverage criteria ($SC$)' coverage goals are \mynewcontent{2}{denoted} as $Goal(SC)$. To keep the property consistency  (T2), for each criterion ($c$) of unselected criteria ($USC$), we select a subset $Goal(c)_{sub}$ from its coverage goals based on the goals' subsumption relationships (see Sec. \ref{subsec:rs}). Finally, we combine $Goal(SC)$ and $\underset{c\in USC}{\cup}Goal(c)_{sub}$ to guide the test case generation process.



\noindent\textbf{Contribution.} In summary, the contribution of this paper includes:

\noindent$\bullet$ To the best of our knowledge, this is the \emph{first} paper that uses coverage correlations to address the coverage decrease caused by combining multiple criteria in SBST.

\noindent$\bullet$ We implement smart selection \mynewcontent{6}{on top of} EvoSuite. It is integrated into three search algorithms (i.e., WS, MOSA, and DynaMOSA). 

\noindent$\bullet$ We conduct experiments on 400 Java classes to compare smart selection and the original combination \mynewcontent{1}{with the $2$-minute time budget}. On average of three algorithms (WS in Sec. \ref{subsec:rq1}, MOSA in Sec. \ref{subsec:rq2}, and DynaMOSA in Sec. \ref{subsec:rq3}), smart selection outperforms the original combination on $77$ ($121$/$78$/$32$) classes, accounting for $65.1\%$ ($85.8\%$/$65\%$/$44.4\%$) of the classes having significant differences between the two approaches. The counterpart data of the $85$ large classes is $34$ ($50$/$35$/$16$), accounting for $86.1\%$ ($98\%$/$87.5\%$/$72.7\%$). Second, we conduct experiments to compare smart selection with/without the subsumption strategy on $173$ classes (Sec. \ref{subsec:rq6}).

\noindent\textbf{Major Extensions.} The article is the extended version of our previous paper~\cite{ASESS}, which was published at the 37th IEEE/ACM International Conference on Automated Software Engineering (ASE). This article introduces the following extended contributions:

\noindent\mynewcontent{1}{$\bullet$ We add the statistical information of our experimental subjects, including their distributions of the number of branches and lines (Sec. \ref{subsec:exp_set}).}

\noindent\mynewcontent{3}{$\bullet$ In Sec. \ref{subsec:cc}, we propose three rules to determine whether two criteria have a coverage correlation so that we can cluster criteria into several groups. To verify our rules, we conduct experiments in Sec. \ref{subsec:rq-4}. The experimental results show that, on average, the Pearson Correlation Coefficient~\cite{pearsonr} of the coverage values of the criteria from the same groups is much higher than that of the criteria from the different groups ($0.88$ versus $0.41$), confirming the effectiveness of our rules.}

\noindent\mynewcontent{3}{$\bullet$ In Sec. \ref{subsec:rc}, we select one criterion that is more representative and effective in guiding the genetic algorithms from each group by analyzing and comparing their fitness functions. We conduct experiments in Sec. \ref{subsec:rq-5} to verify our selection by using two deliberate criteria combinations to guide GAs. The experimental results show that, in most cases, the coverage resulting from smart selection is higher than that of these two criteria combinations, confirming our choice.}

\noindent\mynewcontent{1}{$\bullet$ We investigate how smart selection performs under different search budgets (Sec. \ref{subsec:rq7}).}

\noindent\mynewcontent{6}{$\bullet$ We investigate how smart selection affects bug detection (Sec. \ref{subsec:rq8}).}

\noindent\textbf{Paper Organization.} The rest of this paper is organized as follows: \mynewcontent{3}{In Sec. \ref{sec:bg}, we introduce the background of SBST. Our \mynewcontent{6}{method}, smart selection, is illustrated in Sec. \ref{sec:smart}. Sec. \ref{sec:eval} presents our evaluation of smart selection. We discuss the threats to validity in Sec. \ref{sec:discuss}. We present the related work in Sec. \ref{sec:related} and conclude this paper in Sec. \ref{sec:conclu}.}

\noindent\textbf{Online Artifact.} The online artifact of this paper can be found at: \mynewcontent{6}{\url{https://doi.org/10.5281/zenodo.10440612}}.

\section{Background}\label{sec:bg}
\begin{figure}[htbp]
    \centering
    \includegraphics[width=0.5\textwidth]{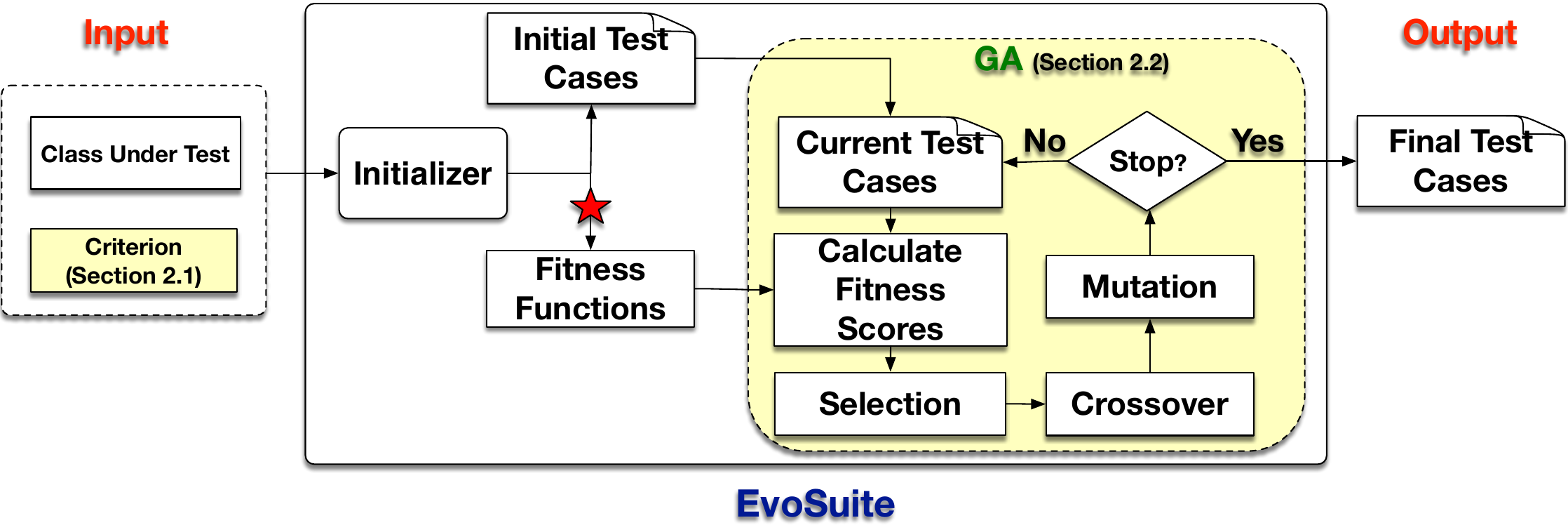}
    \caption{Overview of Unit Tests Generation in EvoSuite}
    \label{fig:evosuite}
\end{figure}
\begin{figure*}[t]
    \centering
    \includegraphics[width=\textwidth]{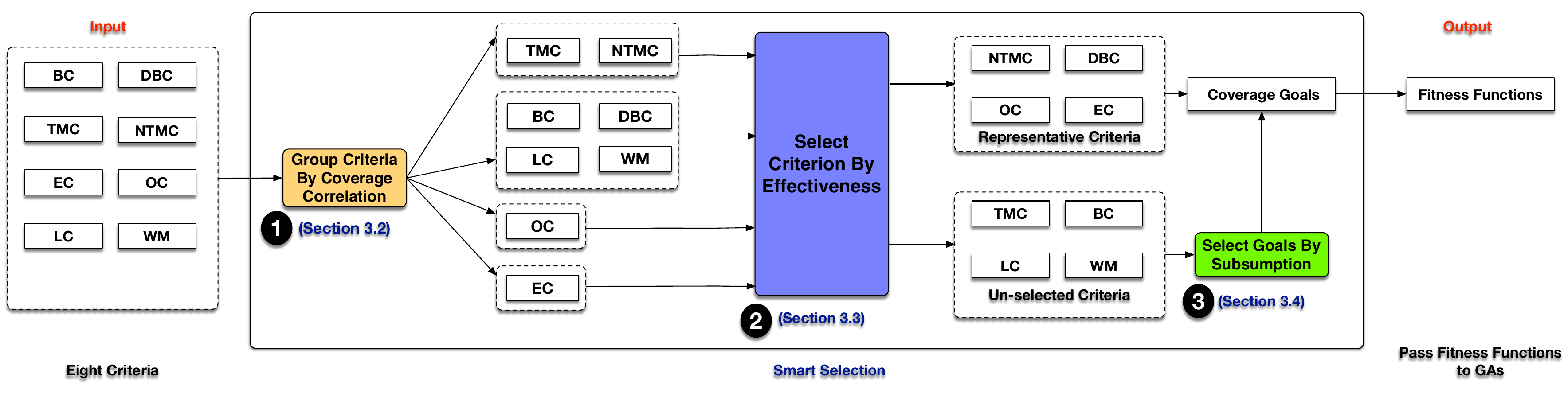}
    \caption{Overview of \textbf{Smart Selection}}
    \label{fig:ss}
\end{figure*}
\noindent\textbf{SBST and EvoSuite} \label{subsec:evosuite}
SBST generates test cases \mynewcontent{2}{using} a genetic algorithm (see Sec. \ref{subsec:ga}) by specifying the coverage criterion (see Sec. \ref{subsec:criterion}). EvoSuite~\cite{FraserEvoSuite} is \mynewcontent{2}{a} state-of-art SBST tool for Java. In this section, we \mynewcontent{2}{use} EvoSuite as an example to illustrate the key idea of SBST. Fig. \ref{fig:evosuite} shows \mynewcontent{2}{an} overview of EvoSuite. The red star in this figure is mentioned later in Sec. \ref{subsec:ss-overview}.

\noindent\textbf{Input.}
Evosuite takes two major inputs: (1) the class under test (CUT) and (2) a coverage criterion (Sec. \ref{subsec:criterion}).

\noindent\textbf{Test Generation.} 
\mynewcontent{2}{The test generation process consists of two main stages:} (1) The initializer extracts all the \mynewcontent{2}{necessary} information \mynewcontent{2}{required} by the genetic algorithm, \mynewcontent{2}{such as method signatures (including names and parameter types)}, from the CUT. Based on \mynewcontent{2}{this} information and the \mynewcontent{2}{coverage criterion}, the initializer generates initial test cases and fitness functions. \mynewcontent{2}{Typically}, each GA requires one or more specific fitness functions. A fitness function \mynewcontent{2}{evaluates} how close a test case covers a coverage goal (e.g., a branch); (2) After a specific genetic algorithm is invoked, it selects test cases based on the scores returned by fitness functions. \mynewcontent{2}{The GA then} creates new test cases \mynewcontent{2}{using} the crossover and mutation operations~\cite{FraserWhole}. \mynewcontent{2}{This process of selecting, mutating, and crossing over test cases continues until all fitness functions reach the optima or the given budget is exhausted.}

\noindent\textbf{Output.}
After running the genetic algorithm, EvoSuite outputs the final test cases.

\subsection{Coverage Criteria}\label{subsec:criterion}
\mynewcontent{6}{In this paper}, we discuss eight criteria as follows. The reason to choose these coverage criteria is that they are EvoSuite’s default criteria and \mynewcontent{2}{have been} widely used in many previous studies~\cite{Rojas2015CombiningMC, GayCombine, CamposCombineDefect}.

\noindent\textbf{Branch Coverage (BC)} 
BC checks the number of branches of conditional statements covered by a test suite. 

\noindent\textbf{Direct Branch Coverage (DBC)}
The only difference between DBC and BC is that a test case must directly invoke a public method to cover its branches. DBC treats others just as BC~\cite{Rojas2015CombiningMC}.

\noindent\textbf{Line Coverage (LC)}
LC checks the number of lines covered by a test suite.

\noindent\textbf{Weak Mutation (WM)}
WM checks how many mutants are detected by a test suite~\cite{HarmanMut, petrovic2021mutation}. A mutant is a variant of the CUT generated by a mutation operator. \mynewcontent{6}{Table~\ref{tab:amo} shows mutation operators integrated into EvoSuite~\cite{EvoSuiteMO}. For example, RC is an operator that replaces a constant value with different values.}
 \begin{table}[!htpb]
 \centering
 \caption{\mynewcontent{6}{EvoSuite's mutation operators}}\label{tab:amo}
 \begin{tabular}
 {@{\extracolsep{\fill}}l|l}
 \hline
    Operator  & Usage  \\ \hline
      RC & Replace a constant value \\ \hline
      RV & Replace a variable \\ \hline
      BOR & Replace a bitwise operator \\ \hline
     UOI & Insert unary operator \\ \hline
     AOR & Replace an arithmetic operator \\ \hline
     ROR & Replace a comparison operator \\ \hline
 \end{tabular}
  \end{table}

\noindent\textbf{Top-Level Method Coverage (TMC)}
TMC checks the number of methods covered by a test suite with a requirement: A public method is covered only when it is directly invoked by test cases.

\noindent\textbf{No-Exception Top-Level Method Coverage (NTMC)}
NTMC is TMC with an extra requirement: A method is not covered if it exits with any exceptions.

\noindent\textbf{Exception Coverage (EC)}
EC checks the number of exceptions triggered by a test suite.

\noindent\textbf{Output Coverage (OC)}
OC measures the diversity of the return values of a method. For example, a \textit{boolean} return variable's value can be \textit{true} or \textit{false}. OC's coverage is $100\%$ only if the test suite captures these two return values.

For each criterion, EvoSuite extracts a group of coverage goals from the CUT and assigns a fitness function to each coverage goal. For example, EvoSuite extracts all branches for BC, e.g., the true branch of a predicate $x==10$. A simplified fitness function of this branch can be the branch distance~\cite{McMinnSbSurvey}, $|x-10|$, which measures the distance between the actual value and the expected value of $x$ in the test case. \mynewcontent{6}{This distance value represents how far a test case is from covering the branch.} More details of these criteria and their fitness functions can be found \mynewcontent{2}{in}~\cite{Rojas2015CombiningMC}.

\subsection{Genetic Algorithms}\label{subsec:ga}
\mynewcontent{6}{In this paper}, we discuss three GAs (i.e., WS, MOSA, DynaMOSA) as follows. All of them are integrated into EvoSuite and perform well in many SBST competitions~\cite{vogl2021evosuite, Panichella2020evosuite, campos2019evosuite}. These algorithms share the same inputs and outputs but differ in how to use fitness functions.

\noindent\textbf{WS.}
WS~\cite{FraserWhole} directly evolves test suites to fit all coverage goals. Consequently, WS can exploit the collateral coverage~\cite{arcuri2011random} and not waste time on infeasible goals (e.g., dead code). Collateral coverage means that a test case generated for one goal can implicitly also cover any number of other coverage goals. Hence,  WS's fitness function is the sum of all goals' fitness functions.

\noindent\textbf{MOSA.}
WS sums fitness scores of all coverage goals as a scalar value. \mynewcontent{2}{However,} this scalar value is less monotonic and continuous than a single goal's fitness score, \mynewcontent{2}{which increases} the probability of being trapped in local optima. To overcome this limit, Panichella et al.~\cite{PanichellaMOSA} formulates SBST as a many-objective optimization problem and propose MOSA, a variant of NSGA-\uppercase\expandafter{\romannumeral2}~\cite{NSGA2}. In general, MOSA maintains a fitness vector for each test case, \mynewcontent{2}{where each item in the fitness vector represents a fitness function value for the test case}. Based on Pareto dominance ~\cite{DebMulti}, MOSA sorts and selects test cases by the fitness vectors.

\noindent\textbf{DynaMOSA.}
Based on MOSA, DynaMOSA~\cite{PanichellaDynaMOSA} adopts \mynewcontent{5}{the} control dependency graph to reduce the coverage goals evolved in search. A goal is selected to be \mynewcontent{2}{part of} the evolving process only when the branch goals it depends on are \mynewcontent{2}{already covered}. Hence, \mynewcontent{2}{DynaMOSA's fitness vectors are often smaller than those of MOSA}. Empirical studies \mynewcontent{2}{have shown} that DynaMOSA outperforms WS and MOSA~\cite{PanichellaDynaMOSA,CamposCombineDefect}.
\subsection{Combining Coverage Criteria}\label{subsec:combine}
With a single criterion (e.g., BC) alone, SBST can generate test cases that reach higher code coverage but fail to meet users' expectations~\cite{Rojas2015CombiningMC}. \mynewcontent{2}{Therefore}, Rojas et al.~\cite{Rojas2015CombiningMC} proposed \mynewcontent{2}{combining} multiple criteria to guide SBST to generate a test suite. \mynewcontent{2}{We use the example of replacing BC with the combination of eight criteria to demonstrate the changes in GAs}. Before the combination, the fitness function of WS is $f_{BC}=\sum \limits_{b \in B}f_{b}$, where B is the set of all branches. The fitness vector of MOSA/DynaMOSA is $[f_{b_{1}},...,f_{b_{n}}].$ After the combination, the fitness function of WS is $f_{BC}+...+f_{OC}$. The fitness vector of MOSA/DynaMOSA is $[f_{b_{1}},...,f_{b_{n}},...,f_{o_{1}},...,f_{o_{m}}],$ where $o_{i}$ is \mynewcontent{6}{an} output coverage goal.
\section{Smart Selection}\label{sec:smart}
\textbf{The Problem and Motivation:} 
The main side-effect of combining multiple criteria is that the generated test suite's coverage decreases for certain criteria. \mynewcontent{2}{This is due to the increase in optimization objectives, which results in a \textit{larger} search space and reduces the search weight of each objective. \mynewcontent{6}{Moreover, some criteria have fitness functions that are not monotonic and continuous, such as LC and WM (see Sec. \ref{subsec:rc}), which makes the search space \textit{more complex}, i.e., it may contain more local optimal points.} Therefore, we propose smart selection to relieve the coverage decrease by providing a smaller and easier search space for GAs.}

\subsection{Overview} \label{subsec:ss-overview}
Fig. \ref{fig:ss} shows the process: \ding{182} group criteria by coverage correlation (Sec. \ref{subsec:cc}), \ding{183} select representative criteria by effectiveness to guide SBST ( Sec. \ref{subsec:rc}), and \ding{184} select representative coverage goals from unselected criteria by subsumption relationships (Sec. \ref{subsec:rs}). The red star in Fig. \ref{fig:evosuite} shows the position of smart selection in EvoSuite.

The inputs are the eight criteria (see Sec. \ref{subsec:criterion}). The output is a subset of fitness functions, i.e., the corresponding fitness functions of our selected coverage goals. This subset is used to guide GAs.

\subsection{Grouping Criteria by Coverage Correlation}\label{subsec:cc}
The first step is clustering these eight criteria. The standard of whether two criteria can be in one group or not is: whether these two criteria have a coverage correlation. Based on this standard, we can divide these criteria into several groups. For two criteria with coverage correlation, if a test suite achieves a high coverage under one of the criteria, then this test suite may also achieve a high coverage under the other criterion. Hence, we can select one of these two criteria to guide SBST. Thus, grouping sets the scope for choosing representative criteria.

We determine that two criteria have a coverage correlation by the following rules:

\noindent$\bullet$\textbf{Rule 1:} If a previous study shows that two criteria have a coverage correlation, we adopt the conclusion:

\noindent\ding{172} \textbf{BC and WM:} Gligoric et al.~\cite{GligoricCorr} find that \enquote{branch coverage performs as well as or better than all other criteria studied, in terms of ability to predict mutation scores}. Their work shows that the average Kendall’s $\tau_{b}$ value~\cite{kendall1938new} of coverage between branch coverage and mutation testing is $0.757$. Hence, we assume that BC and WM have a coverage correlation.

\noindent\ding{173} \textbf{DBC and WM:} Since BC and WM have a coverage correlation, we assume that DBC and WM have a coverage correlation too.

\noindent\ding{174} \textbf{LC and WM:} Gligoric et al.~\cite{GligoricCorr} find that statement coverage~\cite{GligoricCorr, Rojas2015CombiningMC} can be used to predict mutation scores too. Line coverage is an alternative for statement coverage in Java since Java's bytecode instructions may not directly map to source code statements~\cite{Rojas2015CombiningMC}. Hence, we assume that LC and WM have a coverage correlation.

\noindent$\bullet$\textbf{Rule 2:} Two criteria, A and B, have a coverage correlation if they satisfy two conditions: (1) A and B have the same coverage goals; (2) A test covers a goal of A only if it covers the counterpart goal of B and it satisfies A's additional requirements:

\noindent\ding{175} \textbf{DBC and BC:} DBC is BC with an additional requirement (see Sec. \ref{subsec:criterion}).

\noindent\ding{176} \textbf{NTMC and TMC:} NTMC is TMC with an additional requirement (see Sec. \ref{subsec:criterion}).

\noindent$\bullet$\textbf{Rule 3:} Two criteria, A and B, have a coverage correlation if, for an arbitrary test suite, the relationship between these two criteria' coverage (i.e., $C_{A}$ and $C_{B}$) can be formulated as:
    \begin{equation}
    C_{B} = \Theta C_{A},
    \end{equation}
where $\Theta$ is a nonnegative random variable and $E\Theta \approx 1$:

\noindent\ding{177} \textbf{BC and LC:} Intuitively, when a branch is covered, then all lines in that branch are covered. But this is not always true. When a line exits abnormally (e.g., it throws an exception.), the subsequent lines are not covered either. First, we discuss the coverage correlation of branch and line coverage in the absence of abnormal exiting. Let $B$ be the set of branches of the CUT, $L$ be the set of lines, and $T$ be a test suite. For any $b \in B$, let $L_{b}$ be the set of lines \textbf{only} in the branch $b$ (i.e., we don't count the lines in its nested branches). Consequently, $L=\bigcup_{b \in B} L_{b}$. Let $B^{'}$ be the set of covered branches. Let $L^{'}$ be the set of covered lines. The coverage values measured by branch and line coverage are:
    \begin{align}
        C_{Branch} = \frac{|B^{'}|}{|B|}, 
        C_{Line} = \frac{|L^{'}|}{|L|} = \frac{\sum \limits_{b \in B^{'}}|L_{b}|}{\sum \limits_{b \in B}|L_{b}|}.
    \end{align}
    Hence, the relationship of $C_{Branch}$ and $C_{Line}$ is:
    \begin{align}
        \frac{C_{Line}}{C_{Branch}} & = \frac{\sum \limits_{b \in B^{'}}|L_{b}|}{\sum \limits_{b \in B}|L_{b}|} \div \frac{|B^{'}|}{|B|} =  \frac{\sum \limits_{b \in B^{'}}|L_{b}|}{|B^{'}|} \div \frac{\sum \limits_{b \in B}|L_{b}|}{|B|}.
    \end{align}
    Suppose we treat branches with different numbers of lines equally in generating $T$. Then we have:
    \begin{equation}
     \frac{\sum \limits_{b \in B^{'}}|L_{b}|}{|B^{'}|} \approx \frac{\sum \limits_{b \in B}|L_{b}|}{|B|}, 
    \end{equation}
    i.e., 
    \begin{align}
          \frac{C_{Line}}{C_{Branch}} \approx 1.
    \end{align}\label{align:bl}
    As a result, branch coverage and line coverage have a coverage correlation in the absence of abnormal exiting. With abnormal exiting, the coverage measured by line coverage decreases. Assuming that any line can exit abnormally, we can formulate the coverage relationship as:
    \begin{align}
          C_{Line} = \Theta C_{Branch},
    \end{align}
    where $\Theta$ is a random variable. \mynewcontent{6}{In this paper}, instead of analyzing $\Theta$ precisely, we only need to check whether $E\Theta \approx 1$. Previous work~\cite{Rojas2015CombiningMC} shows that, on average, when $78\%$ of branches are covered,  test suites can only find $1.75$ exceptions. Hence, we assume that $E\Theta \approx 1$, i.e., BC and LC have a coverage correlation.

\noindent\ding{178} \textbf{DBC and LC:} We assume that DBC and LC have a coverage correlation since BC and LC have a coverage correlation.

\noindent\textbf{Output.} We cluster the eight criteria into four groups: (1) BC, DBC, LC, and WM; (2) TMC and NTMC; (3) EC; and  (4) OC.
\subsection{Selecting Representative Criterion by Effectiveness to guide SBST}\label{subsec:rc}
In this step,  among the criteria in each group, we select a criterion to represent the others. The criteria within a group differ in the ability to guide SBST. If we only select one criterion with the best effectiveness to guide SBST, SBST will be more efficient in generating unit tests. To select the best criterion to guide SBST in each group, we need to compare \mynewcontent{2}{the effectiveness of the criteria} in guiding SBST. A criterion's effectiveness in guiding SBST largely depends on the continuity of monotonicity of its fitness functions~\cite{LinGraph, lin2020recovering}. Hence, we need to analyze and compare \mynewcontent{2}{the fitness functions of the criteria}.

\noindent\textbf{Group1: BC, DBC, LC, and WM.} 
We use branch coverage as the baseline and divide them into three pairs for discussion. The reason to use branch coverage as the baseline is that branch coverage has been widely used to guide unit test generation~\cite{FraserWhole, PanichellaMOSA, PanichellaDynaMOSA} due to the monotonic continuity of its fitness functions. For a branch goal $b$ and a test case $t$,  its fitness function is~\cite{FraserWhole}:
\begin{equation}
    f_{bc}(b, t)=\left\{
\begin{array}{lcl}
    0 & & {\textnormal{if~the~branch}} \\
    & & {\textnormal{has~been~covered,}} \\
    \nu(d(b, t)) & & {\textnormal{if~the~predicate~has~been}} \\
    & & {\textnormal{executed~at~least~twice,}} \\
    1 & & {\textnormal{otherwise,}}
\end{array}\right.\label{fitness:branch}
\end{equation}
where $\nu(x)$ is a normalizing function in $[0,1]$ (e.g., $\nu(x)=x/(x+1)$). $d(b,t)$ is a function to provide a branch distance to describe how far a test case covers this goal~\cite{McMinnSbSurvey}. To avoid an oscillate situation of a predicate~\cite{FraserWhole}, $f_{bc}(b,t)$ uses $\nu(d(b, t))$ only when a predicate is executed at least twice. \mynewcontent{1}{Note that the final fitness function used by many previous studies~\cite{PanichellaDynaMOSA,vogl2021evosuite} is $f_{bc}$ plus the approach level (AL), the number of control dependencies from a test's execution trace to the target~\cite{PanichellaDynaMOSA}. We omit AL in comparing these criteria\mynewcontent{6}{'s} fitness functions since its usage is the same in different criteria.}

WS uses the sum of all fitness functions as one fitness function (Sec. \ref{subsec:ga}). Hence, for WS, branch coverage's fitness function is:
\begin{equation}
    d_{1}(b,T) = \min{\{f_{bc}(b,t)|t\in T\}}, 
\end{equation}
\begin{equation}
    f_{BC}(T) = \sum_{b\in B}{d_{1}(b, T)}\label{fitness:BC},
\end{equation}
where $B$ denotes all branches of the CUT.

\noindent$\bullet$\textbf{BC vs. LC.}
Based on line coverage's definition (see Sec. \ref{subsec:criterion}), a line $l$'s fitness function can be:
\begin{equation}
f_{lc}(l, t) = \left\{\
\begin{array}{ll}
    0 &  {\textnormal{if~the~line~has~been}} \\
    &  {\textnormal{covered,}} \\
     1  &{\textnormal{otherwise.}} 
\end{array}
\right.\label{fitness:origin-line}
\end{equation}
For WS, line coverage's fitness function is:
\begin{equation}
    f_{LC}(T) = \nu(|L|-|CL|)\label{fitness:origin-LC},
\end{equation}
where $L$ is the set of all lines and $CL$ is the set of covered lines. 

These two fitness functions are not continuous and monotonic since they only tell whether the lines are covered. To overcome this limit, EvoSuite uses branch coverage's fitness functions to augment line coverage's fitness functions~\cite{Rojas2015CombiningMC}. A line $l$'s fitness function is:
\begin{equation}
f_{lc}(l, B, t) = \left\{\
\begin{array}{ll}
    0 &  {\textnormal{if~the~line~has}} \\

    &  {\textnormal{been~covered,}} \\
     1 +\min{\{f_{bc}(b, t)|b \in B\}}  &{\textnormal{otherwise,}} 
\end{array}
\right.\label{fitness:line}
\end{equation}
where $B$ is the set of branches that $l$ depends on~\cite{PanichellaDynaMOSA}. For WS, line coverage's fitness function is:
\begin{equation}
    f_{LC}(T) = \nu(|L|-|CL|) + f_{BC}(T).\label{fitness:LC}
\end{equation}

We call Equation \ref{fitness:origin-line} and \ref{fitness:origin-LC} \textit{def-based} (definition-based) fitness functions and call Equation \ref{fitness:line} and \ref{fitness:LC} \textit{augmented} fitness functions.

Firstly, we compare branch coverage's fitness functions with line coverage's \textit{def-based} fitness functions.  Line coverage's \textit{def-based} fitness functions are not continuous and monotonic since they only tell whether the lines are covered. \mynewcontent{2}{Therefore}, branch coverage \mynewcontent{2}{is more effective than line coverage in guiding} SBST when we use line coverage's \textit{def-based} fitness functions. After the augmentation, line coverage's \textit{def-based} fitness functions disturb the continuity and monotonicity of branch coverage's fitness functions, undermining branch coverage's effectiveness to guide SBST. As a result, BC is better than LC in the effectiveness to guide SBST.

\noindent$\bullet$\textbf{BC vs. WM.}
Based on weak mutation's definition (see Sec. \ref{subsec:criterion}), a mutant's fitness function is:
\begin{equation}
f_{wm}(\mu, t) = \left\{\
\begin{array}{ll}
   1 & {\textnormal{if~mutant~$\mu$}} \\
    & {\textnormal{was not reached,}} \\
    \nu(id(\mu, t)) & {\textnormal{if~mutant~$\mu$}} \\
    & {\textnormal{was~reached,}}
\end{array}\right.\label{fitness:origin-wm}
\end{equation}
where $id(\mu, t)$ is the infection distance function. It describes how distantly a test case triggers a mutant's different state from the source code. Different mutation operators have different infection distance functions~\cite{FraserMutation}. A mutant's fitness function is always $1$ unless a test case reaches it (i.e., the mutated line is covered). Hence, like line coverage, EvoSuite uses the same way to augment weak mutation's fitness functions~\cite{PanichellaDynaMOSA, FraserMutation}. As the conclusion of comparing BC and LC, BC is better than WM in the effectiveness to guide SBST.

\noindent$\bullet$\textbf{BC vs. DBC.}
Direct branch coverage \mynewcontent{2}{(DBC)} is branch coverage with an extra requirement: A test case must directly invoke a public method to cover its branches. Based on branch coverage's fitness function, \mynewcontent{2}{we can obtain DBC’s one}: For a branch in a public method, when the method is not invoked directly, the fitness function always returns 1. Otherwise, the fitness function is the same as branch coverage's one. It is easy for SBST to generate a test case that invokes a public method directly. Hence, \mynewcontent{2}{we consider BC and DBC to be nearly equal in guiding SBST}.

\noindent\textbf{Order of Group1.}
Above all, we get a rough order of this group: (1)BC and DBC; (2) LC and WM. Since we only need one representative, the rough order satisfies our need.

\noindent\textbf{The Representative Criterion of Group1.}
We choose DBC to represent this group instead of BC. The reason is:  When a test case covers a goal of DBC, the test case covers the counterpart of BC. As a result, DBC can fully represent BC. The opposite may not hold.

\noindent\textbf{Group2: TMC and NTMC.} 
Like the relationship between branch coverage and direct branch coverage, no-exce. top-level method coverage is top-level method coverage with an extra requirement: A method must be invoked without triggering exceptions.

\noindent\textbf{The Representative Criterion of Group2.}
We choose NTMC to represent this group. The reason is the same as why we choose DBC to represent group 1: NTMC can fully represent TMC. The opposite does not hold.

\noindent\textbf{Group3: EC and Group 4: OC.} 
Since group 3 only contains EC, we choose EC to represent group 3. Similarly, we choose OC to represent group 4.

\noindent\textbf{Output.}
The representative criteria are DBC, NTMC, EC, and OC. 
\subsection{Selecting Representative Coverage Goals by Subsumption Relationships}\label{subsec:rs}
After selecting the representative criteria in the previous step, there are four unselected criteria: LC, WM, BC, and TMC. To keep property consistency for each unselected criterion, we select a subset from its coverage goals. This subset can represent all properties required by this criterion, ensuring GA archives~\cite{PanichellaDynaMOSA} those tests that fulfill the properties beyond the representative criteria. We have another requirement for these subsets: they should be as small as possible. These unselected criteria' fitness functions are less continuous and monotonic than the ones of the representative criteria (see Sec. \ref{subsec:rc}). \mynewcontent{2}{Therefore}, to minimize the negative effects on guiding SBST, these subsets should be as small as possible.

Two coverage goals, $G_{1}$ and $G_{2}$, having the subsumption relationship denotes that if a test suite covers one coverage goal, it must cover another goal. Specifically, $G_{1}$ subsuming $G_{2}$ represents that if a test suite covers $G_{1}$, it must cover $G_{2}$. According to this definition, for a criterion, if the coverage goals not subsumed by others are covered, all coverage goals are covered. Hence, \mynewcontent{6}{these} coverage goals form the desired subset.

\noindent\textbf{LC.}
For the lines in a basic block, the last line subsumes others. Hence, these last lines of all basic blocks form the desired subset. Since Sec. \ref{subsec:cc} shows that BC/DBC and LC have a strong coverage correlation and DBC is the representative criterion, we do a tradeoff to shrink this subset: We add an integer parameter \textit{lineThreshold}. If a basic block's lines are less than \textit{lineThreshold}, we skip it. \mynewcontent{6}{In this paper}, we set \textit{lineThreshold} as $8$ (Sec. \ref{sec:discuss} discusses \mynewcontent{5}{this decision}).

\noindent\textbf{WM.}
The process to extract the subset from \mynewcontent{5}{all mutants considered by weak mutation} can be divided into three parts: \noindent\ding{172} We select the key operators \mynewcontent{5}{from all of the mutation operators implemented in EvoSuite}; \noindent\ding{173} From the key operators we filter out the \textbf{equal-to-line} operators; \noindent\ding{174} For the remaining operators, we select the subsuming mutants by following the previous work~\cite{GheyiSub}.
  
\noindent\ding{172} \textbf{Select Key Operators.}
Offutt et al.~\cite{OffuttOperators} find that five key operators achieve $99.5\%$ mutation score. They are UOI, AOR, ROR, ABS, and LCR. EvoSuite does not implement LCR (an operator that replaces the logical connectors) and ABS (an operator that inserts absolute values)~\cite{EvoSuiteMO, FraserMutation}. \mynewcontent{6}{In their work~\cite{OffuttOperators}, Offutt et al. discussed $22$ mutation operators, including all of the operators used in EvoSuite (see Table \ref{tab:amo}) except for BOR.} Hence, we select \mynewcontent{6}{four} operators: \mynewcontent{6}{BOR,} UOI, AOR, and ROR (see Table \ref{tab:mo}).
 \begin{table}[!htpb]
 \centering
 \caption{\mynewcontent{6}{Key mutation operators}}\label{tab:mo}
 \begin{tabular}
 {@{\extracolsep{\fill}}l|l}
 \hline
    Operator  & Usage  \\ \hline
     \mynewcontent{6}{BOR} & Replace a bitwise operator \\ \hline
     UOI  & Insert unary operator \\ \hline
     AOR & Replace an arithmetic operator \\ \hline
     ROR & Replace a comparison operator \\ \hline
 \end{tabular}
  \end{table}

\begin{table*}[tbp]
    \centering
    \footnotesize

    \caption{Overview of Java projects and classes in our evaluation}
    \label{tab:classes}

   \begin{tabular}{lllllllll}
\toprule
Project & Package & Classes & \nrcols{3}{Branch} & \nrcols{3}{\mynewcontent{6}{Statement}} \\
\cmidrule(l{1pt}r{1pt}){4-6}
\cmidrule(l{1pt}r{1pt}){7-9}
& & & $25\%$ & mean & $75\%$ & $25\%$ & mean & $75\%$  \\
\midrule

 a4j & \priostrategy{-} & \significant{\num{1}} & \significant{\num{125}} & \significant{\num{125}} & \significant{\num{125}}  & \significant{\num{2038}} & \significant{\num{2038}} & \significant{\num{2038}}  \\

\tabprojectlinespace 
 
 apbsmem & \priostrategy{-} & \significant{\num{1}} & \significant{\num{390}} & \significant{\num{390}} & \significant{\num{390}}  & \significant{\num{4323}} & \significant{\num{4323}} & \significant{\num{4323}}  \\

\tabprojectlinespace 
 
 at-robots2-j & \priostrategy{-} & \significant{\num{1}} & \significant{\num{125}} & \significant{\num{125}} & \significant{\num{125}}  & \significant{\num{832}} & \significant{\num{832}} & \significant{\num{832}}  \\

\tabprojectlinespace 
 
 battlecry & \priostrategy{-} & \significant{\num{2}} & \significant{\num{128}} & \significant{\num{179}} & \significant{\num{230}}  & \significant{\num{1767}} & \significant{\num{1947}} & \significant{\num{2126}}  \\

\tabprojectlinespace 
 
 biff & \priostrategy{-} & \significant{\num{1}} & \significant{\num{817}} & \significant{\num{817}} & \significant{\num{817}}  & \significant{\num{7012}} & \significant{\num{7012}} & \significant{\num{7012}}  \\

\tabprojectlinespace 
 
 caloriecount & \priostrategy{-} & \significant{\num{1}} & \significant{\num{232}} & \significant{\num{232}} & \significant{\num{232}}  & \significant{\num{1425}} & \significant{\num{1425}} & \significant{\num{1425}}  \\

\tabprojectlinespace 
 
 celwars2009 & \priostrategy{-} & \significant{\num{1}} & \significant{\num{360}} & \significant{\num{360}} & \significant{\num{360}}  & \significant{\num{2139}} & \significant{\num{2139}} & \significant{\num{2139}}  \\

\tabprojectlinespace 
 
 checkstyle & \priostrategy{-} & \significant{\num{1}} & \significant{\num{50}} & \significant{\num{50}} & \significant{\num{50}}  & \significant{\num{412}} & \significant{\num{412}} & \significant{\num{412}}  \\

\tabprojectlinespace 
 
 classviewer & \priostrategy{-} & \significant{\num{3}} & \significant{\num{114}} & \significant{\num{154}} & \significant{\num{202}}  & \significant{\num{671}} & \significant{\num{1219}} & \significant{\num{1665}}  \\

\tabprojectlinespace 
 
 commons-cli & \priostrategy{-} & \significant{\num{2}} & \significant{\num{139}} & \significant{\num{146}} & \significant{\num{152}}  & \significant{\num{696}} & \significant{\num{788}} & \significant{\num{880}}  \\

\tabprojectlinespace 
 
 commons-codec & \priostrategy{-} & \significant{\num{1}} & \significant{\num{504}} & \significant{\num{504}} & \significant{\num{504}}  & \significant{\num{2385}} & \significant{\num{2385}} & \significant{\num{2385}}  \\

\tabprojectlinespace 
 
 commons-collections & \priostrategy{-} & \significant{\num{2}} & \significant{\num{122}} & \significant{\num{155}} & \significant{\num{188}}  & \significant{\num{351}} & \significant{\num{426}} & \significant{\num{501}}  \\

\tabprojectlinespace 
 
 commons-lang & \priostrategy{-} & \significant{\num{12}} & \significant{\num{137}} & \significant{\num{353}} & \significant{\num{370}}  & \significant{\num{868}} & \significant{\num{1759}} & \significant{\num{2037}}  \\

\tabprojectlinespace 
 
 commons-math & \priostrategy{-} & \significant{\num{17}} & \significant{\num{75}} & \significant{\num{120}} & \significant{\num{149}}  & \significant{\num{630}} & \significant{\num{1105}} & \significant{\num{1350}}  \\

\tabprojectlinespace 
 
 corina & \priostrategy{-} & \significant{\num{1}} & \significant{\num{55}} & \significant{\num{55}} & \significant{\num{55}}  & \significant{\num{282}} & \significant{\num{282}} & \significant{\num{282}}  \\

\tabprojectlinespace 
 
 dcparseargs & \priostrategy{-} & \significant{\num{1}} & \significant{\num{80}} & \significant{\num{80}} & \significant{\num{80}}  & \significant{\num{571}} & \significant{\num{571}} & \significant{\num{571}}  \\

\tabprojectlinespace 
 
 dsachat & \priostrategy{-} & \significant{\num{2}} & \significant{\num{76}} & \significant{\num{83}} & \significant{\num{90}}  & \significant{\num{868}} & \significant{\num{880}} & \significant{\num{892}}  \\

\tabprojectlinespace 
 
 dvd-homevideo & \priostrategy{-} & \significant{\num{2}} & \significant{\num{60}} & \significant{\num{68}} & \significant{\num{76}}  & \significant{\num{1269}} & \significant{\num{1308}} & \significant{\num{1347}}  \\

\tabprojectlinespace 
 
 feudalismgame & \priostrategy{-} & \significant{\num{1}} & \significant{\num{788}} & \significant{\num{788}} & \significant{\num{788}}  & \significant{\num{4873}} & \significant{\num{4873}} & \significant{\num{4873}}  \\

\tabprojectlinespace 
 
 fim1 & \priostrategy{-} & \significant{\num{1}} & \significant{\num{73}} & \significant{\num{73}} & \significant{\num{73}}  & \significant{\num{1140}} & \significant{\num{1140}} & \significant{\num{1140}}  \\

\tabprojectlinespace 
 
 firebird & \priostrategy{-} & \significant{\num{2}} & \significant{\num{123}} & \significant{\num{147}} & \significant{\num{171}}  & \significant{\num{666}} & \significant{\num{751}} & \significant{\num{836}}  \\

\tabprojectlinespace 
 
 fixsuite & \priostrategy{-} & \significant{\num{1}} & \significant{\num{74}} & \significant{\num{74}} & \significant{\num{74}}  & \significant{\num{631}} & \significant{\num{631}} & \significant{\num{631}}  \\

\tabprojectlinespace 
 
 fps370 & \priostrategy{-} & \significant{\num{1}} & \significant{\num{70}} & \significant{\num{70}} & \significant{\num{70}}  & \significant{\num{668}} & \significant{\num{668}} & \significant{\num{668}}  \\

\tabprojectlinespace 
 
 freemind & \priostrategy{-} & \significant{\num{1}} & \significant{\num{208}} & \significant{\num{208}} & \significant{\num{208}}  & \significant{\num{1621}} & \significant{\num{1621}} & \significant{\num{1621}}  \\

\tabprojectlinespace 
 
 gfarcegestionfa & \priostrategy{-} & \significant{\num{2}} & \significant{\num{87}} & \significant{\num{99}} & \significant{\num{111}}  & \significant{\num{664}} & \significant{\num{826}} & \significant{\num{988}}  \\

\tabprojectlinespace 
 
 glengineer & \priostrategy{-} & \significant{\num{1}} & \significant{\num{115}} & \significant{\num{115}} & \significant{\num{115}}  & \significant{\num{658}} & \significant{\num{658}} & \significant{\num{658}}  \\

\tabprojectlinespace 
 
 guava & \priostrategy{-} & \significant{\num{8}} & \significant{\num{84}} & \significant{\num{143}} & \significant{\num{160}}  & \significant{\num{213}} & \significant{\num{490}} & \significant{\num{742}}  \\

\tabprojectlinespace 
  
 httpanalyzer & \priostrategy{-} & \significant{\num{1}} & \significant{\num{56}} & \significant{\num{56}} & \significant{\num{56}}  & \significant{\num{656}} & \significant{\num{656}} & \significant{\num{656}}  \\

\tabprojectlinespace 
 
 ifx-framework & \priostrategy{-} & \significant{\num{1}} & \significant{\num{72}} & \significant{\num{72}} & \significant{\num{72}}  & \significant{\num{462}} & \significant{\num{462}} & \significant{\num{462}}  \\

\tabprojectlinespace 
 
 inspirento & \priostrategy{-} & \significant{\num{1}} & \significant{\num{95}} & \significant{\num{95}} & \significant{\num{95}}  & \significant{\num{649}} & \significant{\num{649}} & \significant{\num{649}}  \\

\tabprojectlinespace 
 
 io-project & \priostrategy{-} & \significant{\num{1}} & \significant{\num{66}} & \significant{\num{66}} & \significant{\num{66}}  & \significant{\num{282}} & \significant{\num{282}} & \significant{\num{282}}  \\

\tabprojectlinespace 
 
 ipcalculator & \priostrategy{-} & \significant{\num{2}} & \significant{\num{67}} & \significant{\num{79}} & \significant{\num{91}}  & \significant{\num{1164}} & \significant{\num{1252}} & \significant{\num{1340}}  \\

\tabprojectlinespace 
 
 javaml & \priostrategy{-} & \significant{\num{3}} & \significant{\num{51}} & \significant{\num{52}} & \significant{\num{54}}  & \significant{\num{390}} & \significant{\num{409}} & \significant{\num{424}}  \\

\tabprojectlinespace 
 
 javathena & \priostrategy{-} & \significant{\num{1}} & \significant{\num{255}} & \significant{\num{255}} & \significant{\num{255}}  & \significant{\num{2278}} & \significant{\num{2278}} & \significant{\num{2278}}  \\

\tabprojectlinespace 
 
 javaviewcontrol & \priostrategy{-} & \significant{\num{2}} & \significant{\num{760}} & \significant{\num{1298}} & \significant{\num{1835}}  & \significant{\num{3078}} & \significant{\num{4394}} & \significant{\num{5710}}  \\

\tabprojectlinespace 
 
 jclo & \priostrategy{-} & \significant{\num{1}} & \significant{\num{133}} & \significant{\num{133}} & \significant{\num{133}}  & \significant{\num{987}} & \significant{\num{987}} & \significant{\num{987}}  \\

\tabprojectlinespace 
 
 jcvi-javacommon & \priostrategy{-} & \significant{\num{1}} & \significant{\num{61}} & \significant{\num{61}} & \significant{\num{61}}  & \significant{\num{240}} & \significant{\num{240}} & \significant{\num{240}}  \\

\tabprojectlinespace 
 
 jdom & \priostrategy{-} & \significant{\num{5}} & \significant{\num{63}} & \significant{\num{110}} & \significant{\num{91}}  & \significant{\num{510}} & \significant{\num{591}} & \significant{\num{583}}  \\

\tabprojectlinespace 
 
 jfreechart & \priostrategy{-} & \significant{\num{8}} & \significant{\num{102}} & \significant{\num{263}} & \significant{\num{242}}  & \significant{\num{754}} & \significant{\num{1881}} & \significant{\num{1843}}  \\

\tabprojectlinespace 
 
 jiggler & \priostrategy{-} & \significant{\num{3}} & \significant{\num{65}} & \significant{\num{82}} & \significant{\num{96}}  & \significant{\num{465}} & \significant{\num{711}} & \significant{\num{836}}  \\

\tabprojectlinespace 
 
 jipa & \priostrategy{-} & \significant{\num{1}} & \significant{\num{134}} & \significant{\num{134}} & \significant{\num{134}}  & \significant{\num{783}} & \significant{\num{783}} & \significant{\num{783}}  \\

\tabprojectlinespace 
 
 jiprof & \priostrategy{-} & \significant{\num{4}} & \significant{\num{84}} & \significant{\num{451}} & \significant{\num{818}}  & \significant{\num{628}} & \significant{\num{2612}} & \significant{\num{4344}}  \\

\tabprojectlinespace 
 
 jmca & \priostrategy{-} & \significant{\num{3}} & \significant{\num{207}} & \significant{\num{707}} & \significant{\num{961}}  & \significant{\num{1122}} & \significant{\num{2307}} & \significant{\num{2978}}  \\

\tabprojectlinespace 
  
 jopenchart & \priostrategy{-} & \significant{\num{1}} & \significant{\num{92}} & \significant{\num{92}} & \significant{\num{92}}  & \significant{\num{1139}} & \significant{\num{1139}} & \significant{\num{1139}}  \\

\tabprojectlinespace 

hadoop & \priostrategy{org} & \significant{\num{45}} & \significant{\num{70}} & \significant{\num{151}} & \significant{\num{156}}  & \significant{\num{121}} & \significant{\num{271}} & \significant{\num{313}}  \\

\tabprojectlinespace 
 
 hadoop & \priostrategy{org.apache} & \significant{\num{65}} & \significant{\num{70}} & \significant{\num{343}} & \significant{\num{158}}  & \significant{\num{105}} & \significant{\num{565}} & \significant{\num{295}}  \\

\tabprojectlinespace 
 
 hadoop & \priostrategy{org.apache.hadoop} & \significant{\num{40}} & \significant{\num{62}} & \significant{\num{220}} & \significant{\num{194}}  & \significant{\num{133}} & \significant{\num{403}} & \significant{\num{331}}  \\

\tabprojectlinespace 
 
 hadoop & \priostrategy{org.apache.hadoop.crypto} & \significant{\num{1}} & \significant{\num{82}} & \significant{\num{82}} & \significant{\num{82}}  & \significant{\num{337}} & \significant{\num{337}} & \significant{\num{337}}  \\

\tabprojectlinespace 
 
 hadoop & \priostrategy{org.apache.hadoop.mapreduce} & \significant{\num{3}} & \significant{\num{78}} & \significant{\num{90}} & \significant{\num{101}}  & \significant{\num{184}} & \significant{\num{204}} & \significant{\num{218}}  \\

\tabprojectlinespace 
 
 hadoop & \priostrategy{org.apache.hadoop.mapreduce.v2.api} & \significant{\num{3}} & \significant{\num{80}} & \significant{\num{91}} & \significant{\num{110}}  & \significant{\num{136}} & \significant{\num{160}} & \significant{\num{196}}  \\

\tabprojectlinespace 
 
 hadoop & \priostrategy{org.apache.hadoop.security} & \significant{\num{1}} & \significant{\num{74}} & \significant{\num{74}} & \significant{\num{74}}  & \significant{\num{159}} & \significant{\num{159}} & \significant{\num{159}}  \\

\tabprojectlinespace 
 
 hadoop & \priostrategy{org.apache.hadoop.thirdparty.com} & \significant{\num{18}} & \significant{\num{71}} & \significant{\num{108}} & \significant{\num{122}}  & \significant{\num{108}} & \significant{\num{154}} & \significant{\num{174}}  \\

\tabprojectlinespace 
 
 hadoop & \priostrategy{org.apache.hadoop.thirdparty.com.google} & \significant{\num{4}} & \significant{\num{74}} & \significant{\num{114}} & \significant{\num{127}}  & \significant{\num{141}} & \significant{\num{196}} & \significant{\num{239}}  \\

\tabprojectlinespace 
 
 hadoop & \priostrategy{org.apache.hadoop.thirdparty.org} & \significant{\num{1}} & \significant{\num{119}} & \significant{\num{119}} & \significant{\num{119}}  & \significant{\num{163}} & \significant{\num{163}} & \significant{\num{163}}  \\

\tabprojectlinespace 
 
 hadoop & \priostrategy{org.apache.hadoop.yarn} & \significant{\num{22}} & \significant{\num{59}} & \significant{\num{103}} & \significant{\num{117}}  & \significant{\num{99}} & \significant{\num{186}} & \significant{\num{208}}  \\

\tabprojectlinespace 
 
 hadoop & \priostrategy{org.apache.hadoop.yarn.api} & \significant{\num{14}} & \significant{\num{56}} & \significant{\num{107}} & \significant{\num{134}}  & \significant{\num{106}} & \significant{\num{183}} & \significant{\num{212}}  \\

\tabprojectlinespace 
 
 hadoop & \priostrategy{org.apache.hadoop.yarn.server} & \significant{\num{11}} & \significant{\num{55}} & \significant{\num{105}} & \significant{\num{101}}  & \significant{\num{98}} & \significant{\num{193}} & \significant{\num{191}}  \\

\tabprojectlinespace 
 
 hadoop & \priostrategy{org.apache.hadoop.yarn.server.api} & \significant{\num{4}} & \significant{\num{95}} & \significant{\num{117}} & \significant{\num{138}}  & \significant{\num{176}} & \significant{\num{208}} & \significant{\num{242}}  \\

\tabprojectlinespace 
 
 hadoop & \priostrategy{org.apache.hadoop.yarn.server.applicationhistoryservice} & \significant{\num{1}} & \significant{\num{55}} & \significant{\num{55}} & \significant{\num{55}}  & \significant{\num{95}} & \significant{\num{95}} & \significant{\num{95}}  \\

\tabprojectlinespace 
 
 hadoop & \priostrategy{org.apache.hadoop.yarn.server.federation.store} & \significant{\num{1}} & \significant{\num{69}} & \significant{\num{69}} & \significant{\num{69}}  & \significant{\num{109}} & \significant{\num{109}} & \significant{\num{109}}  \\

\tabprojectlinespace 
 
 hadoop & \priostrategy{org.apache.hadoop.yarn.server.nodemanager} & \significant{\num{2}} & \significant{\num{84}} & \significant{\num{91}} & \significant{\num{98}}  & \significant{\num{170}} & \significant{\num{187}} & \significant{\num{204}}  \\

\tabprojectlinespace 
 
 hadoop & \priostrategy{org.apache.hadoop.yarn.server.nodemanager.containermanager} & \significant{\num{4}} & \significant{\num{59}} & \significant{\num{60}} & \significant{\num{63}}  & \significant{\num{99}} & \significant{\num{112}} & \significant{\num{131}}  \\

\tabprojectlinespace

\end{tabular}

\end{table*}

\begin{table*}[tbp]
\ContinuedFloat
    \centering
    \footnotesize
    \caption{Continued}
    \label{tab:classes_2}

   \begin{tabular}{lp{8cm}lllllll}
\toprule
Project & Package & Classes & \nrcols{3}{Branch} & \nrcols{3}{Statement} \\
\cmidrule(l{1pt}r{1pt}){4-6}
\cmidrule(l{1pt}r{1pt}){7-9}
& & & $25\%$ & mean & $75\%$ & $25\%$ & mean & $75\%$  \\
\midrule
 
 hadoop & \priostrategy{org.apache.hadoop.yarn.server.resourcemanager} & \significant{\num{2}} & \significant{\num{106}} & \significant{\num{127}} & \significant{\num{147}}  & \significant{\num{246}} & \significant{\num{281}} & \significant{\num{315}}  \\

\tabprojectlinespace
 
 jsecurity & \priostrategy{-} & \significant{\num{1}} & \significant{\num{170}} & \significant{\num{170}} & \significant{\num{170}}  & \significant{\num{725}} & \significant{\num{725}} & \significant{\num{725}}  \\

\tabprojectlinespace 
 
 lagoon & \priostrategy{-} & \significant{\num{2}} & \significant{\num{63}} & \significant{\num{64}} & \significant{\num{64}}  & \significant{\num{758}} & \significant{\num{841}} & \significant{\num{924}}  \\

\tabprojectlinespace 
 
 lhamacaw & \priostrategy{-} & \significant{\num{1}} & \significant{\num{70}} & \significant{\num{70}} & \significant{\num{70}}  & \significant{\num{714}} & \significant{\num{714}} & \significant{\num{714}}  \\

\tabprojectlinespace 
 
 liferay & \priostrategy{-} & \significant{\num{1}} & \significant{\num{78}} & \significant{\num{78}} & \significant{\num{78}}  & \significant{\num{462}} & \significant{\num{462}} & \significant{\num{462}}  \\

\tabprojectlinespace 
 
 lilith & \priostrategy{-} & \significant{\num{1}} & \significant{\num{134}} & \significant{\num{134}} & \significant{\num{134}}  & \significant{\num{496}} & \significant{\num{496}} & \significant{\num{496}}  \\

\tabprojectlinespace 
 
 newzgrabber & \priostrategy{-} & \significant{\num{2}} & \significant{\num{125}} & \significant{\num{173}} & \significant{\num{220}}  & \significant{\num{1110}} & \significant{\num{1257}} & \significant{\num{1404}}  \\

\tabprojectlinespace 
 
 noen & \priostrategy{-} & \significant{\num{2}} & \significant{\num{68}} & \significant{\num{69}} & \significant{\num{70}}  & \significant{\num{386}} & \significant{\num{456}} & \significant{\num{526}}  \\

\tabprojectlinespace 
 
 objectexplorer & \priostrategy{-} & \significant{\num{1}} & \significant{\num{175}} & \significant{\num{175}} & \significant{\num{175}}  & \significant{\num{987}} & \significant{\num{987}} & \significant{\num{987}}  \\

\tabprojectlinespace 
 
 openhre & \priostrategy{-} & \significant{\num{1}} & \significant{\num{50}} & \significant{\num{50}} & \significant{\num{50}}  & \significant{\num{220}} & \significant{\num{220}} & \significant{\num{220}}  \\

\tabprojectlinespace 
 
 quickserver & \priostrategy{-} & \significant{\num{2}} & \significant{\num{70}} & \significant{\num{71}} & \significant{\num{72}}  & \significant{\num{553}} & \significant{\num{618}} & \significant{\num{683}}  \\

\tabprojectlinespace 
 
 resources4j & \priostrategy{-} & \significant{\num{1}} & \significant{\num{176}} & \significant{\num{176}} & \significant{\num{176}}  & \significant{\num{1539}} & \significant{\num{1539}} & \significant{\num{1539}}  \\

\tabprojectlinespace 
 
 saxpath & \priostrategy{-} & \significant{\num{2}} & \significant{\num{162}} & \significant{\num{269}} & \significant{\num{376}}  & \significant{\num{344}} & \significant{\num{574}} & \significant{\num{804}}  \\

\tabprojectlinespace 
 
 schemaspy & \priostrategy{-} & \significant{\num{1}} & \significant{\num{380}} & \significant{\num{380}} & \significant{\num{380}}  & \significant{\num{2241}} & \significant{\num{2241}} & \significant{\num{2241}}  \\

\tabprojectlinespace 
 
 shop & \priostrategy{-} & \significant{\num{3}} & \significant{\num{100}} & \significant{\num{131}} & \significant{\num{153}}  & \significant{\num{833}} & \significant{\num{998}} & \significant{\num{1145}}  \\

\tabprojectlinespace 
 
 sugar & \priostrategy{-} & \significant{\num{1}} & \significant{\num{51}} & \significant{\num{51}} & \significant{\num{51}}  & \significant{\num{484}} & \significant{\num{484}} & \significant{\num{484}}  \\

\tabprojectlinespace 
 
 summa & \priostrategy{-} & \significant{\num{1}} & \significant{\num{372}} & \significant{\num{372}} & \significant{\num{372}}  & \significant{\num{1906}} & \significant{\num{1906}} & \significant{\num{1906}}  \\

\tabprojectlinespace 
 
 sweethome3d & \priostrategy{-} & \significant{\num{2}} & \significant{\num{271}} & \significant{\num{387}} & \significant{\num{503}}  & \significant{\num{1544}} & \significant{\num{2152}} & \significant{\num{2760}}  \\

\tabprojectlinespace 
 
 trove & \priostrategy{-} & \significant{\num{9}} & \significant{\num{77}} & \significant{\num{137}} & \significant{\num{255}}  & \significant{\num{378}} & \significant{\num{652}} & \significant{\num{852}}  \\

\tabprojectlinespace 
 
 twfbplayer & \priostrategy{-} & \significant{\num{2}} & \significant{\num{94}} & \significant{\num{115}} & \significant{\num{135}}  & \significant{\num{631}} & \significant{\num{839}} & \significant{\num{1047}}  \\

\tabprojectlinespace 
 
 twitter4j & \priostrategy{-} & \significant{\num{6}} & \significant{\num{64}} & \significant{\num{117}} & \significant{\num{115}}  & \significant{\num{399}} & \significant{\num{1493}} & \significant{\num{823}}  \\

\tabprojectlinespace 
 
 vuze & \priostrategy{-} & \significant{\num{1}} & \significant{\num{134}} & \significant{\num{134}} & \significant{\num{134}}  & \significant{\num{675}} & \significant{\num{675}} & \significant{\num{675}}  \\

\tabprojectlinespace 
 
 weka & \priostrategy{-} & \significant{\num{3}} & \significant{\num{257}} & \significant{\num{441}} & \significant{\num{538}}  & \significant{\num{1505}} & \significant{\num{3606}} & \significant{\num{4777}}  \\

\tabprojectlinespace 
 
 wheelwebtool & \priostrategy{-} & \significant{\num{3}} & \significant{\num{116}} & \significant{\num{349}} & \significant{\num{495}}  & \significant{\num{1031}} & \significant{\num{2002}} & \significant{\num{2817}}  \\

\midrule

Overall
& \priostrategy{-}  & \significant{\num{400}} & \significant{\num{67}} & \significant{\num{201}} & \significant{\num{171}}  & \significant{\num{144}} & \significant{\num{725}} & \significant{\num{749}} \\

\bottomrule
\end{tabular}

\end{table*}

\noindent\ding{173} \textbf{Filter out Equal-To-Line Operators.}
For each mutation operator, EvoSuite designs an infection distance function to describe how far a mutant's different state from the source code is triggered~\cite{FraserMutation}. Some infection distance functions always return $0$. For example, UOI only adds $1$ to, subtracts $1$ to, or negates a numerical value, so the infection distance is always $0$. \mynewcontent{2}{Therefore}, if a test case covers the line mutated by UOI, the mutant is killed. We \mynewcontent{2}{refer to} this \mynewcontent{2}{type} of operator an \textbf{equal-to-line} operator. Among three key operators, only UOI is \mynewcontent{2}{an} equal-to-line operator~\cite{FraserMutation}. Since line coverage has been dealt with, we filter out it.

\noindent\ding{174} \textbf{Select Subsuming Mutants.}
\mynewcontent{2}{The remaining operators} are \mynewcontent{6}{BOR,} AOR and ROR. We choose one of the existing approaches~\cite{mresa1999efficiency,JiaHOM,JustRedundant,GdynaMutant,GheyiSub} to select subsuming mutants for them. These approaches can be \mynewcontent{2}{classified} into three categories: (1) Manual analysis: Just et al.~\cite{JustRedundant} \mynewcontent{2}{establish} the subsumption relationships for ROR and LCR by analyzing all possible outputs of their mutants. \mynewcontent{2}{However,} this approach \mynewcontent{6}{cannot} be applied to non-logical operators~\cite{GdynaMutant}; (2) Dynamic analysis: \mynewcontent{2}{Guimar{\~a}es et al.~\cite{GdynaMutant} establish the subsumption relationships by running an exhaustive set of tests}. This approach \mynewcontent{2}{requires} many tests, which we \mynewcontent{6}{cannot} provide; (3) Static analysis: Gheyi and Souza et al.~\cite{GheyiSub, souza2020identifying} encode a theory of subsumption relations in the Z3 theorem prover to identify the subsumption relationships. We adopt this approach because (\textrm{i}) This approach can be applied to \mynewcontent{6}{all the above three operators}; (\textrm{ii}) Using the Z3 prover to identify the subsumption relationships is a once-for-all job\mynewcontent{6}{, and we can directly integrate their results into EvoSuite}.

\noindent\textbf{BC and TMC.}
For a coverage goal of branch coverage, there is a subsuming goal from direct branch coverage (see Sec. \ref{subsec:cc}). As a result, the subset for branch coverage is empty since we select direct branch coverage as the representative (see Sec. \ref{subsec:rc}). Similarly, the subset for top-level method coverage is empty too.

\noindent\textbf{Output.}
For four unselected criteria, we select four subsets of their coverage goals. Two of them are empty. Finally, smart selection joins these subsets with the representative criteria to get their fitness functions for guiding GAs.
\section{Evaluation}\label{sec:eval}
  \begin{figure*}[t]
\centering
\includegraphics[width=1\textwidth]{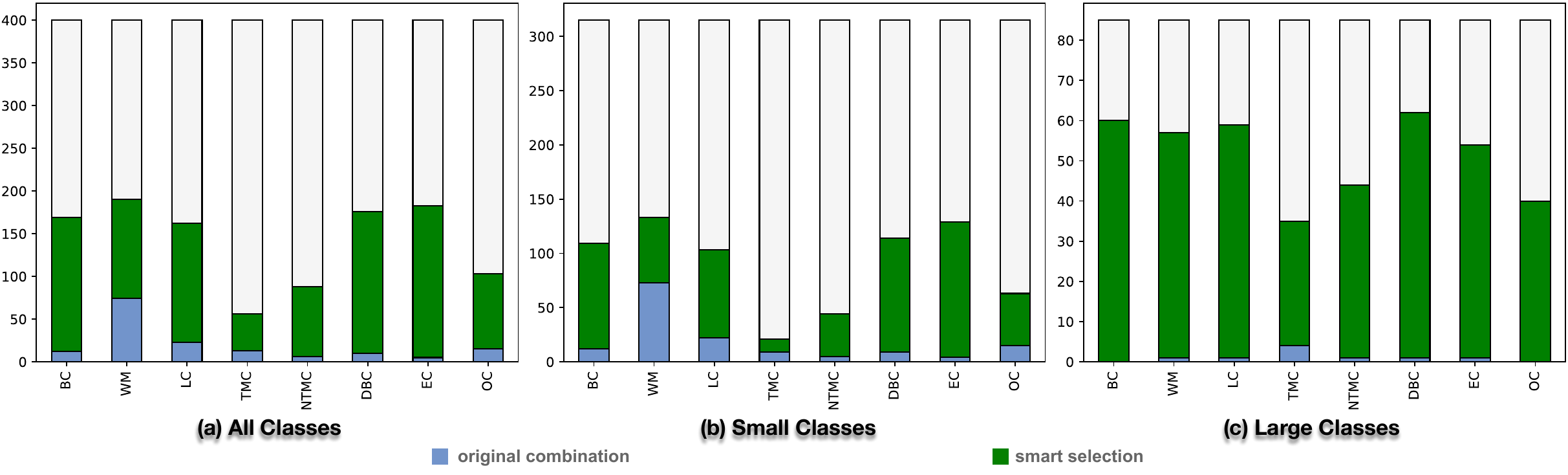}
\caption{\mynewcontent{6}{Significant case summary of smart selection and the original combination with WS}}\label{fig:suite_os}
\end{figure*}
\subsection{Experiment Setting}\label{subsec:exp_set}
The evaluation mainly focuses on the coverage performance of smart selection. Our evaluation aims to answer the following research questions:

\noindent$\bullet$\textbf{RQ1:} How does smart selection perform with WS?

\noindent$\bullet$\textbf{RQ2:} How does smart selection perform with MOSA?

\noindent$\bullet$\textbf{RQ3:} How does smart selection perform with DynaMOSA?

\noindent$\bullet$\mynewcontent{3}{\textbf{RQ4:} Do criteria within the same criteria group exhibit a coverage correlation?}

\noindent$\bullet$\mynewcontent{3}{\textbf{RQ5:} Are representative criteria efficient in guiding SBST?}

\noindent$\bullet$\textbf{RQ6:} How does the subsumption strategy affect the performance of smart selection?

\noindent\mynewcontent{1}{$\bullet$\textbf{RQ7:} How does smart selection perform under different search budgets?}

\noindent\mynewcontent{6}{$\bullet$\textbf{RQ8:} How does smart selection affect detecting faults?}

\noindent\textbf{Environment.} All experiments are conducted on \mynewcontent{1}{three} machines with Intel(R) Core(TM) i9-10900 CPU @ 2.80GHz and 128 GB RAM.

\noindent\textbf{Subjects.} 
We randomly select Java classes from \mynewcontent{5}{two} sources: the benchmark of DynaMOSA~\cite{PanichellaDynaMOSA} and Hadoop~\cite{hadoop}. Following the previous work~\cite{PanichellaMOSA}, the only restriction of randomly selecting classes is that the class must contain at least $50$ branches, aiming to filter out the trivial classes. As a result, we select $400$ classes: $158$ from the benchmark of DynaMOSA and $242$ from Hadoop. \mynewcontent{2}{Table \ref{tab:classes} shows the statistical data of these classes' branches and lines grouped by the projects. The second column, Package, is only for the Hadoop project: Since Hadoop contains too many classes \mynewcontent{5}{(more than half of all classes considered)}, we individually present its Java Packages' statistical data, not the whole project; The third column, Classes, presents the class number of this group (a project or a Package); The fourth column, Branches, contains three sub-columns, showing the 25th percentile, mean, and 75th percentile of the branches of this group's classes. The last column is similar to the fourth column but shows the counterpart data of \mynewcontent{6}{statements}.}
 
\noindent\textbf{Baseline for RQ1-3.} We have two baselines: (1) the original combination, used to be compared with smart selection on each Java class; (2) a single constituent criterion, used to show the data of coverage decrease caused by the above two combination approaches.  A single constituent criterion means that we only use each criterion of these eight criteria (see Sec. \ref{subsec:criterion}) to guide GAs. There is one exception: when the constituent criterion is exception or output coverage, we combine this criterion and branch coverage to guide GAs. The reason is that only exception or output coverage is weak in the effectiveness of guiding the GAs~\cite{Rojas2015CombiningMC, GayCombine}. Branch coverage can guide the GAs to reach more source lines of the CUT, increasing the possibility of triggering exceptions or covering output goals.

\noindent\textbf{Configuration for RQ1-3.}
EvoSuite provides many parameters (e.g., crossover probability, population size~\cite{FraserWhole}) to run the algorithms. In this paper, we adopt EvoSuite's default parameters to run smart selection and other baselines. 

Smart selection introduces a new parameter \textit{lineThreshold} (see Sec. \ref{subsec:rs}). It controls smart selection to skip basic blocks with less than \textit{lineThreshold} lines. We set  \textit{lineThreshold} as $8$. The discussion on this value is in Sec. \ref{sec:discuss}. For each Java class, we run EvoSuite with ten approaches: (1) smart selection, (2) the original combination, and (3) each constituent criterion of all eight criteria. We run each approach for $30$ rounds per Java class, and each run's search budget is \mynewcontent{5}{two} minutes.
\begin{figure*}[t]
\centering
\includegraphics[width=1\textwidth]{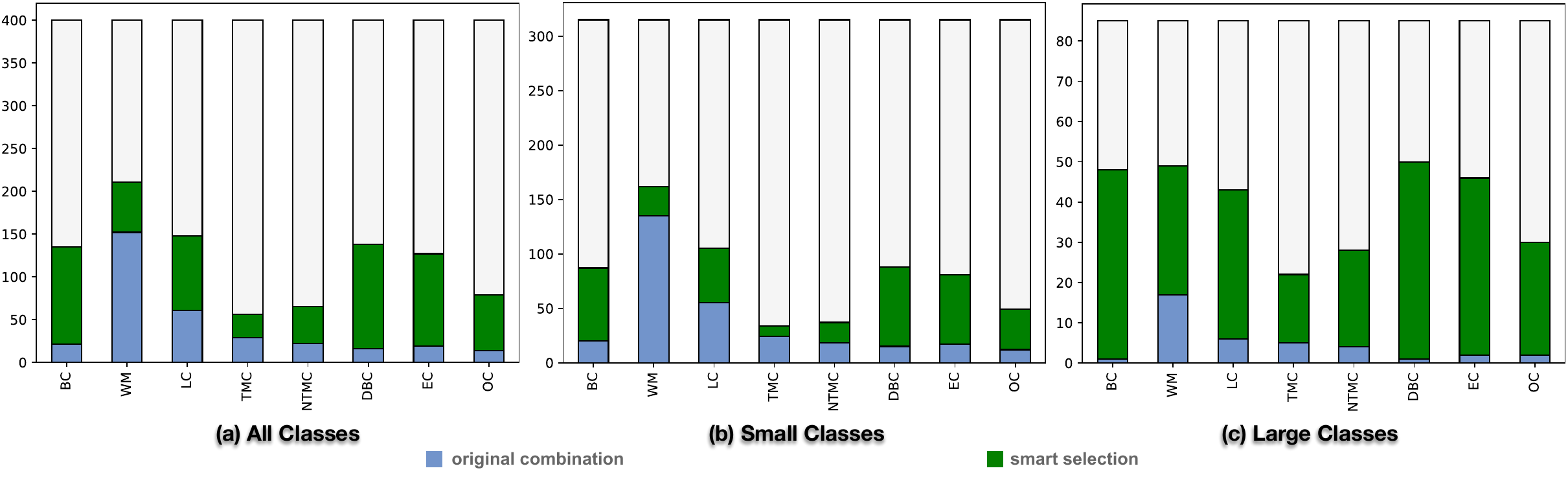}
\caption{\mynewcontent{6}{Significant case summary of smart selection and the original combination with MOSA}}\label{fig:mosa_os}
\end{figure*}

\subsection{RQ1: How does smart selection perform with WS?}\label{subsec:rq1}
\noindent\textbf{Motivation.}
In this RQ, we evaluate smart selection (\textbf{SS}) with WS. First, we compare the performance of SS and the original combination (\textbf{OC}). \mynewcontent{5}{Note that in RQ 1-3 and 5-7, \enquote{performance} denotes coverage. In these RQs, We regard that an approach $A$ outperforms another approach $B$ under a criterion if the test suite generated by $A$ achieves a higher coverage under this criterion than that by $B$.} Next, we use the coverage of each constituent criterion (\textbf{CC}) to show the coverage decrease caused by SS and OC. Furthermore, we show these approaches' differences in the resulting suite sizes (i.e., the number of tests in a test suite).

\noindent\textbf{Methodology.} 
EvoSuite records the coverage for generated unit tests. For each class, we obtain $10$ coverage data sets: One data set records the coverage of the eight criteria when using SS; One data set records the coverage of the eight criteria when using OC; The rest data sets record the coverage when using each CC.

For each Java class, we follow previous research work~\cite{Rojas2015CombiningMC} to use Mann-Whitney U Test to measure the statistical difference between SS and OC. Then, we use the Vargha-Delaney $\hat{A}_{ab}$~\cite{vargha2000critique} to evaluate whether a particular approach $a$ outperforms another approach $b$ ($A_{ab}>0.5$ and the significant value $p$ is smaller than $0.05$).

\noindent\textbf{Result.}
\begin{table}[htbp]
	\centering
	\scriptsize
	\caption{Average coverage results for each approach with WS}
	\label{tab:suite}
\textbf{(a)} All Classes\\
\begin{tabular}{l| l| l |l  }
\hline
approach & SS & OC & CC \\ \hline
BC & 55\% & 53\% & \textcolor[RGB]{0,128,28}{57\%} \\ \hline
WM & \textcolor[RGB]{0,128,28}{59\%} & 57\% & \textcolor[RGB]{0,128,28}{59\%} \\ \hline
LC & \textcolor[RGB]{0,128,28}{60\%} & 58\% & \textcolor[RGB]{0,128,28}{60\%} \\ \hline
TMC & \textcolor[RGB]{0,128,28}{84\%} & 83\% & \textcolor[RGB]{0,128,28}{84\%} \\ \hline
\end{tabular}
\begin{tabular}{l| l| l |l  }
\hline
approach & SS & OC & CC \\ \hline
NTMC & \textcolor[RGB]{0,128,28}{71\%} & 70\% & \textcolor[RGB]{0,128,28}{71\%} \\ \hline
DBC & 55\% & 53\% & \textcolor[RGB]{0,128,28}{56\%} \\ \hline
EC & 15.92 & 14.52 & \textcolor[RGB]{0,128,28}{16.52} \\ \hline
OC & 44\% & 43\% & \textcolor[RGB]{0,128,28}{45\%} \\ \hline
\end{tabular}\\
\textbf{(b)} Small Classes\\
\begin{tabular}{l| l| l |l  }
\hline
approach & SS & OC & CC \\ \hline
BC & 58\% & 57\% & \textcolor[RGB]{0,128,28}{59\%} \\ \hline
WM & \textcolor[RGB]{0,128,28}{62\%} & 61\% & \textcolor[RGB]{0,128,28}{62\%} \\ \hline
LC & \textcolor[RGB]{0,128,28}{62\%} & 61\% & \textcolor[RGB]{0,128,28}{62\%} \\ \hline
TMC & 85\% & 85\% & 85\% \\ \hline
\end{tabular}
\begin{tabular}{l| l| l |l  }
\hline
approach & SS & OC & CC \\ \hline
NTMC & \textcolor[RGB]{0,128,28}{73\%} & 72\% & 72\% \\ \hline
DBC & 57\% & 56\% & \textcolor[RGB]{0,128,28}{58\%} \\ \hline
EC & \textcolor[RGB]{0,128,28}{13.29} & 12.48 & 12.95 \\ \hline
OC & \textcolor[RGB]{0,128,28}{46\%} & 45\% & \textcolor[RGB]{0,128,28}{46\%} \\ \hline
\end{tabular}\\
\textbf{(c)} Large Classes\\
\begin{tabular}{l| l| l |l  }
\hline
approach & SS & OC & CC \\ \hline
BC & 45\% & 41\% & \textcolor[RGB]{0,128,28}{51\%} \\ \hline
WM & \textcolor[RGB]{0,128,28}{48\%} & 44\% & 47\% \\ \hline
LC & \textcolor[RGB]{0,128,28}{50\%} & 47\% & \textcolor[RGB]{0,128,28}{50\%} \\ \hline
TMC & 79\% & 77\% & \textcolor[RGB]{0,128,28}{82\%} \\ \hline
\end{tabular}
\begin{tabular}{l| l| l |l  }
\hline
approach & SS & OC & CC \\ \hline
NTMC & 67\% & 63\% & \textcolor[RGB]{0,128,28}{68\%} \\ \hline
DBC & 45\% & 41\% & \textcolor[RGB]{0,128,28}{50\%} \\ \hline
EC & 25.69 & 22.08 & \textcolor[RGB]{0,128,28}{29.74} \\ \hline
OC & 39\% & 36\% & \textcolor[RGB]{0,128,28}{41\%} \\ \hline
\end{tabular}
\end{table}
\begin{table}[htbp]
    \centering
    \small
    \caption{Average test suite size of each approach with WS}
    \begin{tabular}{l|l|l|l}
    \hline
        approach & SS & OC & CC (Average)  \\ \hline
        size (All Classes) & 51.35 & 47.77 & 31.59 \\ \hline
        size (Small Classes) & 37.27 & 36.39 & 19.43 \\ \hline
        size (Large Classes) & 103.53 & 89.95 & 76.64 \\ \hline
    \end{tabular}
    \label{suite_size_ws}
\end{table}
\noindent$\blacktriangleright$\textbf{All Classes.}

\noindent\textbf{Significant Cases.} Fig. \ref{fig:suite_os} (a) shows the comparison results of SS and OC on all $400$ Java classes. SS outperforms OC on $121$ ($30.3\%$) classes (a.k.a., SS-outperforming classes) on average for each coverage. OC outperforms SS on $20$ ($4.9\%$) classes (a.k.a., OC-outperforming classes). These two approaches have no significant difference on $259$ ($64.8\%$) classes (a.k.a., No-significant classes) on average.

\noindent\textbf{Average Coverage.} Table \ref{tab:suite} (a) shows the average coverage of all classes with three approaches. For exception coverage, the table shows the average number of the triggered exceptions since we can not know the total number of exceptions in a class~\cite{Rojas2015CombiningMC}. The green number represents the highest coverage at a given criterion. SS outperforms OC for eight criteria' coverage. Among three approaches, SS reaches the highest coverage for four criteria. CC reaches the highest coverage for all criteria.

\noindent\textbf{Average Suite Size.} The first row of Table \ref{suite_size_ws} shows the test suites' average sizes of all classes. Compared to CC (average suite size of all constituent criteria), the size of OC increases by $51.2\%$ ($(47.77-31.59)/31.59$). Compared to OC, the size of SS increases by $7.4\%$ ($(51.35-47.77)/47.77$).

\noindent$\blacktriangleright$\textbf{Small Classes.($<$ 200 branches)}

\noindent\textbf{Significant Cases.} Fig. \ref{fig:suite_os} (b) shows the comparison results of SS and OC on $315$ small Java classes. For each criterion, on average, SS-outperforming classes are $71$ ($22.5\%$). OC-outperforming classes are $19$ ($5.9\%$). No-significant classes are $226$ ($71.6\%$).

\noindent\textbf{Average Coverage.} Table \ref{tab:suite} (b) shows the average coverage of small classes. SS outperforms OC for seven criteria' coverage. SS reaches the highest coverage for five criteria.

\noindent\textbf{Average Suite Size.} The second row of Table \ref{suite_size_ws} shows the average suite sizes of small classes. Compared to CC, the size of OC increases by $87.3\%$. Compared to OC, the size of SS increases by $2.4\%$.

\noindent$\blacktriangleright$\textbf{Large Classes. ($\geq$ 200 branches)}

\noindent\textbf{Significant Cases.} Fig. \ref{fig:suite_os} (c) shows the comparison results of SS and OC on $85$ large Java classes. For each criterion, on average, SS-outperforming classes are $50$ ($59.1\%$). The number of OC-outperforming classes is $1$ ($1.3\%$). No-significant classes are $34$ ($39.6\%$).
 
\noindent\textbf{Average Coverage.} Table \ref{tab:suite} (c) shows the average coverage of large classes. SS outperforms OC for eight criteria' coverage. SS reaches the highest coverage for two criteria.

\noindent\textbf{Average Suite Size.} The third row of Table \ref{suite_size_ws} shows the average suite sizes of large classes. Compared to CC, the size of OC increases by $17.4\%$. Compared to OC, the size of SS increases by $15.1\%$.

\noindent\textbf{Analysis.}
SS outperforms OC statistically, especially on the large classes. There is one exception: On the small classes, the number ($73$) of OC-outperforming classes in weak mutation is more than the counterpart number ($60$) (see Fig. \ref{fig:suite_os} (b)). On average, each of those $73$ classes has $82$ branches and $321$ mutants, while each of those $60$ classes has $115$ branches and $381$ mutants. It also supports that SS outperforms OC on the large classes. Furthermore, in most cases, the average coverage of CC is higher than the one of OC. SS narrows the coverage gap between them. For example, the biggest gap is $10\%$, happening in the large classes' branch coverage. SS narrows the gap by $4\%$ (see Table \ref{tab:suite} (c)). These facts confirm that combining criteria offers more objectives for optimization, affecting the efficacy of GAs. The \mynewcontent{5}{larger} classes bring more objectives, leading to a higher impact. The suite size increase brought by OC/SS to CC is significant, confirming the experimental results of the work proposing OC~\cite{Rojas2015CombiningMC}. The main reason is that the GA (not only WS but also MOSA/DynaMOSA) needs more tests for more goals. With the coverage increase, the suite size of SS is also greater than OC. Compared with the extent of the suite increase brought by OC to CC, \mynewcontent{5}{we regard that increase as reasonable}.
\mytcbox{Answer to RQ1}{With WS, SS outperforms OC statistically, especially on the large classes (i.e., the classes with \mynewcontent{5}{no fewer than} 200 branches).}
\subsection{RQ2: How does smart selection perform with MOSA?}\label{subsec:rq2}
\begin{figure*}[t]
\centering
\includegraphics[width=1\textwidth]{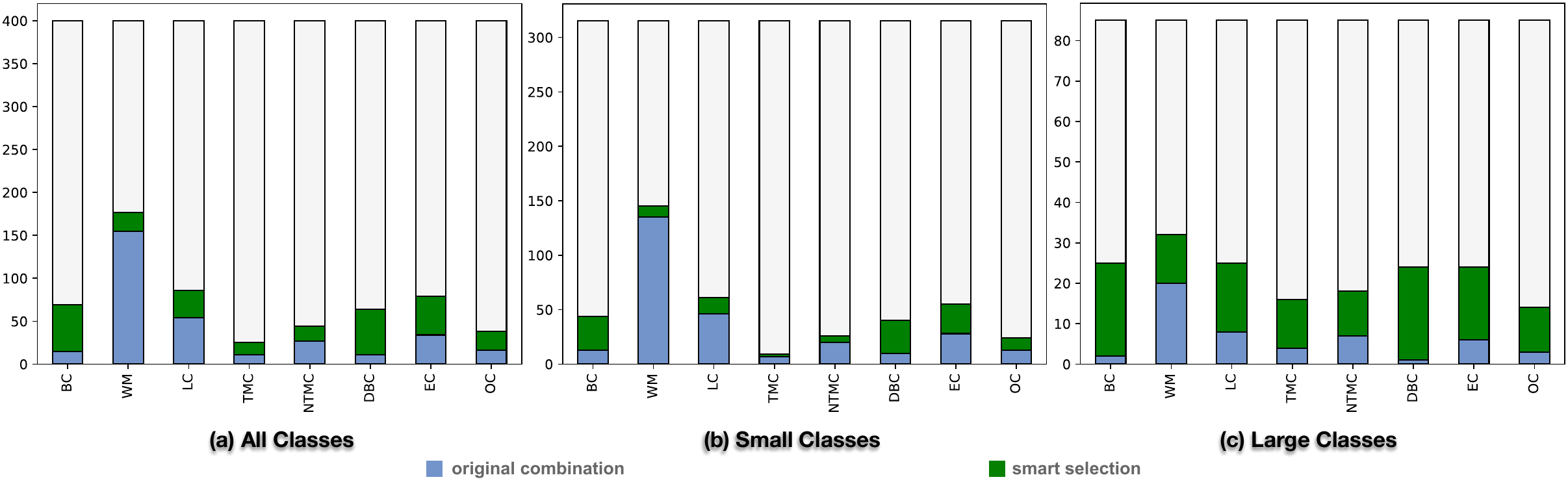}
\caption{\mynewcontent{6}{Significant case summary of smart selection and the original combination with DynaMOSA}}\label{fig:dynamosa_os}
\end{figure*}
\noindent\textbf{Motivation.}
In this RQ, we evaluate smart selection with MOSA.

\noindent\textbf{Methodology.}
The methodology is the same as RQ1's.

\noindent\textbf{Result.}
\begin{table}[htbp]
	\centering
	\scriptsize
	\caption{Average coverage results for each approach with MOSA}
	\label{tab:mosa}
\textbf{(a)} All Classes\\
\begin{tabular}{l| l| l |l  }
\hline
approach & SS & OC & CC \\ \hline
BC & 57\% & 56\% & \textcolor[RGB]{0,128,28}{58\%} \\ \hline
WM & 60\% & 60\% & 60\% \\ \hline
LC & \textcolor[RGB]{0,128,28}{61\%} & 60\% & \textcolor[RGB]{0,128,28}{61\%} \\ \hline
TMC & \textcolor[RGB]{0,128,28}{84\%} & 83\% & 82\% \\ \hline
\end{tabular}
\begin{tabular}{l| l| l |l  }
\hline
approach & SS & OC & CC \\ \hline
NTMC & \textcolor[RGB]{0,128,28}{71\%} & \textcolor[RGB]{0,128,28}{71\%} & 69\% \\ \hline
DBC & \textcolor[RGB]{0,128,28}{57\%} & 55\% & \textcolor[RGB]{0,128,28}{57\%} \\ \hline
EC & \textcolor[RGB]{0,128,28}{16.95} & 16.15 & 16.41 \\ \hline
OC & 44\% & 44\% & \textcolor[RGB]{0,128,28}{45\%} \\ \hline
\end{tabular}\\
\textbf{(b)} Small Classes\\
\begin{tabular}{l| l| l |l  }
\hline
approach & SS & OC & CC \\ \hline
BC & 59\% & 58\% & \textcolor[RGB]{0,128,28}{60\%} \\ \hline
WM & 62\% & 62\% & 62\% \\ \hline
LC & 63\% & 63\% & 63\% \\ \hline
TMC & \textcolor[RGB]{0,128,28}{85\%} & \textcolor[RGB]{0,128,28}{85\%} & 84\% \\ \hline

\end{tabular}
\begin{tabular}{l| l| l |l  }
\hline
approach & SS & OC & CC \\ \hline
NTMC & \textcolor[RGB]{0,128,28}{73\%} & 72\% & 71\% \\ \hline
DBC & 58\% & 58\% & \textcolor[RGB]{0,128,28}{59\%} \\ \hline
EC & \textcolor[RGB]{0,128,28}{13.28} & 13.07 & 12.73 \\ \hline
OC & 45\% & 45\% & \textcolor[RGB]{0,128,28}{46\%} \\ \hline
\end{tabular}\\
\textbf{(c)} Large Classes\\
\begin{tabular}{l| l| l |l  }
\hline
approach & SS & OC & CC \\ \hline
BC & 49\% & 46\% & \textcolor[RGB]{0,128,28}{51\%} \\ \hline
WM & \textcolor[RGB]{0,128,28}{52\%} & 49\% & 51\% \\ \hline
LC & \textcolor[RGB]{0,128,28}{54\%} & 52\% & 53\% \\ \hline
TMC & \textcolor[RGB]{0,128,28}{79\%} & 78\% & 77\% \\ \hline
\end{tabular}
\begin{tabular}{l| l| l |l  }
\hline
approach & SS & OC & CC \\ \hline
NTMC & \textcolor[RGB]{0,128,28}{66\%} & 64\% & 64\% \\ \hline
DBC & 50\% & 46\% & \textcolor[RGB]{0,128,28}{51\%} \\ \hline
EC & \textcolor[RGB]{0,128,28}{30.54} & 27.53 & 30.03 \\ \hline
OC & 40\% & 38\% & \textcolor[RGB]{0,128,28}{42\%} \\ \hline
\end{tabular}
\end{table}
\begin{table}[htbp]
    \centering
    \small
    \caption{Average test suite size of each approach with MOSA}
    \begin{tabular}{l|l|l|l}
    \hline
        approach & SS & OC & CC (Average)  \\ \hline
        size (All Classes) & 57.03 & 54.46 & 31.47 \\ \hline
        size (Small Classes) & 38.27 & 38.83 & 19.85 \\ \hline
        size (Large Classes) & 126.56 & 112.38 & 74.53 \\ \hline
    \end{tabular}
    \label{suite_size_mosa}
\end{table}
\noindent$\blacktriangleright$\textbf{All Classes.}

\noindent\textbf{Significant Cases.} Fig. \ref{fig:mosa_os} (a) shows the comparison results of SS and OC on all $400$ Java classes. For each criterion, on average, SS-outperforming classes are $78$ ($19.5\%$). OC-outperforming classes are $42$ ($10.4\%$). No-significant classes are $280$ ($70.1\%$).

\noindent\textbf{Average Coverage.} Table \ref{tab:mosa} (a) shows the average coverage of all classes. SS outperforms OC for five criteria' coverage. Among three approaches, SS reaches five criteria' highest coverage.

\noindent\textbf{Average Suite Size.} The first row of Table \ref{suite_size_mosa} shows the average suite sizes of all classes. Compared to CC, the size of OC increases by $73.1\%$. Compared to OC, the size of SS increases by $4.7\%$.

\noindent$\blacktriangleright$\textbf{Small Classes.}

\noindent\textbf{Significant Cases.} Fig. \ref{fig:mosa_os} (b) shows the comparison results of SS and OC on $315$ small Java classes. For each criterion, on average, SS-outperforming classes are $43$ ($13.8\%$). OC-outperforming classes are $37$ ($11.7\%$). No-significant classes are $235$ ($74.5\%$).

\noindent\textbf{Average Coverage.} Table \ref{tab:mosa} (b) shows the average coverage of small classes. SS outperforms OC for three criteria' coverage. SS reaches three criteria' highest coverage.

\noindent\textbf{Average Suite Size.} The second row of Table \ref{suite_size_mosa} shows the average suite sizes of small classes. Compared to CC, the size of OC increases by $95.6\%$. OC is nearly equal to SS.

\noindent$\blacktriangleright$\textbf{Large Classes.}

\noindent\textbf{Significant Cases.} Fig. \ref{fig:mosa_os} (c) shows the comparison results of SS and OC on $85$ large Java classes. For each criterion, on average, SS-outperforming classes are $35$ ($40.9\%$). OC-outperforming classes are $5$ ($5.6\%$). No-significant classes are $46$ ($53.5\%$).

\noindent\textbf{Average Coverage.} Table \ref{tab:mosa} (c) shows the average coverage of large classes. SS outperforms OC for eight criteria' coverage. SS reaches five criteria' highest coverage.

\noindent\textbf{Average Suite Size.} The third row of Table \ref{suite_size_mosa} shows the average suite sizes of large classes. Compared to CC, the size of OC increases by $50.7\%$. Compared to OC, the size of SS increases by $12.6\%$.

\noindent\textbf{Analysis.}
SS outperforms OC on the large classes like WS. But the advantage of SS is unnoticeable on the small classes. The coverage gap between CC and OC is not significant as the gap in WS. SS nearly bridges this gap. The biggest gap is $5\%$, happening in branch coverage of the large classes. SS narrows this gap by $3\%$. The suite size gap between SS and OC is smaller than on WS, which is consistent with the fact that SS and OC have a smaller coverage gap on MOSA. These facts show the advantage of multi-objective approaches (e.g., MOSA) over single-objective approaches (e.g., WS)~\cite{KnowlesSingle, PanichellaMOSA, BrockhoffMulti}. However, the advantage of SS on the large classes indicates that too many objectives also affect the multi-objective algorithms.
\mytcbox{Answer to RQ2}{With MOSA, SS outperforms OC statistically on the large classes. Smart selection has only a slight advantage on the small classes.}
\subsection{RQ3: How does smart selection perform with DynaMOSA?}\label{subsec:rq3}
\noindent\textbf{Motivation.}
In this RQ, we evaluate smart selection with DynaMOSA.

\noindent\textbf{Methodology.}
The methodology is the same as RQ1's. \mynewcontent{3}{Note that the difference is that we add BC to the guiding criteria no matter whether the approach contains BC. For example, if the constituent criterion is LC, we will combine BC with it to guide DynaMOSA. Similarly, we add BC to the representative group (DBC, LC, EC, OC) selected by smart selection. The reason is that DynaMOSA can not run without BC since it needs the goals of branch coverage to build the control dependency graph~\cite{PanichellaDynaMOSA}.}

\noindent\textbf{Result.}
\begin{table}[htbp]
	\centering
	\scriptsize
	\caption{Average coverage results for each approach with DynaMOSA}
	\label{tab:dynamosa}
	\textbf{(a)} All Classes\\
\begin{tabular}{l| l| l |l}
\hline
approach& SS & OC & CC \\ \hline
BC & 58\% & 58\% & 58\% \\ \hline
WM & 60\% & 61\% & \textcolor[RGB]{0,128,28}{62\%} \\ \hline
LC & 61\% & \textcolor[RGB]{0,128,28}{62\%} & \textcolor[RGB]{0,128,28}{62\%} \\ \hline
TMC & \textcolor[RGB]{0,128,28}{83\%} & \textcolor[RGB]{0,128,28}{83\%} & 81\% \\ \hline
\end{tabular}
\begin{tabular}{l| l| l |l  }
\hline
approach & SS & OC & CC \\ \hline
NTMC & \textcolor[RGB]{0,128,28}{71\%} & \textcolor[RGB]{0,128,28}{71\%} & 70\% \\ \hline
DBC & 57\% & 57\% & \textcolor[RGB]{0,128,28}{58\%} \\ \hline
EC & \textcolor[RGB]{0,128,28}{17.15} & 17.14 & 16.64 \\ \hline
OC & 45\% & 45\% & 45\% \\ \hline
\end{tabular}\\
\textbf{(b)} Small Classes\\
\begin{tabular}{l| l| l |l  }
\hline
approach & SS & OC & CC \\ \hline
BC & \textcolor[RGB]{0,128,28}{60\%} & 59\% & \textcolor[RGB]{0,128,28}{60\%} \\ \hline
WM & 63\% & 63\% & \textcolor[RGB]{0,128,28}{64\%} \\ \hline
LC & 63\% & \textcolor[RGB]{0,128,28}{64\%} & \textcolor[RGB]{0,128,28}{64\%} \\ \hline
TMC & 84\% & \textcolor[RGB]{0,128,28}{85\%} & 82\% \\ \hline
\end{tabular}
\begin{tabular}{l| l| l |l  }
\hline
approach & SS & OC & CC \\ \hline
NTMC & 72\% & \textcolor[RGB]{0,128,28}{73\%} & 72\% \\ \hline
DBC & 59\% & 59\% & 59\% \\ \hline
EC & \textcolor[RGB]{0,128,28}{13.42} & \textcolor[RGB]{0,128,28}{13.42} & 12.81 \\ \hline
OC & 46\% & 46\% & 46\% \\ \hline
\end{tabular}\\
\textbf{(c)} Large Classes\\
\begin{tabular}{l| l| l |l  }
\hline
approach & SS & OC & CC \\ \hline
BC & 51\% & 51\% & \textcolor[RGB]{0,128,28}{53\%} \\ \hline
WM & 53\% & 52\% & \textcolor[RGB]{0,128,28}{54\%} \\ \hline
LC & 54\% & 54\% & \textcolor[RGB]{0,128,28}{55\%} \\ \hline
TMC & \textcolor[RGB]{0,128,28}{79\%} & \textcolor[RGB]{0,128,28}{79\%} & 78\% \\ \hline
\end{tabular}
\begin{tabular}{l| l| l |l  }
\hline
approach & SS & OC & CC \\ \hline
NTMC & \textcolor[RGB]{0,128,28}{66\%} & \textcolor[RGB]{0,128,28}{66\%} & 65\% \\ \hline
DBC & 51\% & 50\% & \textcolor[RGB]{0,128,28}{53\%} \\ \hline
EC & \textcolor[RGB]{0,128,28}{30.98} & 30.93 & 30.81 \\ \hline
OC & 41\% & 41\% & \textcolor[RGB]{0,128,28}{42\%} \\ \hline
\end{tabular}
\end{table}
\begin{table}[htbp]
    \centering
    \small
    \caption{Average test suite size of each approach with DynaMOSA}
    \begin{tabular}{l|l|l|l}
    \hline
        approach & SS & OC & CC (Average)  \\ \hline
        size (All Classes) & 61.13 & 60.59 & 39.2 \\ \hline
        size (Small Classes) & 38.9 & 39.59 & 23.71 \\ \hline
        size (Large Classes) & 143.51 & 138.44 & 96.58 \\ \hline
    \end{tabular}
    \label{suite_size_dynamosa}
\end{table}
\noindent$\blacktriangleright$\textbf{All Classes.}

\noindent\textbf{Significant Cases.} Fig. \ref{fig:dynamosa_os} (a) shows the comparison results of SS and OC on all $400$ Java classes. For each criterion, on average, SS-outperforming classes are $32$ ($8.1\%$). OC-outperforming classes are $40$ ($10.1\%$). No-significant classes are $328$ ($81.8\%$).

\noindent\textbf{Average Coverage.} Table \ref{tab:dynamosa} (a) shows the average coverage of all classes with three approaches. SS outperforms OC for one criterion's coverage, i.e., exception coverage. Among three approaches, SS reaches the highest coverage for three criteria.

\noindent\textbf{Average Suite Size.} The first row of Table \ref{suite_size_dynamosa} shows the average suite sizes of all classes. Compared to CC, the size of OC increases by $54.7\%$. OC is nearly equal to SS.

\noindent$\blacktriangleright$\textbf{Small Classes.}

\noindent\textbf{Significant Cases.} Fig. \ref{fig:dynamosa_os} (b) shows the comparison results of SS and OC on $315$ small Java classes. For each criterion, on average, SS-outperforming classes are $17$ ($5.2\%$). OC-outperforming classes are $34$ ($10.8\%$). No-significant classes are $265$ ($84\%$).

\noindent\textbf{Average Coverage.} Table \ref{tab:dynamosa} (b) shows the average coverage of small classes. SS outperforms OC for one criterion's coverage (branch coverage). SS reaches two criteria' highest coverage.

\noindent\textbf{Average Suite Size.} The second row of Table \ref{suite_size_dynamosa} shows the average suite sizes of small classes. Compared to CC, the size of OC increases by $66.9\%$. OC is nearly equal to SS.

\noindent$\blacktriangleright$\textbf{Large Classes.}

\noindent\textbf{Significant Cases.} Fig. \ref{fig:dynamosa_os} (c) shows the comparison results of SS and OC on $85$ large Java classes. For each criterion, on average, SS-outperforming classes are $16$ ($18.7\%$). OC-outperforming classes are $6$ ($7.5\%$). No-significant classes are $63$ ($73.8\%$).

\noindent\textbf{Average Coverage.} Table \ref{tab:dynamosa} (c) shows the average coverage of large classes. SS outperforms OC for three criteria' coverage. SS reaches three criteria' highest coverage.

\noindent\textbf{Average Suite Size.} The third row of Table \ref{suite_size_dynamosa} shows the average suite sizes of large classes. Compared to CC, the size of OC increases by $43.3\%$. Compared to OC, the size of SS increases by $3.7\%$.

\noindent\textbf{Analysis.}
SS still outperforms OC on the large classes, but not as obvious as WS and MOSA. In addition, SS is almost the same or slightly worse than OC on the small classes. Furthermore, the coverage gaps among the three approaches are not significant. The gap in the suite size between SS and OC is slight as in the coverage. One reason is that DynaMOSA selects the uncovered goals into the iteration process only when its branch dependencies are covered (see Sec. \ref{subsec:ga}). Hence, the number of optimization objectives is reduced. Therefore, an increase in the goals has a much smaller impact on DynaMOSA's coverage performance than on WS and MOSA. \mynewcontent{3}{Another reason is that we add BC to SS's guiding criteria because DynaMOSA can only run with BC. Consequently, SS's reduction in optimization objectives is undermined. Due to the same cause, we combine BC with an arbitrary constituent criterion as the guiding criteria, thus improving CC's ability to guide DynaMOSA. As a result, the differences between the three approaches are narrowed.}
\mytcbox{Answer to RQ3}{With DynaMOSA, the coverage of SS and OC is close for most criteria.}
\begin{figure*}[htbp]
    \centering
    \includegraphics[width=0.8\textwidth]{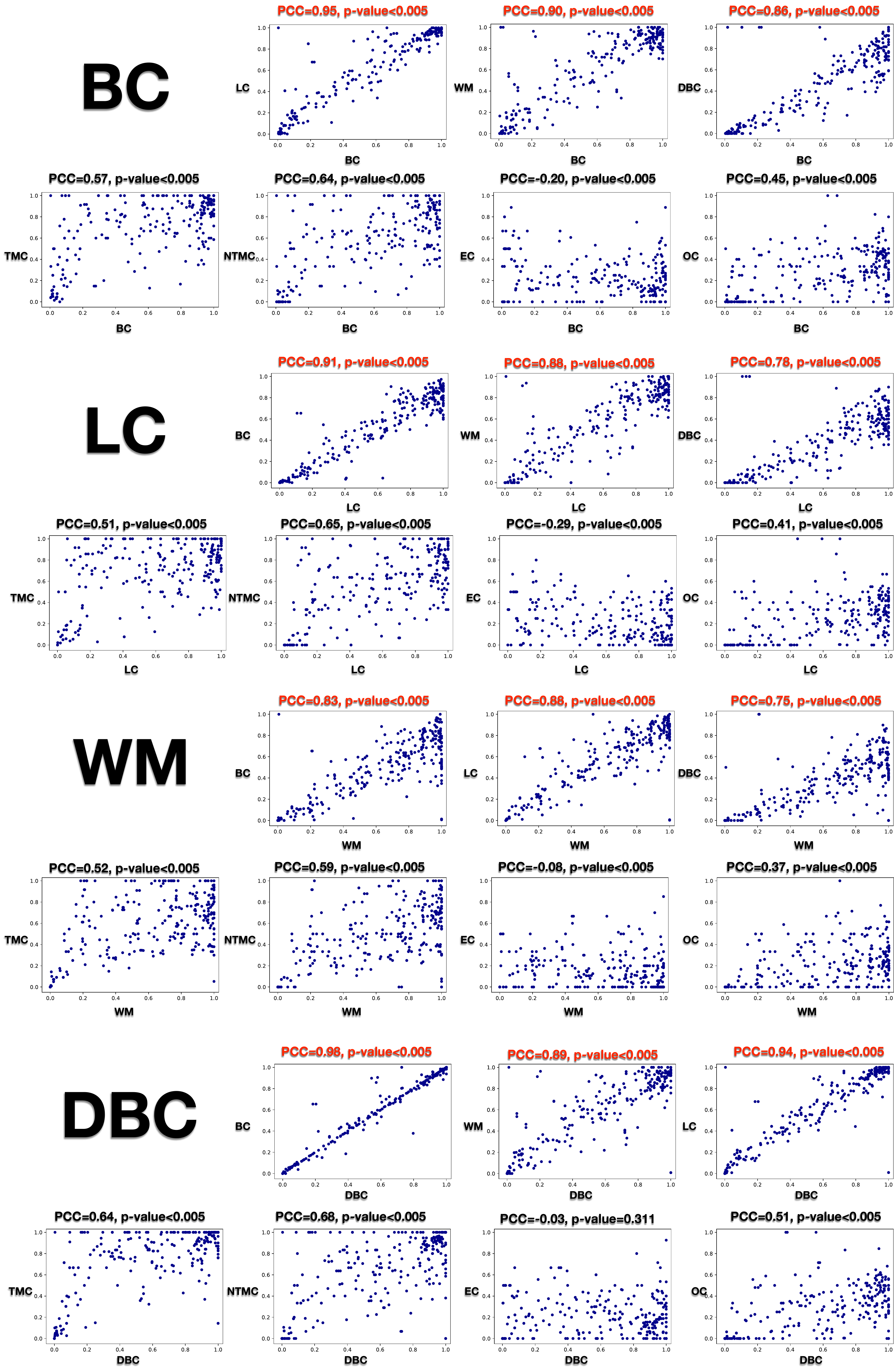}
    \caption{\mynewcontent{6}{Coverage correlation data for BC, LC, WM, and DBC when the GA is WS, including the Pearson Correlation Coefficient (PCC) with its $p$-value (The text is marked with red color when PPC $\geq 0.7$) and the scatter plot of data points in which the X-axis of the scatter plot represents the coverage values of the criterion guiding GA, and the Y-axis represents that of the criterion to be measured for its coverage}}
  \label{fig:suite_corr}
\end{figure*}
\begin{figure*}[htbp]
    \centering
    \includegraphics[width=0.8\textwidth]{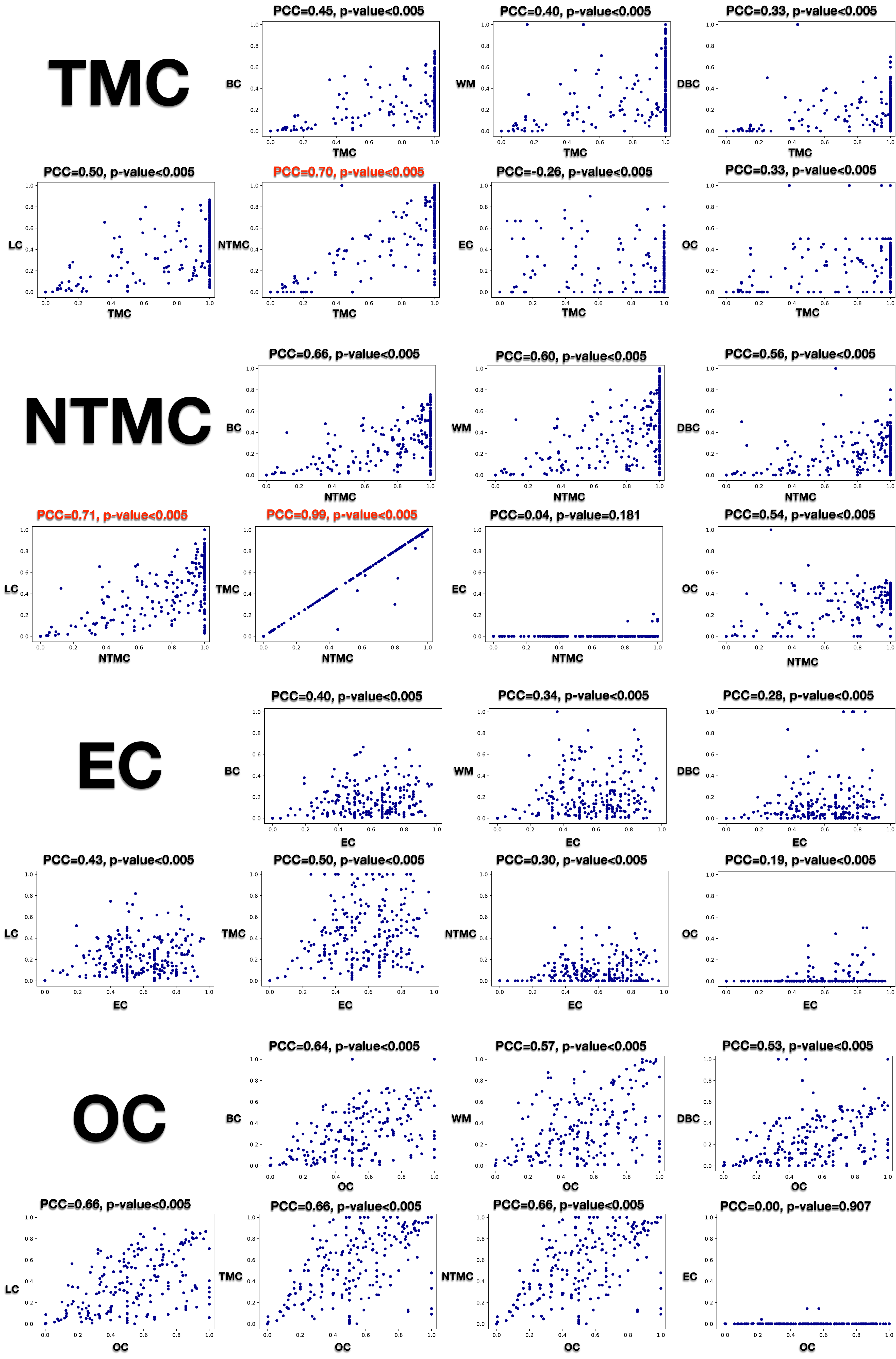}
    \caption{\mynewcontent{6}{Coverage correlation data for TMC, NTMC, EC, and OC when the GA is WS}}
  \label{fig:suite_corr_tmc_ntmc_ec_oc}
\end{figure*}
\begin{figure*}[htbp]
    \centering
    \includegraphics[width=0.8\textwidth]{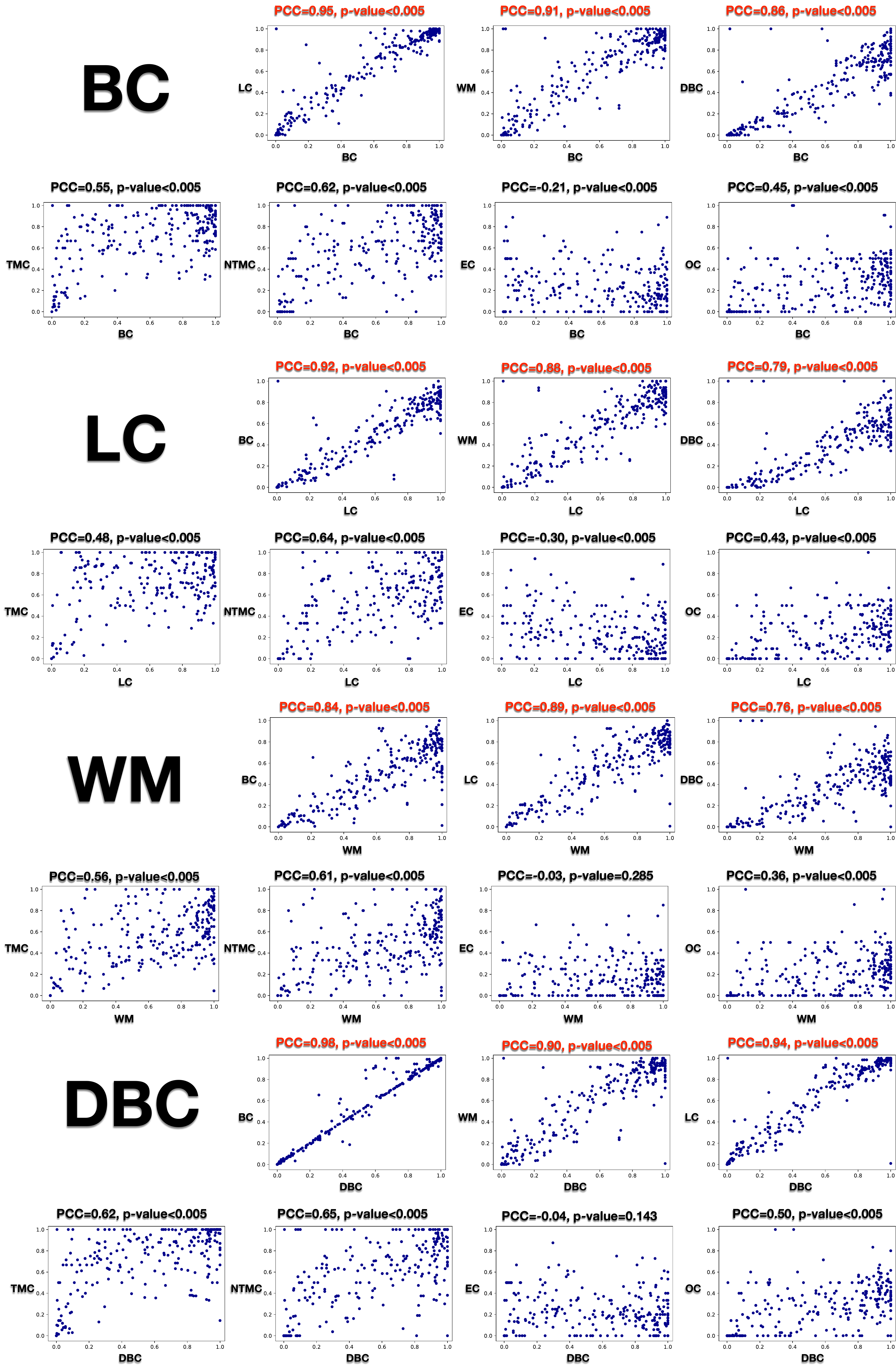}
    \caption{\mynewcontent{6}{Coverage correlation data for BC, LC, WM, and DBC when the GA is MOSA}}
  \label{fig:mosa_corr}
\end{figure*}
\begin{figure*}[htbp]
    \centering
    \includegraphics[width=0.8\textwidth]{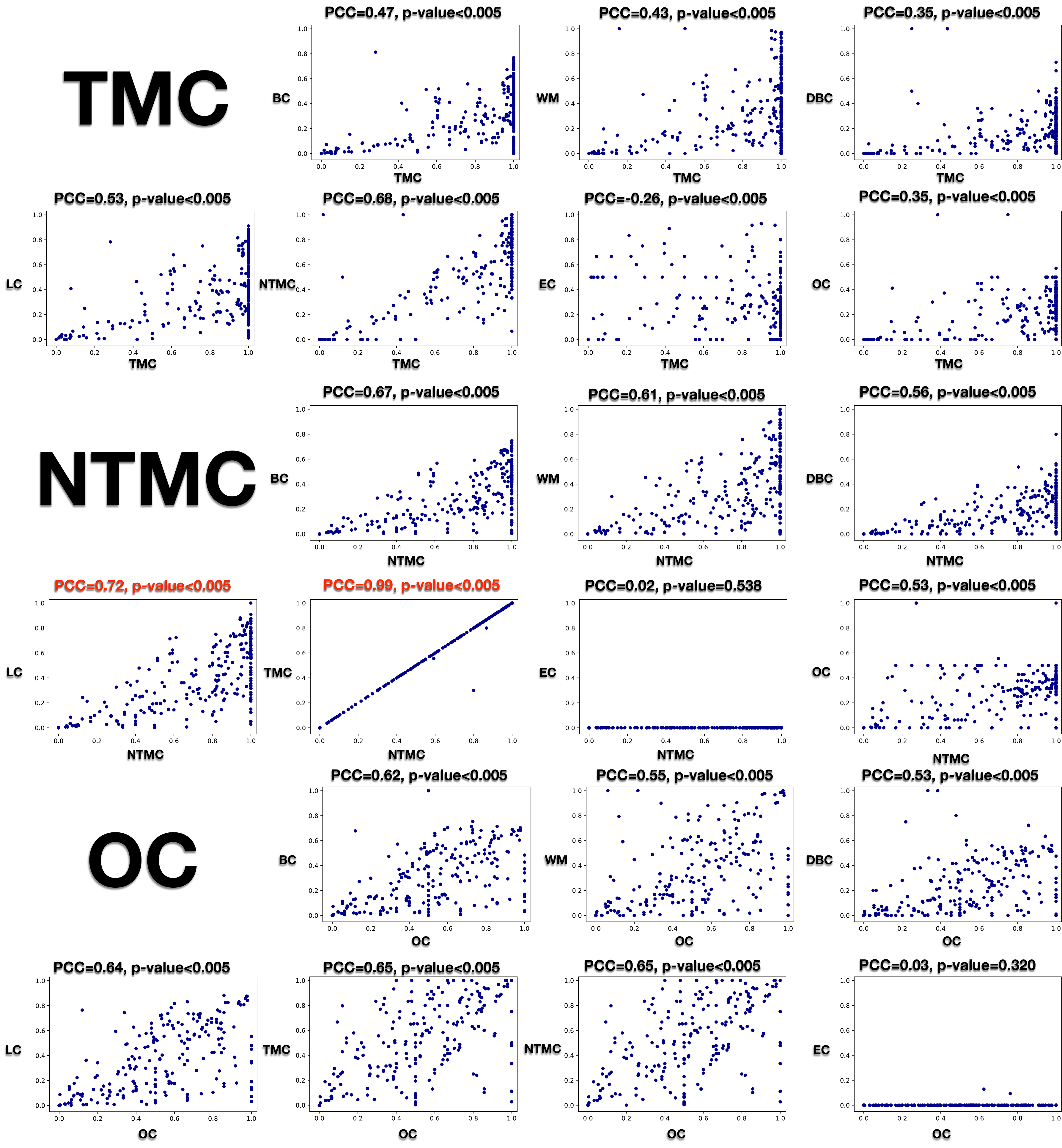}
    \caption{\mynewcontent{6}{Coverage correlation data for TMC, NTMC, and OC when the GA is MOSA}}
  \label{fig:mosa_corr_tmc_ntmc_oc}
\end{figure*}
\subsection{\mynewcontent{3}{\textbf{RQ4:} Do criteria within the same criteria group exhibit a coverage correlation?}}\label{subsec:rq-4}
\begin{figure*}[t]
\centering
\includegraphics[width=1\textwidth]{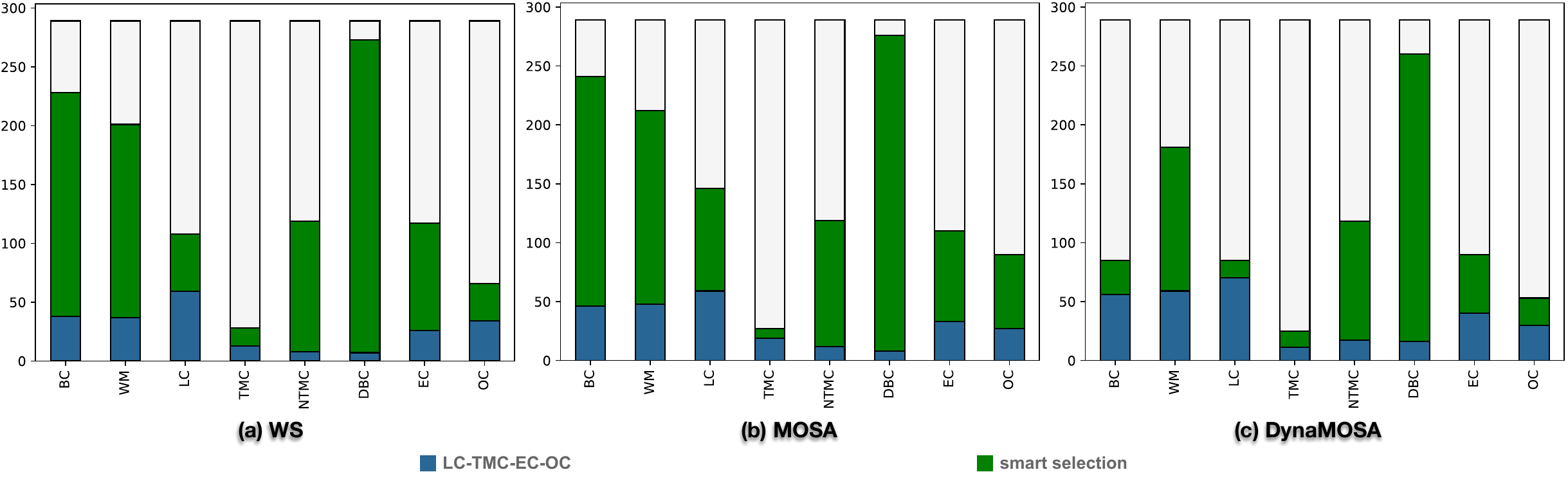}
\caption{\mynewcontent{6}{Significant case summary of smart selection and LT (i.e., LC-TMC-EC-OC)}}\label{fig:lteo}
\end{figure*}
\noindent\textbf{Motivation.}
One fundamental assumption of smart selection is that the criteria within the same criteria group have a strong coverage correlation (see Sec. \ref{subsec:cc}) so that we can choose a representative criterion to represent a group, thus reducing the goals optimized by GAs. In this RQ, we aim to verify this assumption.

\noindent\textbf{Methodology.}
In this RQ, \mynewcontent{5}{we iteratively run EvoSuite}, and each iteration has a specific four-tuple configuration, including a Java class, a GA, a criterion to guide GA, and a search budget. The Java class has $400$ options (i.e., $400$ Java classes used in RQ1-3), the GA has two options (WS and MOSA), the criterion has eight options (BC, LC, DBC, WM, TMC, NTMC, EC, and OC), and the budget has four options: $120$ seconds, $300$ seconds, $480$ seconds, and $600$ seconds. After running EvoSuite, we record the coverage score of all eight coverage criteria to calculate the correlation of a pair of criteria. Note that (1) DynaMOSA cannot run without BC (see Sec. \ref{subsec:rq3}). In other words, we can not calculate the correlation of a pair of criteria without the interference of BC. As a result, DynaMOSA is not used in this RQ; (2) MOSA can not run only with EC because EvoSuite can not provide any of EC's fitness functions at the beginning of running GA since EvoSuite does not know any exceptions thrown in the subject under test until GA randomly generates a unit test catching one. However, MOSA needs at least one fitness function to build Pareto Fronts~\cite{PanichellaMOSA}. As a result, when the GA is MOSA, the criterion to guide GA \mynewcontent{6}{has only} seven options, i.e., there are $8\times 7=64$ criterion pairs when the GA is WS while the counterpart number is $7\times 7=49$ when the GA is MOSA; (3) The origin coverage value provided by EvoSuite for EC is the number of exceptions triggered by the test suite. We normalize it as a value ranging from $0$ to $1$ (like other criteria) by dividing the origin value by the maximum number of exceptions triggered among all suites for the same Java class. After running all configurations on EvoSuite, we obtain $400\times 4=1600$ data points to estimate the correlation for each criterion pair and algorithm. Based on these points, we calculate the Pearson Correlation Coefficient (PCC) and the significant value $p$~\cite{cohen2009pearson}. PCC ranges from $-1$ to $+1$, where $0$ indicates no correlation. Correlation values of $-1$ or $+1$ suggest a perfect linear relationship. Positive values indicate that when x increases, y also increases. Conversely, negative values imply that as x increases, y decreases~\cite{pearsonr}.

\noindent\textbf{Result.}
\noindent$\blacktriangleright$\textbf{WS.}
\mynewcontent{6}{Fig.~\ref{fig:suite_corr} and~\ref{fig:suite_corr_tmc_ntmc_ec_oc} show} the coverage value point scatter plot, PCC, and the $p$-value of each criterion pair (A, B) when using criterion A to guide WS.

\noindent\textbf{Coverage correlation of the same groups' criteria.} 
There are $14$ criterion pairs in the same groups: (BC, DBC), (DBC, BC), (BC, LC), (LC, BC), (BC, WM), (WM, BC), (DBC, LC), (LC, DBC), (DBC, WM), (WM, DBC), (LC, WM), (WM, LC), (TMC, NTMC), and (NTMC, TMC). Among them, the minimum PCC is $0.70$ (TMC, NTMC), the maximum two are $0.99$ (NTMC, TMC) and $0.98$ (DBC, BC), and both the mean and median values are closely $0.88$. Besides, all of the $p$-values are smaller than $0.005$.

The criterion pairs among them we are most concerned with are (DBC, BC), (DBC, LC), (DBC, WM), and (NTMC, TMC) since we use DBC to represent (BC, DBC, LC, WM) and NTMC to represent (TMC, NTMC) (see Sec. \ref{subsec:rc}). The PCCs of (DBC, BC), (DBC, LC), (DBC, WM), and (NTMC, TMC) are $0.98$, $0.94$, $0.89$, and $0.99$, respectively.

\noindent\textbf{Coverage correlation of the different groups' criteria.} 
There are $56-14=42$ criterion pairs in the different groups. Among them, the minimum PCC is $-0.29$ (LC, EC), the maximum is $0.71$ (NTMC, LC), and the mean/median value is $0.40$/$0.50$.

\noindent$\blacktriangleright$\textbf{MOSA.}
\mynewcontent{6}{Fig.~\ref{fig:mosa_corr} and~\ref{fig:mosa_corr_tmc_ntmc_oc} show} the coverage value point scatter plot, PCC, and the $p$-value of each criterion pair (A, B) when using criterion A to guide MOSA.

\noindent\textbf{Coverage correlation of the same groups' criteria.} 
Among the $14$ criterion pairs in the same groups, the minimum PCC is $0.68$ (TMC, NTMC), the maximum two are $0.99$ (NTMC, TMC) and $0.98$ (DBC, BC), and the mean/median value is $0.88$/$0.89$. All of the $p$-values are smaller than $0.005$.

The PCCs of the criterion pairs we are most concerned with, (DBC, BC), (DBC, LC), (DBC, WM), and (NTMC, TMC), are $0.98$, $0.94$, $0.90$, and $0.99$, respectively.

\noindent\textbf{Coverage correlation of the different groups' criteria.} 
There are $49-14=35$ pairs in the different groups. Among them, the minimum PCC is $-0.30$ (LC, EC), the maximum is $0.72$ (NTMC, LC), and the mean/median value is $0.42$/$0.53$.

\noindent$\blacktriangleright$\textbf{Difference between WS and MOSA.} 
For a criterion pair, assume that $\text{PCC}_{\text{WS}}$/$\text{PCC}_{\text{MOSA}}$ is the PCC value when the GA is WS/MOSA, and $|\text{PCC}_{\text{WS}}-\text{PCC}_{\text{MOSA}}|$ is the difference between WS and MOSA. The min, max, mean, and median differences are $0.0011$ (NTMC, TMC)/$0.049$ (WM, EC)/$0.013$/$0.011$, respectively.

\noindent\textbf{Analysis.}
Firstly, the coverage correlation of a criterion pair from the same group is significantly higher than that of a criterion pair from different groups, confirming the effectiveness of our criteria grouping. Secondly, the difference in the same criterion pair's correlation with different GAs is tiny, showing the correlation's independence to GAs.
\mytcbox{Answer to RQ4}{The average Pearson Correlation Coefficient of a criterion pair's coverage values from the same group ($0.88$) is significantly higher than that of a pair from different groups ($0.41$), confirming our criteria grouping's effectiveness.}
\begin{figure*}[t]
\centering
\includegraphics[width=1\textwidth]{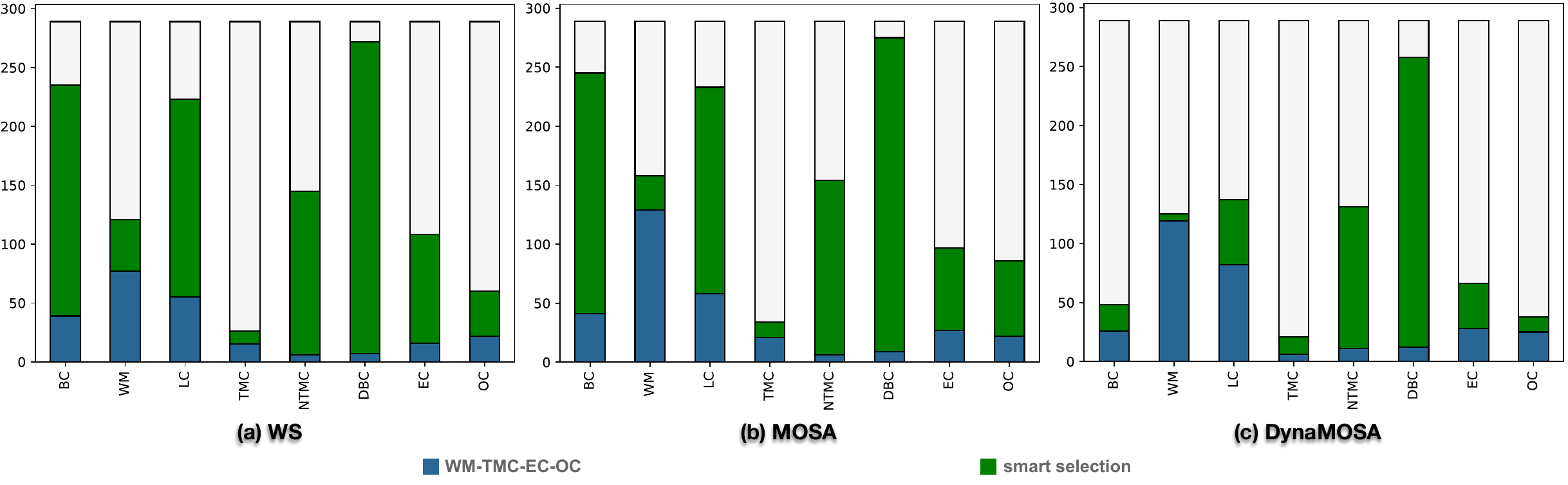}
\caption{\mynewcontent{6}{Significant case summary of smart selection and WT (i.e., WM-TMC-EC-OC)}}\label{fig:wteo}
\end{figure*}
\begin{figure*}[t]
\centering
\includegraphics[width=1\textwidth]{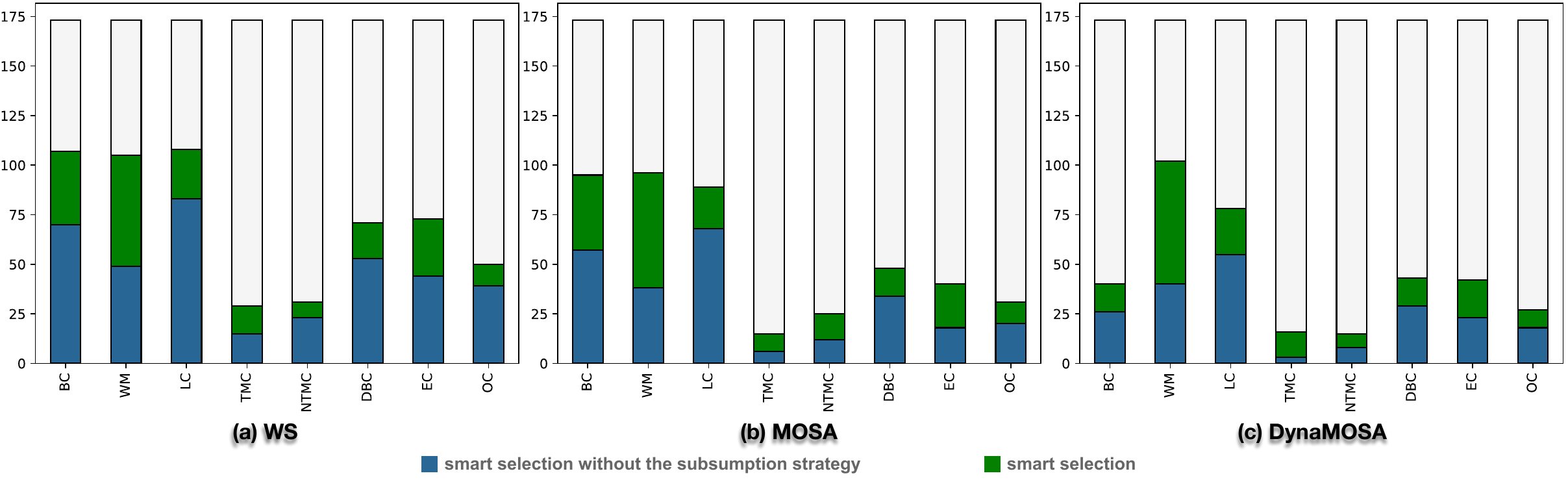}
\caption{\mynewcontent{6}{Significant case summary of smart selection and smart selection with subsumption strategy}}\label{fig:rs}
\end{figure*}
\subsection{\mynewcontent{3}{\textbf{RQ5:} Are representative criteria efficient in guiding SBST?}}\label{subsec:rq-5}
In Sec. \ref{subsec:rc}, we choose four criteria (DBC, NTMC, EC, and OC) to present four groups ((BC, DBC, LC, WM), (TMC, NTMC), (EC), and (OC)), respectively. Regardless of the two groups with only one criterion, \mynewcontent{5}{the main assumptions leading us} to choose DBC and NTMC are (1) DBC is more effective in guiding GAs than EC and OC, and (2) DBC/NTMC can represent BC/TMC, but the opposite does not hold. In this RQ, we aim to verify these assumptions.

\noindent\textbf{Methodology.}
We choose two new criteria combinations to guide GAs as two new approaches: (LC, TMC, EC, OC) (a.k.a, LT, i.e., LC to present group 1 and TMC to present group 2) and (WM, TMC, EC, OC) (a.k.a, WT). Then, we follow RQ1-3's methodology to compare smart selection with LTEO and WTEO with three algorithms: WS, MOSA, and DynaMOSA. Similar to RQ3 (Sec. \ref{subsec:rq3}), we add BC to the guiding criteria since DynaMOSA can not run without BC.

\noindent\textbf{Subjects.}
We take the $400$ classes as the experimental subjects. However, there are $111$ classes on which EvoSuite with at least one of six configurations ($2$ combinations $\times 3$ GAs) crashed. As a result, there are remaining $289$ classes as this RQ's subjects, which is still a large sample according to the previous studies~\cite{arcuri2014hitchhiker, PanichellaMOSA, PanichellaDynaMOSA}.

\begin{table}[htbp]
	\centering
	\caption{Average coverage and size results for smart selection, LT, and WT}
	\label{tab:rq5}
	\footnotesize{\textbf{(a)} WS  (Suite Size: SS (47.63), LT (40.94), WT (41.91))}\\
 \scriptsize
\begin{tabular}{l| l| l |l  }
\hline
approach & SS & LT & WT \\ \hline
BC & \textcolor[RGB]{0,128,28}{56\%} & 52\% & 51\% \\ \hline
WM & \textcolor[RGB]{0,128,28}{60\%} & 58\% & 59\% \\ \hline
LC & \textcolor[RGB]{0,128,28}{60\%} & \textcolor[RGB]{0,128,28}{60\%} & 58\% \\ \hline
TMC & 84\% & 84\% & 84\% \\ \hline

\end{tabular}
\begin{tabular}{l| l| l |l  }
\hline
approach & SS & LT & WT \\ \hline
NTMC & \textcolor[RGB]{0,128,28}{72\%} & 69\% & 68\% \\ \hline
DBC & \textcolor[RGB]{0,128,28}{56\%} & 40\% & 40\% \\ \hline
EC & \textcolor[RGB]{0,128,28}{16.12} & 15.51 & 15.56 \\ \hline
OC & 44\% & 44\% & 44\% \\ \hline

\end{tabular}\\
\footnotesize{\textbf{(b)} MOSA (Suite Size: SS (53.66), LT (45.16), WT (46.10))}\\
 \scriptsize
\begin{tabular}{l| l| l |l  }
\hline
approach & SS & LT & WT \\ \hline
BC & \textcolor[RGB]{0,128,28}{58\%} & 53\% & 52\% \\ \hline
WM & \textcolor[RGB]{0,128,28}{61\%} & 59\% & \textcolor[RGB]{0,128,28}{61\%} \\ \hline
LC & \textcolor[RGB]{0,128,28}{62\%} & 61\% & 59\% \\ \hline
TMC & 84\% & 84\% & 84\% \\ \hline

\end{tabular}
\begin{tabular}{l| l| l |l  }
\hline
approach & SS & LT & WT \\ \hline
NTMC & \textcolor[RGB]{0,128,28}{72\%} & 69\% & 67\% \\ \hline
DBC & \textcolor[RGB]{0,128,28}{57\%} & 41\% & 41\% \\ \hline
EC & \textcolor[RGB]{0,128,28}{17.18} & 16.84 & 16.89 \\ \hline
OC & \textcolor[RGB]{0,128,28}{45\%} & 44\% & 44\% \\ \hline

\end{tabular}\\
\footnotesize{\textbf{(c)} DynaMOSA (Suite Size: SS (57.09), LT (52.54), WT (55.30))}\\
 \scriptsize
\begin{tabular}{l| l| l |l  }
\hline
approach & SS & LT & WT \\ \hline
BC & 58\% & \textcolor[RGB]{0,128,28}{59\%} & 58\% \\ \hline
WM & \textcolor[RGB]{0,128,28}{62\%} & 61\% & \textcolor[RGB]{0,128,28}{62\%} \\ \hline
LC & 62\% & 62\% & \textcolor[RGB]{0,128,28}{63\%} \\ \hline
TMC & 84\% & 84\% & 84\% \\ \hline

\end{tabular}
\begin{tabular}{l| l| l |l  }
\hline
approach & SS & LT & WT \\ \hline
NTMC & \textcolor[RGB]{0,128,28}{72\%} & 70\% & 69\% \\ \hline
DBC & \textcolor[RGB]{0,128,28}{58\%} & 47\% & 47\% \\ \hline
EC & 17.36 & 17.44 & \textcolor[RGB]{0,128,28}{17.56} \\ \hline
OC & 45\% & 45\% & 45\% \\ \hline

\end{tabular}
\end{table}
\noindent\textbf{Result.}
\noindent$\blacktriangleright$\textbf{WS.}

\noindent\textbf{Significant Cases.}
Fig. \ref{fig:lteo} (a) and \ref{fig:wteo} (a) show the significant case summary for SS versus LT and SS versus WT, respectively, when the GA is WS. On average of all criteria, SS-outperforming-LT classes are $115$ ($40\%$), LT-outperforming-SS classes are $28$ ($10\%$), and no-significant classes are $146$ ($50\%$); SS-outperforming-WT classes are $119$ ($41\%$), WM-outperforming-SS classes are $30$ ($10\%$), and no-significant classes are $140$ ($48\%$).

\noindent\textbf{Average Coverage.}
Table \ref{tab:rq5} (a) shows the average coverage for SS, LT, and WT. SS achieves the highest coverage on all criteria. Notably, SS is higher than LT and WT by $16\%$ on direct branch coverage.

\noindent$\blacktriangleright$\textbf{MOSA.}

\noindent\textbf{Significant Cases.}
Fig. \ref{fig:lteo} (b) and \ref{fig:wteo} (b) show the significant case summary for SS versus LT and SS versus WT, respectively, when the GA is MOSA. On average of all criteria, SS-outperforming-LT classes are $121$ ($42\%$), LT-outperforming-SS classes are $32$ ($11\%$), and no-significant classes are $136$ ($47\%$); SS-outperforming-WT classes are $121$ ($42\%$), WM-outperforming-SS classes are $39$ ($13\%$), and no-significant classes are $129$ ($45\%$).

\noindent\textbf{Average Coverage.}
Table \ref{tab:rq5} (b) shows the average coverage for SS, LT, and WT. Like the GA being WS, SS achieves the highest coverage on all criteria; SS is higher than LT and WT by $16\%$ on direct branch coverage.

\noindent$\blacktriangleright$\textbf{DynaMOSA.}

\noindent\textbf{Significant Cases.}
Fig. \ref{fig:lteo} (c) and \ref{fig:wteo} (c) show the significant case summary for SS versus LT and SS versus WT, respectively, when the GA is DynaMOSA. On average of all criteria, SS-outperforming-LT classes are $75$ ($26\%$), LT-outperforming-SS classes are $40$ ($13\%$), and no-significant classes are $177$ ($61\%$); SS-outperforming-WT classes are $64$ ($22\%$), WM-outperforming-SS classes are $39$ ($13\%$), and no-significant classes are $186$ ($64\%$).

\noindent\textbf{Average Coverage.}
Table \ref{tab:rq5} (c) shows the average coverage for SS, LT, and WT. Unlike the GA being WS and MOSA, the coverage values of the three approaches are close on all criteria except for direct branch coverage, on which  SS is higher than LT and WT by $11\%$.

\noindent\textbf{Analysis.}
Smart selection outperforms two criteria combinations (LC, TMC, EC, OC) and (WM, TMC, EC, OC) on almost all criteria (except for EC, OC, and TMC) when the GA is WS or MOSA. This result shows that the representative criteria (mainly DBC) selected from smart selection are more efficient in guiding GAs. The differences between smart selection and two criteria combinations on all criteria (except for DBC) are tiny when the GA is DynaMOSA. The main reason is that We add BC to these two criteria combinations since DynaMOSA cannot run without BC, thus improving the ability to guide DynaMOSA. However, smart selection still outperforms them on DBC by $11\%$, confirming our assumption in Sec. \ref{subsec:rc} that DBC can represent BC, but the opposite does not hold.
\mytcbox{Answer to RQ5}{In most cases, smart selection is better than two criteria combinations (LC, TMC, EC, OC) and (WM, TMC, EC, OC), confirming that the criteria selected from the criteria groups are more efficient in guiding GA and more representative than the other criteria in the same criteria group.}
\subsection{RQ6: How does the subsumption strategy affect the performance of smart selection?}\label{subsec:rq6}
\noindent\textbf{Motivation.}
 We select the representative goals from line coverage and weak mutation by the subsumption relationships (see Sec \ref{subsec:rs}). We need to test how it affects the performance of smart selection.

\noindent\textbf{Subjects.}
From $400$ classes, we select those classes that satisfy this condition: The subsumption strategy can find at least one line coverage goal and one mutant. As a result, $173$ classes are selected.

\noindent\textbf{Methodology.}
We take smart selection without the subsumption strategy (SSWS) as a new approach. To compare SS and SSWS, we follow RQ1's methodology.

\begin{table}[htbp]
	\centering
	\footnotesize
	\caption{Average coverage and size results for smart selection and smart selection without the subsumption strategy}
	\label{tab:rq6}
	\footnotesize{\textbf{(a)} WS (Suite Size: SS (45.18), SSWS (48.08))}\\
\begin{tabular}{l| l| l  }
\hline
approach & SS & SSWS \\ \hline
BC & 47\% & \textcolor[RGB]{0,128,28}{48\%} \\ \hline
WM & 52\% & \textcolor[RGB]{0,128,28}{53\%} \\ \hline
LC & 51\% & \textcolor[RGB]{0,128,28}{53\%} \\ \hline
TMC & \textcolor[RGB]{0,128,28}{79\%} & 78\% \\ \hline
\end{tabular}
\begin{tabular}{l| l| l  }
\hline
approach & SS & SSWS \\ \hline
NTMC & 63\% & 63\% \\ \hline
DBC & 46\% & \textcolor[RGB]{0,128,28}{48\%} \\ \hline
EC & 16.43 & \textcolor[RGB]{0,128,28}{17.46} \\ \hline
OC & 38\% & \textcolor[RGB]{0,128,28}{39\%} \\ \hline
\end{tabular}\\
\footnotesize{\textbf{(b)} MOSA (Suite Size: SS (52.08), SSWS (51.66))}\\
\begin{tabular}{l| l| l  }
\hline
approach & SS & SSWS \\ \hline
BC & 49\% & 49\% \\ \hline
WM & 53\% & 53\% \\ \hline
LC & 53\% & \textcolor[RGB]{0,128,28}{54\%} \\ \hline
TMC & \textcolor[RGB]{0,128,28}{79\%} & 78\% \\ \hline
\end{tabular}
\begin{tabular}{l| l| l  }
\hline
approach & SS & SSWS \\ \hline
NTMC & 63\% & 63\% \\ \hline
DBC & 49\% & 49\% \\ \hline
EC & 18.19 & \textcolor[RGB]{0,128,28}{18.32} \\ \hline
OC & 38\% & \textcolor[RGB]{0,128,28}{39\%} \\ \hline
\end{tabular}\\
\footnotesize{\textbf{(c)} DynaMOSA (Suite Size: SS (57.1), SSWS (54.79))}\\
\begin{tabular}{l| l| l  }
\hline
approach & SS & SSWS \\ \hline
BC & 50\% & 50\% \\ \hline
WM & 54\% & 54\% \\ \hline
LC & 53\% & \textcolor[RGB]{0,128,28}{54\%} \\ \hline
TMC & \textcolor[RGB]{0,128,28}{79\%} & 78\% \\ \hline
\end{tabular}
\begin{tabular}{l| l| l  }
\hline
approach & SS & SSWS \\ \hline
NTMC & 63\% & 63\% \\ \hline
DBC & 49\% & 49\% \\ \hline
EC & 18.26 & \textcolor[RGB]{0,128,28}{18.68} \\ \hline
OC & 39\% & 39\% \\ \hline
\end{tabular}
\end{table}
\noindent\textbf{Result.}
\noindent$\blacktriangleright$\textbf{WS.}

\noindent\textbf{Significant Cases.} Fig. \ref{fig:rs} (a) shows the comparison results of SS and SSWS on $173$ classes with WS. For each criterion, on average, SS-outperforming classes are $25$ ($14.5\%$).  SSWS-outperforming classes are $47$ ($27.2\%$). No-significant classes are $101$ ($58.3\%$).

\noindent\textbf{Average Coverage.} Table \ref{tab:rq6} (a) shows the average coverage for WS. SS outperforms SSWS on top-level method coverage. SSWS outperforms SS on six criteria.

\noindent$\blacktriangleright$\textbf{MOSA.}

\noindent\textbf{Significant Cases.} Fig. \ref{fig:rs} (b) shows the results with GA being MOSA. On average, SS-outperforming classes are $23$ ($13.3\%$). SSWS-outperforming classes are $32$ ($18.5\%$). No-significant classes are $118$ ($68.2\%$). 

\noindent\textbf{Average Coverage.} Table \ref{tab:rq6} (b) shows the average coverage for MOSA. SS outperforms SSWS for one top-level method coverage. SSWS outperforms SS on three criteria.

\noindent$\blacktriangleright$\textbf{DynaMOSA.}

\noindent\textbf{Significant Cases.} Fig. \ref{fig:rs} (c) shows the results with GA being DynaMOSA. On average, SS-outperforming classes are $20$ ($11.6\%$). SSWS-outperforming classes are $25$ ($14.5\%$). No-significant classes are $128$ ($73.9\%$).

\noindent\textbf{Average Coverage.} Table \ref{tab:rq6} (c) shows the average coverage for DynaMOSA. SS outperforms SSWS on top-level method coverage. SSWS outperforms SS on two criteria.

\noindent\textbf{Analysis.} SSWS outperforms slightly SS for most criteria on WS, confirming that an increase in the objectives has a much bigger impact on WS than on MOSA and DynaMOSA. Furthermore, the results are different on line coverage and weak mutation for which SS adds subsets. For three algorithms, SSWS is better in line coverage in terms of the outperforming classes and average coverage. Contrarily, SS is better in weak mutation in terms of the outperforming classes. It indicates that the coverage correlation between (direct) branch coverage and line coverage is stronger than the one between (direct) branch coverage and weak mutation. As for the suite size, Table \ref{tab:rq6} shows that SS and SSWS are similar. Unexpectedly, SS outperforms SSWS on top-level method coverage. We analyze some classes qualitatively. For example, there is a public method named \textit{compare} in an inner class of the class \textit{org.apache.hadoop.mapred}~\cite{HadoopCompareMethod}. The results show that $88$ out of $90$ test suites generated by SS cover this top-level method goal, while only $1$ out of $90$ test suites generated by SSWS covers this goal. We find that this method contains $8$ lines, $2$ branches, and $3$ output goals. EvoSuite skips the branches and output goals in the inner class (lines, methods, and mutants are kept). This class has $350$ branches, $48$ methods, and $10$ output goals. SS selects $14$ lines and $216$ mutants for this class ($0$ lines and $10$ mutants for this method). As a result, if a test directly invokes this method, under SS, at most $(1+10)$ out of $638$ ($2\%$) goals are closer to being covered; under SSWS, the number is $1$ out of $408$ ($0.2\%$). It explains why all algorithms with SSWS tend to ignore this method goal since the gain is tiny. We find that this scenario is common in those large classes that contain short and branch-less methods.
\begin{figure*}[t]
\centering
\includegraphics[width=1\textwidth]{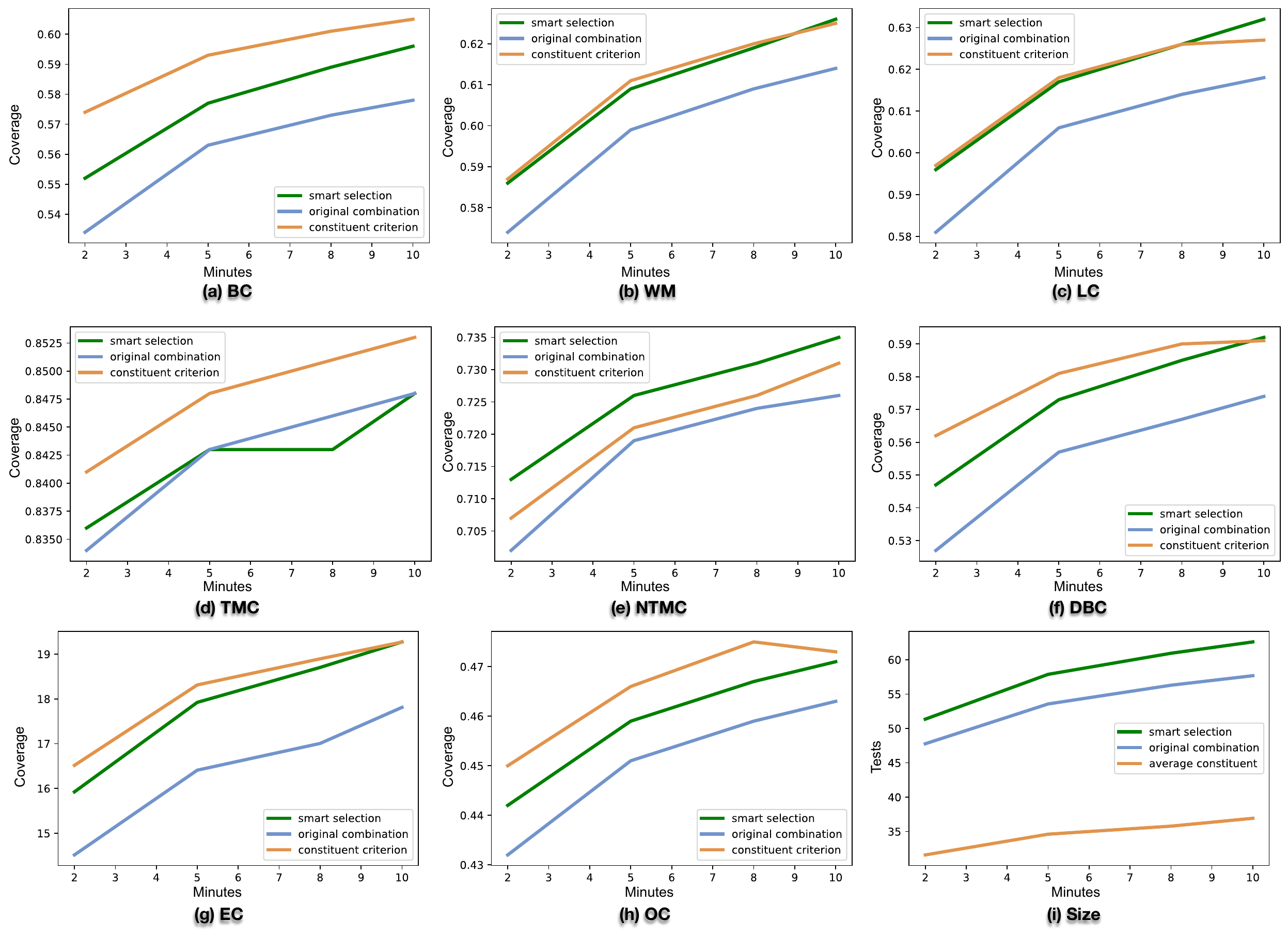}
\caption{\mynewcontent{6}{The coverage and size of three strategies on WS under the various budgets (2-10 minutes)}}\label{fig:budget_suite_all}
\end{figure*}
\begin{figure*}[t]
\centering
\includegraphics[width=1\textwidth]{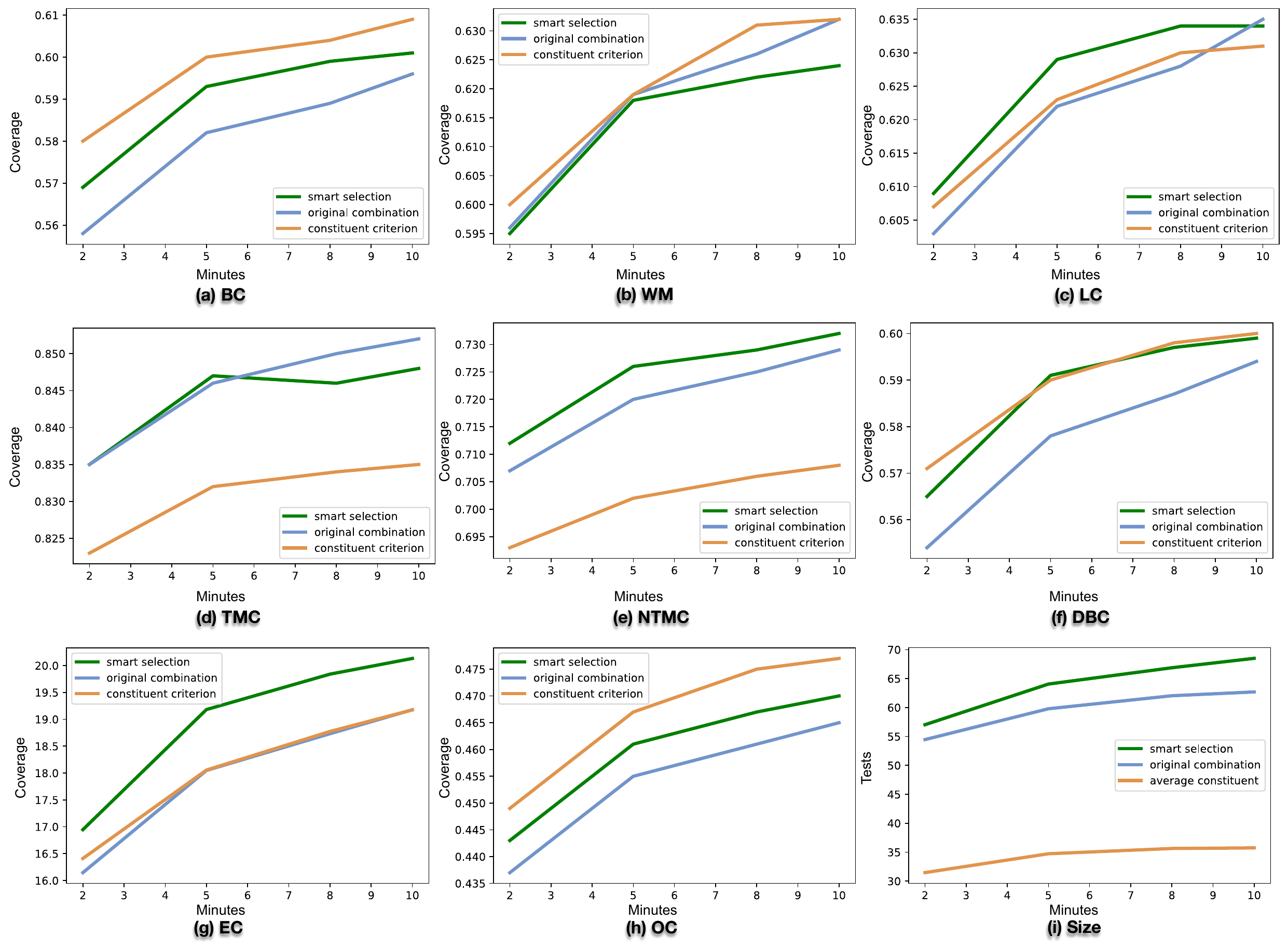}
\caption{\mynewcontent{6}{The coverage and size of three strategies on MOSA under the various budgets (2-10 minutes)}}\label{fig:budget_mosa_all}
\end{figure*}
\begin{figure*}[t]
\centering
\includegraphics[width=1\textwidth]{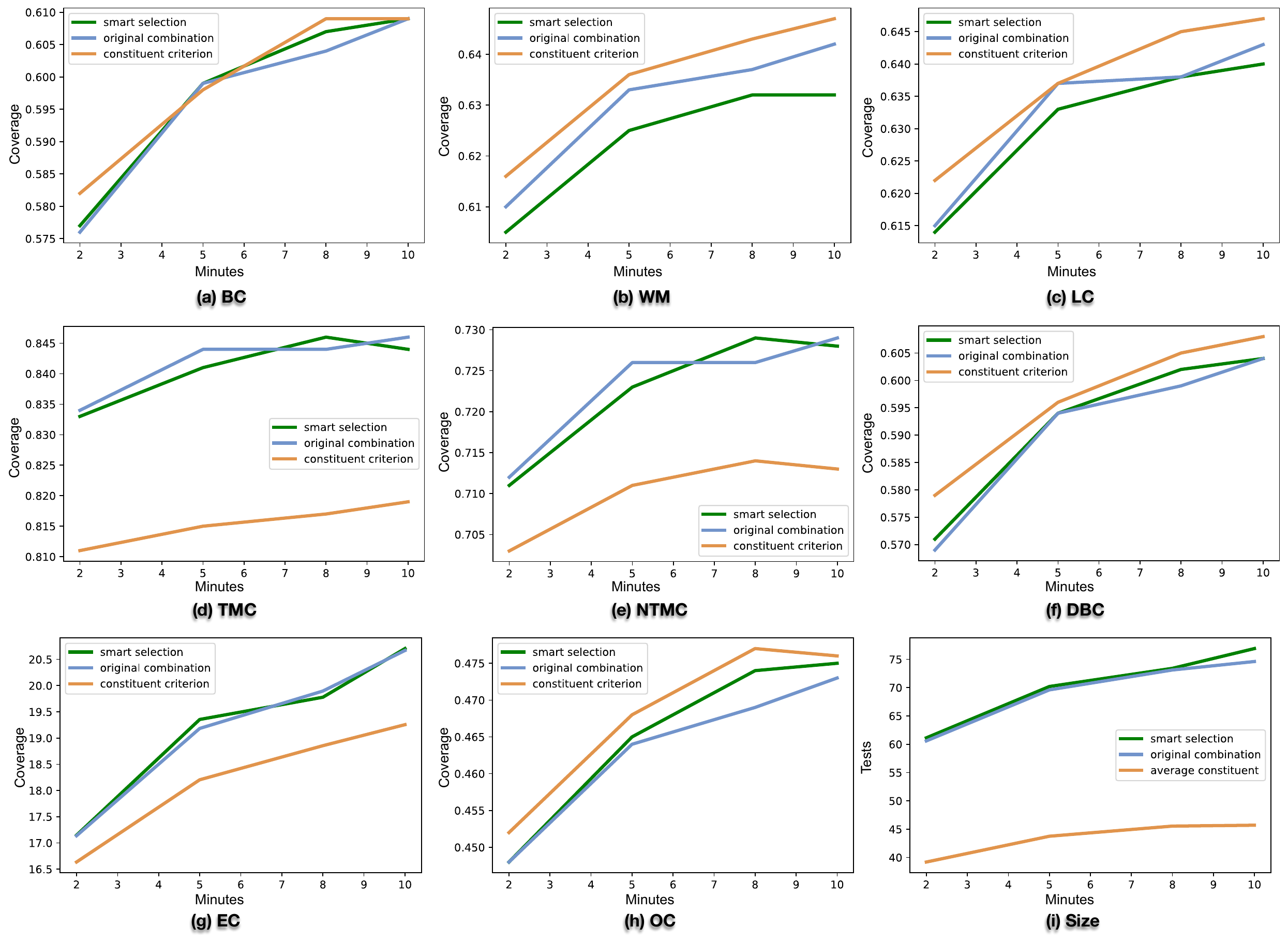}
\caption{\mynewcontent{6}{The coverage and size of three strategies on DynaMOSA under the various budgets (2-10 minutes)}}\label{fig:budget_dynamosa_all}
\end{figure*}
\mytcbox{Answer to RQ6}{Smart selection without the subsumption strategy outperforms slightly smart selection in most criteria on WS (except for WM and TMC). Smart selection outperforms slightly smart selection without the subsumption strategy in WM and TMC on three algorithms.}
\subsection{\mynewcontent{1}{RQ7: How does smart selection perform under different search budgets?}}\label{subsec:rq7}
\noindent\textbf{Motivation.}
In all the above RQs, we evaluate smart selection and other baselines under a fixed search budget, i.e., 2 minutes. In this RQ, we aim to answer how smart selection varies and the gaps among three strategies (i.e., smart selection, the original combination, and the constituent criterion) vary under different search budgets.

\noindent\textbf{Subjects.}
Similar to RQ1-3, we select the $400$ Java classes as the subjects under test.

\noindent\textbf{Configuration.}
Similar to RQ1-3, we run three genetic algorithms (WS, MOSA, and DynaMOSA) with smart selection and another two strategies but under three more search budgets: $5$, $8$, and $10$ minutes. Following the previous study~\cite{Rojas2015CombiningMC}, we repeat $5$ times per Java class for the $10$-minute budget. We repeat  $10$ times per Java class for the $5$-minute and $8$-minute budgets. Regardless of the search budget difference, this RQ's running configuration of EvoSuite is the same as that of RQ1-3.

\noindent\textbf{Result.}

\noindent$\blacktriangleright$\textbf{WS.}

\noindent Fig. \ref{fig:budget_suite_all} shows the average coverage and suite sizes of three strategies on WS under the various budgets. In Sec. \ref{subsec:cc}, the first step of smart selection, we cluster the eight criteria into four groups: (1) BC, DBC, LC, and WM; (2) TMC and NTMC; (3) EC; and  (4) OC. We present the data by groups because the coverage of criteria in different groups changes differently as the search budget increases.

\noindent\textbf{\noindent\ding{172}BC-DBC-LC-WM.}
Fig. \ref{fig:budget_suite_all} (a), (b), (c), and (f) show the coverage change of BC, WM, LC, and DBC. From them, we observe that (1) the coverage of three strategies increases similarly as the budget increases. For example, from 2 to 10 minutes, the BC increase of smart selection/the original combination/the constituent criterion is $4\%$/$4\%$/$3\%$. (2) The coverage gap keeps stable from 2 to 10 minutes. For example,  when the budget is \mynewcontent{5}{two} minutes, the BC gap between smart selection and the original combination is $2\%$, and the gap between the constituent criterion and smart selection is $2\%$. When the budget is $10$ minutes, the gap between smart selection and the original combination is $2\%$ too, and the gap between the constituent criterion and smart selection is $1\%$; (3) Compared to the other two criteria, the LC and WM coverage of the constituent criterion and smart selection is close but keeps higher than the original combination by nearly $2\%$ as the budget changes.

\noindent\textbf{\noindent\ding{173}TMC-NTMC.}
Fig. \ref{fig:budget_suite_all} (d) and (e) show the coverage change of TMC and NTMC. As the budget increases, (1) the coverage increase of the three strategies is tiny. For example, from $2$ to $10$ minutes, the TMC increase of smart selection/the original combination/the constituent criterion is $0.7\%$/$1.5\%$/$1.7\%$; (2)  The coverage gap keeps tiny. For example, the highest coverage of TMC is $85.3\%$, reached by the constituent criterion with the $10$-minute budget. The lowest coverage is $83\%$, reached by the original combination with the $2$-minute budget. The difference is only $2.3\%$. For comparison, the corresponding data of BC is $7\%$.

\noindent\textbf{\noindent\ding{174}\ding{175}EC and OC.}
Fig. \ref{fig:budget_suite_all} (g) and (h) show the coverage change of EC and OC. The coverage changes of them are similar. As the budget increases, the coverage of the constituent criterion and smart selection is close, and there are stable coverage gaps between the original combination and the other two strategies.

\noindent\textbf{Size.}
Fig. \ref{fig:budget_suite_all} (i) shows the suite size change. The gaps between the three strategies are stable  from $2$ to $10$ minutes. For all budgets, smart selection generates an average of nearly $5$ more tests than the original combination. The original combination generates $19$ more tests than the average constituent criterion.

\noindent\textbf{Analysis.} 
For most criteria, smart selection outperforms the original combination no matter how the budget changes. The situations of comparing the constituent criterion and the others vary on different criteria.

\noindent\textbf{\noindent\ding{172}BC-DBC-LC-WM.}
Firstly, the data of LC and WM shows that the coverage of smart selection and the constituent criterion is close and outperforms that of the original combination. This fact confirms the two fundamental assumptions of our methodology: (1) The negative impact of increasing the optimization goals of LC and WM on the algorithm WS exists, even if the budget is increased to \num{10} minutes; (2) The representation of BC/DBC to LC and WM is significant. After smart selection removes most of the goals of LC and WM, their coverage is stably higher than that of the original combination. Secondly, the data of BC and DBC shows that the constituent criterion outperforms smart selection and smart selection outperforms the original combination. This fact indicates that the other remaining fitness functions  (e.g., the ones of OC) still hinder BC/DBC's fitness functions, undermining their effectiveness in guiding WS.

\noindent\textbf{\noindent\ding{173}TMC-NTMC.} 
Firstly, although the coverage of the three strategies is not the same, the differences between them are tiny compared to group \num{1}. Secondly, this group has significantly higher coverage than the other groups. For example, the lowest coverage of NTMC is \num{0.7} (see Fig. \ref{fig:budget_suite_all} (e)), but the highest coverage of BC is nearly \num{0.6} (see Fig. \ref{fig:budget_suite_all} (a)). These facts indicate that the goals of TMC/NTMC are easier to be covered than the others since they only require that a method is invoked (without exceptions). As a result, their coverage is more robust when combining multiple criteria.

\noindent\textbf{\noindent\ding{174}\ding{175}EC and OC.}
For coverage of EC and OC, the difference between the constituent criterion and smart selection is tiny, and they outperform the original combination. Hence, combining multiple criteria negatively impacts their coverage, although their own fitness functions are weak in guiding genetic algorithms.

\noindent\textbf{Size.}
Firstly, with the budget and coverage increase, the three strategies' suite sizes also increase. Secondly, the sizes of the constituent criterion are much smaller than the sizes of the other two since the latter two strategies need to generate more tests for more goals brought by combining multiple criteria.

\noindent$\blacktriangleright$\textbf{MOSA.}

\noindent  Fig. \ref{fig:budget_mosa_all} shows the average coverage and suite sizes of three strategies on MOSA under the various budgets. Similar to WS, we present data by groups.

\noindent\textbf{\noindent\ding{172}BC-DBC-LC-WM.}
Fig. \ref{fig:budget_mosa_all} (a), (b), (c), and (f) show the coverage change of BC, WM, LC, and DBC. As the budget increases, (1) for BC and DBC, the coverage of smart selection and the constituent criterion is close and outperforms slightly that of the original combination; (2) For LC, the coverage gaps between the three strategies are tiny; (2) For WM, the coverage gap between smart selection and the other two gradually increases. The coverage of smart selection is nearly $1\%$ behind when the budget is \num{10} minutes.

\noindent\textbf{\noindent\ding{173}TMC-NTMC.}
Fig. \ref{fig:budget_mosa_all} (d) and (e) show the coverage change of TMC and NTMC. As the budget increases, (1) the coverage increase of the three strategies is tiny. For instance, from \num{2} to \num{10} minutes, the TMC coverage increase of smart selection/the original combination/the constituent criterion is $1.3\%/1.7\%/1.2\%$; (2) The coverage of smart selection and the original combination is close and stably outperforms that of the constituent criterion.

\noindent\textbf{\noindent\ding{174}\ding{175}EC and OC.}
Fig. \ref{fig:budget_mosa_all} (g) and (h) show the coverage change of EC and OC. As the budget increases, for EC, the coverage of the original combination and the constituent criterion is close and smart selection outperforms them by nearly one exception. For OC, the constituent criterion outperforms smart selection, and smart selection outperforms the original combination. 

\noindent\textbf{Size.}
Fig. \ref{fig:budget_mosa_all} (i) shows the suite size change. For all budgets, smart selection generates an average of nearly $4$ more tests than the original combination. The original combination generates $25$ more tests than the average constituent criterion.

\noindent\textbf{Analysis.} 
Similar to WS, for most criteria, smart selection outperforms the original combination no matter how the budget changes. The situations of comparing the constituent criterion and the others vary on different criteria.

\noindent\textbf{\noindent\ding{172}BC-DBC-LC-WM.}
Firstly, the data of BC and DBC shows that the coverage of smart selection and the constituent criterion outperforms that of the original combination, even when the budget increases to \num{10} minutes. This fact confirms our findings in RQ2 (see Sec. \ref{subsec:rq2}): Too many objectives also affect the multi-objective algorithms, e.g., MOSA. Secondly, the data of WM shows that as the budget increases, the coverage of the original combination and the constituent criterion is close, and the lag of smart selection is increasing. On the contrary, the data of LC shows that The coverage lead of smart selection is obvious under most budgets. These facts confirm our findings in RQ6 (see \ref{subsec:rq6}): The coverage correlation between (direct) branch coverage and line coverage is stronger than the one between  (direct) branch coverage and weak mutation.

\noindent\textbf{\noindent\ding{173}TMC-NTMC.} 
Firstly, the coverage differences between smart selection and the original combination are tiny. This fact confirms that the coverage of TMC and NTMC is robust when combing multiple criteria. Secondly, smart selection and the original combination stably outperform the constituent criterion as the budget increases. This could be due to the weak guidance of TMC and NTMC's fitness functions since such a fitness function can only tell MOSA whether a method is covered, downgrading MOSA to the random search. As a result, when combining fitness functions with better guidance (e.g., ones of BC/DBC), MOSA has a chance to cover more methods.

\noindent\textbf{\noindent\ding{174}\ding{175}EC and OC.}
For both EC and OC, smart selection outperforms the original combination, confirming that combining more criteria hurts their coverage. Secondly, for EC, smart selection outperforms the constituent criterion, contrary to the case of OC. This fact indicates that the fitness function of OC is better than that of EC in guiding GAs.

\noindent\textbf{Size.}
Likewise, the three strategies' suite sizes also increase as the budget and coverage increase. Secondly, the test sizes of the constituent criterion are much smaller than the other two since more criteria call for more tests.

\noindent$\blacktriangleright$\textbf{DynaMOSA.}

\noindent  Fig. \ref{fig:budget_dynamosa_all} shows the average coverage and suite sizes of three strategies on MOSA under the various budgets. Similar to WS and MOSA, we present data by groups.

\noindent\textbf{\noindent\ding{172}BC-DBC-LC-WM.}
Fig. \ref{fig:budget_dynamosa_all} (a), (b), (c), and (f) show the coverage change of BC, WM, LC, and DBC. As the budget increases, (1) for BC and DBC, the coverage of three strategies is close; (2) For LC and WM, the constituent criterion outperforms the original combination, and the original combination outperforms smart selection. Note that the advantage of the constituent criterion/the original combination over the original combination/smart selection in LC is tinier than in WM.

\noindent\textbf{\noindent\ding{173}TMC-NTMC.}
Fig. \ref{fig:budget_dynamosa_all} (d) and (e) show the coverage change of TMC and NTMC. Like MOSA, (1) the coverage increase of the three strategies is tiny; and (2) the coverage of smart selection and the original combination is close and outperforms that of the constituent criterion.

\noindent\textbf{\noindent\ding{174}\ding{175}EC and OC.}
Fig. \ref{fig:budget_dynamosa_all} (g) and (h) show the coverage change of EC and OC. For EC, the performances of smart selection and the original combination are close and are better than that of the constituent criterion by nearly one exception. For OC, the performances of the three strategies keep close, and the constituent criterion/smart selection maintains a slight advantage  (nearly $0.3\%/0.2\%$) over smart selection/the original combination.

\noindent\textbf{Size.}
Fig. \ref{fig:budget_dynamosa_all} (i) shows the suite size change. The sizes of generated tests by smart selection and the original combination are nearly equal for all budgets. They generate nearly $25$ more tests than the average constituent criterion.

\noindent\textbf{Analysis.}
Compared to WS and MOSA, the comparing situation of the three strategies on DynaMOSA is different. For most criteria, the performance of the original combination is close to or slightly better than the other two strategies, showing the better robustness of DynaMOSA over WS and MOSA when facing combining multiple criteria.

\noindent\textbf{\noindent\ding{172}BC-DBC-LC-WM.}
Firstly, the data of BC and DBC shows that the coverage of the three strategies is close. Secondly, the data of WM and LC shows that the constituent criterion outperforms the original combination, and the original combination outperforms smart selection. These facts confirm our findings in RQ3 (see Sec. \ref{subsec:rq3}): DynaMOSA is more robust in handling combining multiple criteria. Especially the fact that the original combination outperforms smart selection for WM and LC shows that the fitness functions of LC and WM removed by smart selection but kept by the original combination help DynaMOSA to cover the corresponding goals of LC and WM. Furthermore,  we notice that the constituent criterion outperforms the original combination for WM and LC. This fact indicates that when the fitness functions of a criterion (e.g., WM and LC) have weak guidance for GAs, combining multiple criteria still harms DynaMOSA, which is potential because more criteria lower the search weight of each fitness function.

\noindent\textbf{\noindent\ding{173}TMC-NTMC.} 
Like MOSA, the coverage gaps between smart selection and the original combination are tiny. Secondly, smart selection and the original combination stably outperform the constituent criterion as the budget increases. This could be due to the fact that a fitness function of TMC/NTMC can only tell GAs whether a method is covered, downgrading them to the random search. As a result, when combining fitness functions with better guidance (e.g., ones of BC/DBC), DynaMOSA has a chance to cover more methods.

\noindent\textbf{\noindent\ding{174}\ding{175}EC and OC.}
Firstly, the situation of EC is like TMC/NTMC: the performances of smart selection and the original combination are close and better than the constituent criterion, which is potential because, similar to TMC/NTMC, EC's fitness functions can only tell GAs whether a method is covered, downgrading them to the random search. Combining multiple criteria brings a more extensive search space for DynaMOSA, increasing its possibility of catching exceptions, e.g., killing a mutant may trigger an exception. Secondly, the situation of OC is like BC/DBC: the coverage of the three strategies is close, confirming the better robustness of DynaMOSA over WS and MOSA when facing more coverage criteria.

\noindent\textbf{Size.}
Like WS and MOSA, the three strategies' suite sizes also increase as the budget and coverage increase. Secondly, the test sizes of the constituent criterion are much smaller than the other two since more criteria call for more tests. Furthermore, DynaMOSA's size gap between smart selection and the original combination is smaller than that of WS and MOSA since the coverage of these two strategies is close, showing DynaMOSA's robustness again.

\mytcbox{Answer to RQ7}{From $2$ to $10$ minutes:
\begin{enumerate}
\item \textbf{WS:} For most criteria, the coverage rank of the three strategies is (1) CC, (2) SS, and (3) OC.
\item \textbf{MOSA:} Like WS, for most criteria, OC obtains the lowest coverage.
\item \textbf{DynaMOSA:} For most criteria, the coverage of the three strategies is close.
\end{enumerate}}
\begin{figure*}[t]
\centering
\includegraphics[width=1\textwidth]{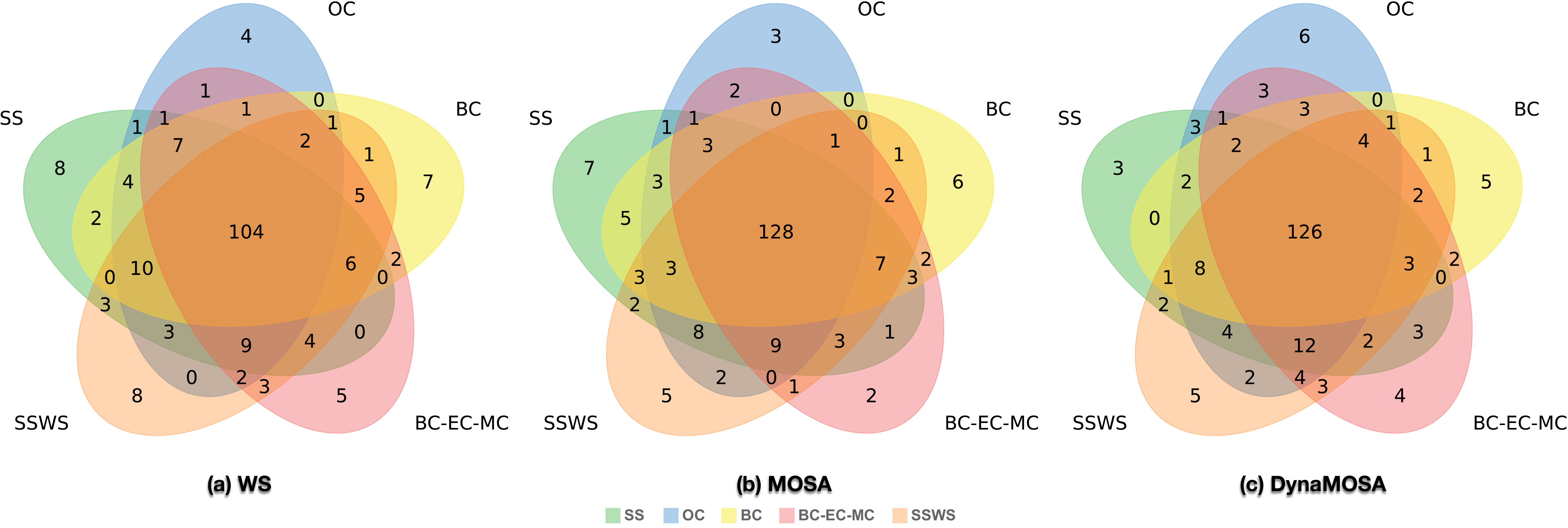}
\caption{\mynewcontent{6}{Overlapping detected bugs of different strategies under three algorithms (SS: smart selection; OC: original combination; BC: branch coverage; BC-EC-MC: the combination of branch, exception, and method coverage; SSWS: smart selection without subsumption strategy)}}\label{fig:bug_venn}
\end{figure*}
\subsection{\mynewcontent{6}{RQ8: How does smart selection affect detecting faults?}}\label{subsec:rq8}
\noindent\mynewcontent{6}{\textbf{Motivation.}
In all previous RQs, we assess the coverage performance of smart selection and other baselines. In this RQ, we focus on their performance in detecting real faults.}
\begin{table*}[tbp]
    \centering
    \tiny

    \caption{\mynewcontent{6}{Bug detection summary for each strategy with WS}}\label{tab:ws_bugs}
       \begin{tabular}{llccc|ccc|ccc|ccc|ccc}
\toprule
Project & Bugs & \nrcols{3}{Smart Selection} & \nrcols{3}{Original Combination}  & \nrcols{3}{Branch Coverage} & \nrcols{3}{BC-EC-MC} & \nrcols{3}{SSWS} \\
\cmidrule(l{1pt}r{1pt}){3-5}
\cmidrule(l{1pt}r{1pt}){6-8}
\cmidrule(l{1pt}r{1pt}){9-11}
\cmidrule(l{1pt}r{1pt}){12-14}
\cmidrule(l{1pt}r{1pt}){15-17}
 & & Detected & Detect$\%$ & Fail$\%$ & Detected & Detect$\%$ & Fail$\%$ & Detected & Detect$\%$ & Fail$\%$ & Detected & Detect$\%$ & Fail$\%$ & Detected & Detect$\%$ & Fail$\%$  \\
\midrule

 Chart  & \significant{\num{24}} & \significant{\num{19}} & \significant{\num{0.57}} & \significant{\num{0.05}}  & \significant{\num{17}} & \significant{\num{0.57}} & \significant{\num{0.07}} & \significant{\num{18}} & \significant{\num{0.56}} & \significant{\num{0.05}} & \significant{\num{15}} & \significant{\num{0.48}} & \significant{\num{0.06}} & \significant{\num{16}} & \significant{\num{0.55}} & \significant{\num{0.10}}  \\

\tabprojectlinespace

 Cli  & \significant{\num{28}} & \significant{\num{14}} & \significant{\num{0.42}} & \significant{\num{0.01}}  & \significant{\num{14}} & \significant{\num{0.41}} & \significant{\num{0.03}} & \significant{\num{15}} & \significant{\num{0.41}} & \significant{\num{0.01}} & \significant{\num{12}} & \significant{\num{0.38}} & \significant{\num{0.00}} & \significant{\num{15}} & \significant{\num{0.38}} & \significant{\num{0.00}}  \\

\tabprojectlinespace

 Closure  & \significant{\num{38}} & \significant{\num{8}} & \significant{\num{0.16}} & \significant{\num{0.17}}  & \significant{\num{9}} & \significant{\num{0.14}} & \significant{\num{0.14}} & \significant{\num{5}} & \significant{\num{0.07}} & \significant{\num{0.06}} & \significant{\num{10}} & \significant{\num{0.15}} & \significant{\num{0.21}} & \significant{\num{7}} & \significant{\num{0.12}} & \significant{\num{0.16}}  \\

\tabprojectlinespace

 Compress  & \significant{\num{29}} & \significant{\num{13}} & \significant{\num{0.27}} & \significant{\num{0.08}}  & \significant{\num{13}} & \significant{\num{0.26}} & \significant{\num{0.05}} & \significant{\num{15}} & \significant{\num{0.29}} & \significant{\num{0.03}} & \significant{\num{11}} & \significant{\num{0.23}} & \significant{\num{0.06}} & \significant{\num{9}} & \significant{\num{0.25}} & \significant{\num{0.05}}  \\

\tabprojectlinespace

 Csv  & \significant{\num{8}} & \significant{\num{7}} & \significant{\num{0.67}} & \significant{\num{0.03}}  & \significant{\num{6}} & \significant{\num{0.71}} & \significant{\num{0.09}} & \significant{\num{6}} & \significant{\num{0.57}} & \significant{\num{0.00}} & \significant{\num{6}} & \significant{\num{0.64}} & \significant{\num{0.00}} & \significant{\num{6}} & \significant{\num{0.69}} & \significant{\num{0.09}}  \\

\tabprojectlinespace

 Gson  & \significant{\num{14}} & \significant{\num{5}} & \significant{\num{0.36}} & \significant{\num{0.01}}  & \significant{\num{5}} & \significant{\num{0.36}} & \significant{\num{0.01}} & \significant{\num{5}} & \significant{\num{0.31}} & \significant{\num{0.01}} & \significant{\num{5}} & \significant{\num{0.36}} & \significant{\num{0.03}} & \significant{\num{5}} & \significant{\num{0.32}} & \significant{\num{0.04}}  \\

\tabprojectlinespace

 JacksonCore  & \significant{\num{22}} & \significant{\num{13}} & \significant{\num{0.41}} & \significant{\num{0.04}}  & \significant{\num{11}} & \significant{\num{0.37}} & \significant{\num{0.04}} & \significant{\num{14}} & \significant{\num{0.51}} & \significant{\num{0.05}} & \significant{\num{11}} & \significant{\num{0.46}} & \significant{\num{0.08}} & \significant{\num{13}} & \significant{\num{0.46}} & \significant{\num{0.01}}  \\

\tabprojectlinespace

 JacksonDatabind  & \significant{\num{75}} & \significant{\num{30}} & \significant{\num{0.27}} & \significant{\num{0.11}}  & \significant{\num{27}} & \significant{\num{0.23}} & \significant{\num{0.12}} & \significant{\num{22}} & \significant{\num{0.20}} & \significant{\num{0.11}} & \significant{\num{26}} & \significant{\num{0.26}} & \significant{\num{0.14}} & \significant{\num{31}} & \significant{\num{0.27}} & \significant{\num{0.13}}  \\

\tabprojectlinespace

 JacksonXml  & \significant{\num{6}} & \significant{\num{2}} & \significant{\num{0.23}} & \significant{\num{0.10}}  & \significant{\num{2}} & \significant{\num{0.23}} & \significant{\num{0.02}} & \significant{\num{2}} & \significant{\num{0.33}} & \significant{\num{0.03}} & \significant{\num{2}} & \significant{\num{0.32}} & \significant{\num{0.07}} & \significant{\num{2}} & \significant{\num{0.28}} & \significant{\num{0.07}}  \\

\tabprojectlinespace

 Jsoup  & \significant{\num{58}} & \significant{\num{21}} & \significant{\num{0.28}} & \significant{\num{0.06}}  & \significant{\num{20}} & \significant{\num{0.31}} & \significant{\num{0.04}} & \significant{\num{20}} & \significant{\num{0.27}} & \significant{\num{0.01}} & \significant{\num{22}} & \significant{\num{0.29}} & \significant{\num{0.06}} & \significant{\num{22}} & \significant{\num{0.29}} & \significant{\num{0.06}}  \\

\tabprojectlinespace

 JxPath  & \significant{\num{12}} & \significant{\num{6}} & \significant{\num{0.38}} & \significant{\num{0.28}}  & \significant{\num{6}} & \significant{\num{0.30}} & \significant{\num{0.13}} & \significant{\num{7}} & \significant{\num{0.40}} & \significant{\num{0.09}} & \significant{\num{7}} & \significant{\num{0.33}} & \significant{\num{0.22}} & \significant{\num{5}} & \significant{\num{0.22}} & \significant{\num{0.17}}  \\

\tabprojectlinespace

 Lang  & \significant{\num{44}} & \significant{\num{23}} & \significant{\num{0.40}} & \significant{\num{0.04}}  & \significant{\num{19}} & \significant{\num{0.35}} & \significant{\num{0.02}} & \significant{\num{22}} & \significant{\num{0.44}} & \significant{\num{0.01}} & \significant{\num{24}} & \significant{\num{0.50}} & \significant{\num{0.01}} & \significant{\num{29}} & \significant{\num{0.48}} & \significant{\num{0.03}}  \\

\tabprojectlinespace

 Math  & \significant{\num{2}} & \significant{\num{1}} & \significant{\num{0.45}} & \significant{\num{0.00}}  & \significant{\num{1}} & \significant{\num{0.50}} & \significant{\num{0.00}} & \significant{\num{1}} & \significant{\num{0.50}} & \significant{\num{0.00}} & \significant{\num{1}} & \significant{\num{0.50}} & \significant{\num{0.00}} & \significant{\num{1}} & \significant{\num{0.50}} & \significant{\num{0.00}}  \\

\midrule

 Overall  & \significant{\num{360}} & \significant{\num{162}} & \significant{\num{0.33}} & \significant{\num{0.08}}  & \significant{\num{150}} & \significant{\num{0.31}} & \significant{\num{0.07}} & \significant{\num{152}} & \significant{\num{0.32}} & \significant{\num{0.05}} & \significant{\num{152}} & \significant{\num{0.33}} & \significant{\num{0.09}} & \significant{\num{161}} & \significant{\num{0.33}} & \significant{\num{0.08}}  \\

\bottomrule
\end{tabular}
\end{table*}
\begin{table*}[tbp]
    \centering
    \tiny

    \caption{\mynewcontent{6}{Bug detection summary for each strategy with MOSA}}\label{tab:mosa_bugs}
       \begin{tabular}{llccc|ccc|ccc|ccc|ccc}
\toprule
Project & Bugs & \nrcols{3}{Smart Selection} & \nrcols{3}{Original Combination}  & \nrcols{3}{Branch Coverage} & \nrcols{3}{BC-EC-MC} & \nrcols{3}{SSWS} \\
\cmidrule(l{1pt}r{1pt}){3-5}
\cmidrule(l{1pt}r{1pt}){6-8}
\cmidrule(l{1pt}r{1pt}){9-11}
\cmidrule(l{1pt}r{1pt}){12-14}
\cmidrule(l{1pt}r{1pt}){15-17}
 & & Detected & Detect$\%$ & Fail$\%$ & Detected & Detect$\%$ & Fail$\%$ & Detected & Detect$\%$ & Fail$\%$ & Detected & Detect$\%$ & Fail$\%$ & Detected & Detect$\%$ & Fail$\%$  \\
\midrule

 Chart  & \significant{\num{24}} & \significant{\num{18}} & \significant{\num{0.66}} & \significant{\num{0.05}}  & \significant{\num{16}} & \significant{\num{0.56}} & \significant{\num{0.05}} & \significant{\num{14}} & \significant{\num{0.50}} & \significant{\num{0.02}} & \significant{\num{17}} & \significant{\num{0.63}} & \significant{\num{0.07}} & \significant{\num{19}} & \significant{\num{0.63}} & \significant{\num{0.04}}  \\

\tabprojectlinespace

 Cli  & \significant{\num{28}} & \significant{\num{17}} & \significant{\num{0.40}} & \significant{\num{0.05}}  & \significant{\num{16}} & \significant{\num{0.39}} & \significant{\num{0.01}} & \significant{\num{16}} & \significant{\num{0.41}} & \significant{\num{0.00}} & \significant{\num{14}} & \significant{\num{0.38}} & \significant{\num{0.00}} & \significant{\num{16}} & \significant{\num{0.43}} & \significant{\num{0.01}}  \\

\tabprojectlinespace

 Closure  & \significant{\num{38}} & \significant{\num{10}} & \significant{\num{0.14}} & \significant{\num{0.17}}  & \significant{\num{6}} & \significant{\num{0.13}} & \significant{\num{0.16}} & \significant{\num{6}} & \significant{\num{0.07}} & \significant{\num{0.06}} & \significant{\num{9}} & \significant{\num{0.11}} & \significant{\num{0.18}} & \significant{\num{9}} & \significant{\num{0.13}} & \significant{\num{0.18}}  \\

\tabprojectlinespace

 Compress  & \significant{\num{29}} & \significant{\num{16}} & \significant{\num{0.34}} & \significant{\num{0.07}}  & \significant{\num{14}} & \significant{\num{0.36}} & \significant{\num{0.08}} & \significant{\num{15}} & \significant{\num{0.35}} & \significant{\num{0.03}} & \significant{\num{16}} & \significant{\num{0.43}} & \significant{\num{0.07}} & \significant{\num{14}} & \significant{\num{0.33}} & \significant{\num{0.11}}  \\

\tabprojectlinespace

 Csv  & \significant{\num{8}} & \significant{\num{5}} & \significant{\num{0.57}} & \significant{\num{0.09}}  & \significant{\num{6}} & \significant{\num{0.59}} & \significant{\num{0.03}} & \significant{\num{6}} & \significant{\num{0.61}} & \significant{\num{0.00}} & \significant{\num{6}} & \significant{\num{0.59}} & \significant{\num{0.04}} & \significant{\num{6}} & \significant{\num{0.65}} & \significant{\num{0.05}}  \\

\tabprojectlinespace

 Gson  & \significant{\num{14}} & \significant{\num{6}} & \significant{\num{0.34}} & \significant{\num{0.02}}  & \significant{\num{5}} & \significant{\num{0.36}} & \significant{\num{0.04}} & \significant{\num{5}} & \significant{\num{0.33}} & \significant{\num{0.05}} & \significant{\num{5}} & \significant{\num{0.35}} & \significant{\num{0.02}} & \significant{\num{5}} & \significant{\num{0.36}} & \significant{\num{0.02}}  \\

\tabprojectlinespace

 JacksonCore  & \significant{\num{22}} & \significant{\num{14}} & \significant{\num{0.53}} & \significant{\num{0.05}}  & \significant{\num{15}} & \significant{\num{0.58}} & \significant{\num{0.05}} & \significant{\num{14}} & \significant{\num{0.43}} & \significant{\num{0.08}} & \significant{\num{12}} & \significant{\num{0.40}} & \significant{\num{0.07}} & \significant{\num{14}} & \significant{\num{0.58}} & \significant{\num{0.04}}  \\

\tabprojectlinespace

 JacksonDatabind  & \significant{\num{75}} & \significant{\num{35}} & \significant{\num{0.27}} & \significant{\num{0.16}}  & \significant{\num{30}} & \significant{\num{0.28}} & \significant{\num{0.11}} & \significant{\num{27}} & \significant{\num{0.25}} & \significant{\num{0.12}} & \significant{\num{27}} & \significant{\num{0.25}} & \significant{\num{0.15}} & \significant{\num{30}} & \significant{\num{0.28}} & \significant{\num{0.14}}  \\

\tabprojectlinespace

 JacksonXml  & \significant{\num{6}} & \significant{\num{2}} & \significant{\num{0.33}} & \significant{\num{0.02}}  & \significant{\num{2}} & \significant{\num{0.30}} & \significant{\num{0.03}} & \significant{\num{2}} & \significant{\num{0.33}} & \significant{\num{0.07}} & \significant{\num{2}} & \significant{\num{0.32}} & \significant{\num{0.12}} & \significant{\num{2}} & \significant{\num{0.32}} & \significant{\num{0.02}}  \\

\tabprojectlinespace

 Jsoup  & \significant{\num{58}} & \significant{\num{26}} & \significant{\num{0.32}} & \significant{\num{0.05}}  & \significant{\num{23}} & \significant{\num{0.28}} & \significant{\num{0.05}} & \significant{\num{24}} & \significant{\num{0.30}} & \significant{\num{0.02}} & \significant{\num{23}} & \significant{\num{0.33}} & \significant{\num{0.07}} & \significant{\num{24}} & \significant{\num{0.26}} & \significant{\num{0.04}}  \\

\tabprojectlinespace

 JxPath  & \significant{\num{12}} & \significant{\num{8}} & \significant{\num{0.41}} & \significant{\num{0.20}}  & \significant{\num{8}} & \significant{\num{0.28}} & \significant{\num{0.17}} & \significant{\num{10}} & \significant{\num{0.57}} & \significant{\num{0.13}} & \significant{\num{10}} & \significant{\num{0.52}} & \significant{\num{0.27}} & \significant{\num{7}} & \significant{\num{0.48}} & \significant{\num{0.33}}  \\

\tabprojectlinespace

 Lang  & \significant{\num{44}} & \significant{\num{29}} & \significant{\num{0.48}} & \significant{\num{0.05}}  & \significant{\num{22}} & \significant{\num{0.45}} & \significant{\num{0.04}} & \significant{\num{27}} & \significant{\num{0.53}} & \significant{\num{0.02}} & \significant{\num{23}} & \significant{\num{0.45}} & \significant{\num{0.01}} & \significant{\num{28}} & \significant{\num{0.52}} & \significant{\num{0.05}}  \\

\tabprojectlinespace

 Math  & \significant{\num{2}} & \significant{\num{1}} & \significant{\num{0.50}} & \significant{\num{0.00}}  & \significant{\num{1}} & \significant{\num{0.50}} & \significant{\num{0.00}} & \significant{\num{1}} & \significant{\num{0.50}} & \significant{\num{0.00}} & \significant{\num{1}} & \significant{\num{0.50}} & \significant{\num{0.00}} & \significant{\num{1}} & \significant{\num{0.50}} & \significant{\num{0.00}}  \\

\midrule

 Overall  & \significant{\num{360}} & \significant{\num{187}} & \significant{\num{0.36}} & \significant{\num{0.09}}  & \significant{\num{164}} & \significant{\num{0.35}} & \significant{\num{0.07}} & \significant{\num{167}} & \significant{\num{0.34}} & \significant{\num{0.05}} & \significant{\num{165}} & \significant{\num{0.35}} & \significant{\num{0.09}} & \significant{\num{175}} & \significant{\num{0.37}} & \significant{\num{0.09}}  \\

\bottomrule
\end{tabular}
\end{table*}
\begin{table*}[tbp]
    \centering
    \tiny

    \caption{\mynewcontent{6}{Bug detection summary for each strategy with DynaMOSA}}\label{tab:dynamosa_bugs}
       \begin{tabular}{llccc|ccc|ccc|ccc|ccc}
\toprule
Project & Bugs & \nrcols{3}{Smart Selection} & \nrcols{3}{Original Combination}  & \nrcols{3}{Branch Coverage} & \nrcols{3}{BC-EC-MC} & \nrcols{3}{SSWS} \\
\cmidrule(l{1pt}r{1pt}){3-5}
\cmidrule(l{1pt}r{1pt}){6-8}
\cmidrule(l{1pt}r{1pt}){9-11}
\cmidrule(l{1pt}r{1pt}){12-14}
\cmidrule(l{1pt}r{1pt}){15-17}
 & & Detected & Detect$\%$ & Fail$\%$ & Detected & Detect$\%$ & Fail$\%$ & Detected & Detect$\%$ & Fail$\%$ & Detected & Detect$\%$ & Fail$\%$ & Detected & Detect$\%$ & Fail$\%$  \\
\midrule

 Chart  & \significant{\num{24}} & \significant{\num{18}} & \significant{\num{0.65}} & \significant{\num{0.06}}  & \significant{\num{21}} & \significant{\num{0.68}} & \significant{\num{0.05}} & \significant{\num{16}} & \significant{\num{0.52}} & \significant{\num{0.02}} & \significant{\num{20}} & \significant{\num{0.67}} & \significant{\num{0.08}} & \significant{\num{20}} & \significant{\num{0.64}} & \significant{\num{0.05}}  \\

\tabprojectlinespace

 Cli  & \significant{\num{28}} & \significant{\num{16}} & \significant{\num{0.44}} & \significant{\num{0.02}}  & \significant{\num{15}} & \significant{\num{0.44}} & \significant{\num{0.01}} & \significant{\num{16}} & \significant{\num{0.41}} & \significant{\num{0.00}} & \significant{\num{15}} & \significant{\num{0.40}} & \significant{\num{0.00}} & \significant{\num{15}} & \significant{\num{0.39}} & \significant{\num{0.00}}  \\

\tabprojectlinespace

 Closure  & \significant{\num{38}} & \significant{\num{9}} & \significant{\num{0.14}} & \significant{\num{0.21}}  & \significant{\num{10}} & \significant{\num{0.16}} & \significant{\num{0.20}} & \significant{\num{6}} & \significant{\num{0.05}} & \significant{\num{0.06}} & \significant{\num{10}} & \significant{\num{0.11}} & \significant{\num{0.18}} & \significant{\num{9}} & \significant{\num{0.10}} & \significant{\num{0.19}}  \\

\tabprojectlinespace

 Compress  & \significant{\num{29}} & \significant{\num{14}} & \significant{\num{0.33}} & \significant{\num{0.08}}  & \significant{\num{15}} & \significant{\num{0.36}} & \significant{\num{0.08}} & \significant{\num{13}} & \significant{\num{0.30}} & \significant{\num{0.03}} & \significant{\num{15}} & \significant{\num{0.38}} & \significant{\num{0.04}} & \significant{\num{14}} & \significant{\num{0.36}} & \significant{\num{0.08}}  \\

\tabprojectlinespace

 Csv  & \significant{\num{8}} & \significant{\num{6}} & \significant{\num{0.64}} & \significant{\num{0.05}}  & \significant{\num{7}} & \significant{\num{0.63}} & \significant{\num{0.05}} & \significant{\num{6}} & \significant{\num{0.61}} & \significant{\num{0.01}} & \significant{\num{5}} & \significant{\num{0.62}} & \significant{\num{0.09}} & \significant{\num{5}} & \significant{\num{0.51}} & \significant{\num{0.05}}  \\

\tabprojectlinespace

 Gson  & \significant{\num{14}} & \significant{\num{5}} & \significant{\num{0.34}} & \significant{\num{0.01}}  & \significant{\num{5}} & \significant{\num{0.36}} & \significant{\num{0.01}} & \significant{\num{6}} & \significant{\num{0.34}} & \significant{\num{0.06}} & \significant{\num{5}} & \significant{\num{0.36}} & \significant{\num{0.06}} & \significant{\num{5}} & \significant{\num{0.36}} & \significant{\num{0.03}}  \\

\tabprojectlinespace

 JacksonCore  & \significant{\num{22}} & \significant{\num{13}} & \significant{\num{0.54}} & \significant{\num{0.10}}  & \significant{\num{13}} & \significant{\num{0.52}} & \significant{\num{0.05}} & \significant{\num{13}} & \significant{\num{0.45}} & \significant{\num{0.06}} & \significant{\num{14}} & \significant{\num{0.48}} & \significant{\num{0.07}} & \significant{\num{15}} & \significant{\num{0.53}} & \significant{\num{0.07}}  \\

\tabprojectlinespace

 JacksonDatabind  & \significant{\num{75}} & \significant{\num{29}} & \significant{\num{0.28}} & \significant{\num{0.14}}  & \significant{\num{33}} & \significant{\num{0.31}} & \significant{\num{0.13}} & \significant{\num{29}} & \significant{\num{0.27}} & \significant{\num{0.12}} & \significant{\num{32}} & \significant{\num{0.29}} & \significant{\num{0.15}} & \significant{\num{30}} & \significant{\num{0.30}} & \significant{\num{0.13}}  \\

\tabprojectlinespace

 JacksonXml  & \significant{\num{6}} & \significant{\num{2}} & \significant{\num{0.33}} & \significant{\num{0.00}}  & \significant{\num{2}} & \significant{\num{0.32}} & \significant{\num{0.02}} & \significant{\num{2}} & \significant{\num{0.30}} & \significant{\num{0.03}} & \significant{\num{2}} & \significant{\num{0.33}} & \significant{\num{0.03}} & \significant{\num{2}} & \significant{\num{0.33}} & \significant{\num{0.08}}  \\

\tabprojectlinespace

 Jsoup  & \significant{\num{58}} & \significant{\num{24}} & \significant{\num{0.31}} & \significant{\num{0.04}}  & \significant{\num{23}} & \significant{\num{0.30}} & \significant{\num{0.06}} & \significant{\num{20}} & \significant{\num{0.28}} & \significant{\num{0.02}} & \significant{\num{27}} & \significant{\num{0.33}} & \significant{\num{0.07}} & \significant{\num{27}} & \significant{\num{0.29}} & \significant{\num{0.05}}  \\

\tabprojectlinespace

 JxPath  & \significant{\num{12}} & \significant{\num{10}} & \significant{\num{0.51}} & \significant{\num{0.23}}  & \significant{\num{8}} & \significant{\num{0.46}} & \significant{\num{0.32}} & \significant{\num{10}} & \significant{\num{0.57}} & \significant{\num{0.17}} & \significant{\num{6}} & \significant{\num{0.38}} & \significant{\num{0.29}} & \significant{\num{9}} & \significant{\num{0.44}} & \significant{\num{0.25}}  \\

\tabprojectlinespace

 Lang  & \significant{\num{44}} & \significant{\num{25}} & \significant{\num{0.49}} & \significant{\num{0.03}}  & \significant{\num{28}} & \significant{\num{0.55}} & \significant{\num{0.02}} & \significant{\num{22}} & \significant{\num{0.48}} & \significant{\num{0.02}} & \significant{\num{22}} & \significant{\num{0.46}} & \significant{\num{0.01}} & \significant{\num{28}} & \significant{\num{0.51}} & \significant{\num{0.03}}  \\

\tabprojectlinespace

 Math  & \significant{\num{2}} & \significant{\num{1}} & \significant{\num{0.50}} & \significant{\num{0.00}}  & \significant{\num{1}} & \significant{\num{0.50}} & \significant{\num{0.00}} & \significant{\num{1}} & \significant{\num{0.50}} & \significant{\num{0.00}} & \significant{\num{1}} & \significant{\num{0.50}} & \significant{\num{0.00}} & \significant{\num{1}} & \significant{\num{0.50}} & \significant{\num{0.00}}  \\

\midrule

 Overall  & \significant{\num{360}} & \significant{\num{172}} & \significant{\num{0.37}} & \significant{\num{0.09}}  & \significant{\num{181}} & \significant{\num{0.39}} & \significant{\num{0.09}} & \significant{\num{160}} & \significant{\num{0.34}} & \significant{\num{0.05}} & \significant{\num{174}} & \significant{\num{0.37}} & \significant{\num{0.09}} & \significant{\num{180}} & \significant{\num{0.37}} & \significant{\num{0.09}}  \\

\bottomrule
\end{tabular}
\end{table*}

\noindent\mynewcontent{6}{\textbf{Subjects.}
From the Defects4J dataset~\cite{defects4j}, we randomly sample $360$ bugs as this RQ's experimental subjects. These bugs can be found in our artifact.}

\noindent\mynewcontent{6}{\textbf{Baselines.}
We compare smart selection with four baselines: the original combination, branch coverage, a combination of branch, exception, and method coverage (BC-EC-MC), and smart selection without the subsumption strategy (SSWS, i.e., the combination of DBC, NTMC, EC, and OC). Note that, unlike top-level method coverage (TMC) and no-exception top-level coverage method coverage (NTMC), method coverage (MC) merely requires that a method is invoked while executing a test. We chose BC-EC-MC as a baseline because previous studies~\cite{GayCombine, salahirad2019choosing} have demonstrated its effectiveness in bug detection. As a result, we have five criterion strategies for executing genetic algorithms.}

\noindent\mynewcontent{6}{\textbf{Configuration.}
Defects4J provides two versions of CUT for each bug: a buggy version and a fixed version. For each of the three genetic algorithms (WS, MOSA, and DynaMOSA), we run it with each of five criterion strategies and a $2$-minute budget on the fixed version to get a test suite. Then, we use a script provided by Defects4J to run the suite on two versions and check whether the bug is detected. The outcome can be one of three options: \enquote{detected} (the suite passes on the fixed version but fails on the buggy version), \enquote{not\_detected} (the suite passes on both versions), and \enquote{fail} (the suite fails on the fixed version). The \enquote{fail} result is due to the flaky tests generated by EvoSuite~\cite{Shamshiri2015Do}. Due to the randomness of GAs, we repeat $10$ times for each criterion strategy.}

\noindent\mynewcontent{6}{\textbf{Result.}}

\noindent$\blacktriangleright$\mynewcontent{6}{\textbf{WS.}
Table~\ref{tab:ws_bugs} presents the number of detected bugs (the Detected column), the proportion of suites that detected bugs to all non-failing suites (the Detect$\%$ column), and the proportion of failing suites to all suites (the Fail$\%$ column) for five criterion strategies with the genetic algorithm being WS. Among these strategies, smart selection and SSWS detect the most bugs ($162$ and $161$, respectively), while the other three strategies detect nearly $150$ bugs each. The differences in detection and failure rates among the five strategies are marginal. Fig.~\ref{fig:bug_venn} (a) illustrates the overlap of bugs detected by these strategies. $104$ bugs are detected by them all, while each strategy independently detects between $4$ and $8$ bugs.}

\noindent$\blacktriangleright$\mynewcontent{6}{\textbf{MOSA.}
Table~\ref{tab:mosa_bugs} shows the bug summary for five criterion strategies, with the genetic algorithm being MOSA. Smart selection detects the most bugs ($187$), while the second most effective strategy (SSWS) detects 12 fewer bugs. The other three strategies each detect approximately $165$ bugs. Notably, MOSA with smart selection detects the most bugs among all genetic algorithms and criterion strategies. Similar to WS, the differences in detection and failure rates among the strategies are tiny. Meanwhile, the detection rates are slightly higher than that of WS. Fig.~\ref{fig:bug_venn} (b) illustrates the overlap of bugs. $128$ bugs are detected by them all, while each strategy independently detects between $2$ and $7$ bugs.}

\noindent$\blacktriangleright$\mynewcontent{6}{\textbf{DynaMOSA.}
Table~\ref{tab:dynamosa_bugs} shows the bug summary for five criterion strategies, with the genetic algorithm being DynaMOSA. The original combination and SSWS detect the most bugs ($181$ and $180$, respectively), whereas smart selection ranks fourth with $172$ detected bugs. The detection and failure rates of the strategies are close to that of MOSA. Fig.~\ref{fig:bug_venn} (c) illustrates the overlap of bugs. $126$ bugs are detected by them all, while each strategy independently detects between $3$ and $6$ bugs.}

\noindent\mynewcontent{6}{\textbf{Analysis.}
Firstly, WS significantly underperforms compared to MOSA and DynaMOSA in bug detection across all criterion strategies. This trend positively correlates with their respective coverage performances. For instance, under the BC criterion strategy, WS yields an average/median coverage of $64.7\%$/$74.4\%$, compared to MOSA and DynaMOSA, which achieve $67.2\%$/$77.8\%$ and $67.5\%$/$79.2\%$ respectively. Introducing more criteria does not significantly improve WS's bug detection capability as it does for the other two algorithms. WS with smart selection detects $10$ additional bugs compared to WS with BC, while MOSA with smart selection detects $20$ more bugs than MOSA with BC. WS with the original combination detects $2$ fewer bugs than WS with BC, while DynaMOSA with the original combination detects $21$ more bugs than DynaMOSA with BC. This is primarily because adding more criteria drastically reduces WS's coverage, counteracting guidance from additional criteria. For instance, under the original combination criterion strategy, WS's average/median coverage drops to $57.1\%$/$63.8\%$ ($-7.6\%$/$-10.6\%$), thereby hampering its ability to cover more code for bug detection. With the smart selection criterion strategy, WS's average/median coverage is $60.0\%$/$67.7\%$ ($-4.7\%$/$-6.7\%$). Therefore, we conclude that when the GA is WS, the superior performance of smart selection over other criterion strategies in bug detection is due to its smaller coverage gap and the guidance of more criteria.}

\mynewcontent{6}{Secondly, additional criteria enhance the bug detection capabilities of both MOSA and DynaMOSA. Smart selection and SSWS enable MOSA to identify $20$ and $8$ extra bugs, respectively. The original combination, SSWS, BC-EC-MC, and smart selection helps DynaMOSA uncover $21$, $20$, $14$, and $12$ additional bugs, respectively.} 
\begin{javacodebug}{Chart-10}{bug:chart10}
// buggy version
 public String generateToolTipFragment(String toolTipText) {
    return " title=\"" + toolTipText
        + "\" alt=\"\"";
}
// fixed version
 public String generateToolTipFragment(String toolTipText) {
    return " title=\"" 
        + ImageMapUtilities.htmlEscape(toolTipText)
        + "\" alt=\"\"";
}
\end{javacodebug}

\mynewcontent{6}{The bug \textit{Chart-10} serves as an illustrative example. Code~\ref{bug:chart10} presents both the buggy and fixed versions of the CUT, as well as the root cause. Specifically, the input is not encoded in the buggy version. This bug can be detected by either passing an input string containing escapable characters (such as a quotation mark) or by passing \inlinecode{null}. The latter will cause the \inlinecode{htmlEscape} method to throw an exception. Both MOSA and DynaMOSA can detect this bug using any multi-criterion strategy, while branch coverage alone fails to do so. The inability of branch coverage to detect this bug stems from its limitation: once the single line in the method under test is covered, all branch coverage requirements are deemed fulfilled. Conversely, with exception coverage or weak mutation, MOSA and DynaMOSA can generate tests beyond branch coverage, thereby enabling bug detection.}

\mynewcontent{6}{Thirdly, With MOSA, smart selection outperforms other multiple-criterion strategies in bug detection. Initially, we compare the original combination and smart selection. Smart selection detects $31$ bugs missed by the original combination, whereas the counterpart number is eight. For the fixed versions of these 31 bugs, smart selection's average branch coverage exceeds that of the original combination by $4\%$. For the eight bugs, the average coverage gap between the two strategies is nearly $1\%$. These data confirm the conclusions of RQ2 (see Sec.~\ref{subsec:rq2}), suggesting that multiple criteria also impact MOSA. By reducing goals through smart selection, MOSA achieves superior coverage, enabling the detection of more bugs. Next, we contrast smart selection with SSWS and BE-EC-MC, both of which have fewer coverage goals than smart selection. Smart selection identifies $16$ bugs overlooked by both SSWS and BC-EC-MC. For these $16$ bugs, the average branch coverage of smart selection, SSWS, and BE-EC-MC is $70\%$, $68\%$, and $73\%$, respectively, implying that smart selection does not have a coverage advantage. However, we discovered that $12$ out of the $16$ bugs are functional, not runtime errors (e.g., division by zero error), which can only be detected by assertion failures, not by triggering exceptions. Therefore, the most desired criterion for detecting these bugs is not that of SSWS and BC-EC-MC (e.g., branch coverage and exception coverage) but weak mutation. Consequently, smart selection can detect these bugs, while the other two strategies cannot.}
\begin{javacodebug}{Gson-13}{bug:gson13}
// buggy version
if (last == NUMBER_CHAR_DIGIT && fitsInLong 
    && (value != Long.MIN_VALUE || negative)) {
    peeked = PEEKED_LONG;
}
// fixed version
if (last == NUMBER_CHAR_DIGIT && fitsInLong 
    && (value != Long.MIN_VALUE || negative) 
    && (value!=0 || false==negative)) {
    peeked = PEEKED_LONG; 
}
\end{javacodebug}

\mynewcontent{6}{
We use the bug \textit{Gson-13} \cite{buggson13} as an illustrative example. Code~\ref{bug:gson13} displays the crucial part of both the buggy and fixed versions of CUT (the class \textit{JsonReader}). This bug arises because "\inlinecode{-1}" in JSON should be interpreted as a float value rather than an integer value (i.e., \inlinecode{PEEKED\_LONG} in the code). Without a mutant from the ROR operator of weak mutation, which has a different branch condition from the original CUT, MOSA fails to generate a test that detects this bug. Consequently, both SSWS and BC-EC-MC miss it. Note that even though the original combination includes weak mutation, it does not catch this bug either. This could be attributed to the excessive number of coverage goals (e.g., $678$ branches and $723$ mutants) in this Java class, which hampers the original combination's coverage and bug detection performance. For instance, its average branch coverage on this class is $65\%$, $6\%$ lower than the $71\%$ achieved by smart selection.}

\mynewcontent{6}{Lastly, with DynaMOSA as the GA, the original combination surpasses other multiple-criterion strategies in bug detection. This result confirms the findings of RQ3 (see Sec.~\ref{subsec:rq3}), indicating that DynaMOSA is more robust than WS and MOSA when handling multiple criteria. Surprisingly, MOSA with smart selection detects more bugs than DynaMOSA with the original combination ($187$ versus $181$). This discrepancy can be attributed to the inherent randomness of GAs, explained by two reasons. First, the gap is not substantial compared to other differences discussed previously. Second, we found that among the $20$ bugs detected by MOSA with smart selection but missed by DynaMOSA with the original combination, $16$ bugs require tests to meet complex conditions or to generate complex objects for detection, which are typical challenges in SBST~\cite{Shamshiri2015Do}. As a result, DynaMOSA fails to produce such tests. Although MOSA does detect these bugs, only about one in ten of its runs can detect each of them on average. Therefore, improving GAs to detect these bugs could be a potential direction for future research.}
\mytcbox{Answer to RQ8}{\mynewcontent{6}{Among all genetic algorithms and criterion strategies, MOSA with smart selection detects the most bugs.}}

\section{\mynewcontent{6}{Threats to Validity}}\label{sec:discuss}
\noindent\mynewcontent{6}{\textbf{Construct Validity.} In Sec. \ref{subsec:cc}, we apply three rules to ascertain whether two criteria exhibit a coverage correlation. One rule involves adopting the conclusions from the study by Gligoric et al.~\cite{GligoricCorr}, which investigates the efficacy of various coverage criteria in predicting mutation scores. They assessed criteria (including branch and statement coverage) on various Java and C programs using manual and automated tests. Using three correlation coefficients, they gauged the criteria's accuracy in predicting the suites' mutation scores. Their findings indicate that the average Kendall's $\tau_{b}$ values~\cite{kendall1938new} for coverage correlation between branch, statement coverage, and mutation testing all exceed $0.7$. This suggests a strong coverage correlation with mutation testing. Consequently, in this paper, we consider that both branch coverage and line coverage (a substitute for statement coverage) have a coverage correlation with weak mutation. Potential threats exist, however, due to the following reasons: (1) the subjects used in their research differ from ours, with some of their programs being in C, while all of ours are JAVA classes; (2) their evaluation did not involve one EvoSuite mutation operator, RV (Replace a variable). Nevertheless, considering the large number of mutants generated by mutation operators on a subject, we maintain that the statistical coverage correlations still hold. For instance, the average number of mutants generated for our experimental subjects is $378$. Furthermore, our experiment on measuring coverage correlations among criteria (see Sec. \ref{subsec:rq-4}) also confirms the strong coverage correlation between branch/line coverage and weak mutation.}

\noindent\mynewcontent{6}{\textbf{Internal Validity.} The first threat to internal validity is the concrete value of a new parameter introduced by smart selection: \textit{lineThreshold} (see Sec. \ref{subsec:rs}). In handling line coverage, smart selection skips those basic blocks (BBs) with lines less than \textit{lineThreshold}. The larger the value of this parameter, the more BBs we skip. Without considering the dead code, (direct) branch coverage fails to capture the following lines only when a certain line in a BB exits abnormally. Previous work~\cite{Rojas2015CombiningMC} shows that, on average, when $78\%$ of branches are covered, only $1.75$ exceptions are found. It indicates that (direct) branch coverage can capture most properties of line coverage. Therefore, to minimize the impacts of line goals on SBST, we prefer a larger \textit{lineThreshold}. After statistics on the benchmarks used in DynaMOSA~\cite{PanichellaDynaMOSA}, we find that $50\%$ of the BBs have less than $8$ lines. As a result, we set it to $8$. Another threat arises from the randomness of genetic algorithms. To mitigate this risk, we replicated each approach $30$ times for every class with a search budget of $2$ minutes in RQ1, 2, 3, 5, and 6. Similarly, with budgets of 5, 8, and 10 minutes, we repeated each approach $10$, $10$, and $5$ times, respectively, in RQ7. In RQ8, we replicated each approach $10$ times for every bug. Furthermore, the third threat involves potential bugs introduced through our implementation of smart selection. The most significant defect that could compromise our experimental results would be incorrect statistical results generated by EvoSuite, implying that smart selection did not enhance coverage. To mitigate this risk, we selected several generated test suites and verified their coverage using external coverage tools, such as JaCoCo and the coverage tool embedded in IntelliJ IDEA. The results from these tools corroborate the coverage increase facilitated by smart selection.}

\noindent\mynewcontent{6}{\textbf{External Validity.} One threat to external validity comes from the experimental subjects. We chose 158 Java classes from the benchmark of DynaMOSA~\cite{PanichellaDynaMOSA}.~\cite{PanichellaDynaMOSA} was published in 2018. Many classes have already become obsolete. Some projects are no longer maintained~\cite{LinGraph}. To reduce this risk, we choose 242 classes at random from Hadoop~\cite{hadoop}, thereby increasing the diversity of the dataset. Furthermore, we have implemented smart selection solely in the Java language and assessed it on Java classes. Therefore, our findings may not apply to other languages. However, the foundational principles of smart selection are not exclusive to Java and could potentially be applied to other programming languages.}
\section{Related Work}\label{sec:related}
In this section, we introduce related studies on (1) SBST and (2) coverage criteria combination in SBST.

\noindent\textbf{SBST.} 
SBST formulates test cases generation as an optimization problem. Miller et al.~\cite{miller1976automatic} proposed the first SBST technique to generate test data for functions with inputs of float type. SBST techniques have been widely used in various objects under test~\cite{FraserEvoSuite, arcuri2018evomaster, castelein2018search, gambi2019automatically, wang2021ML,Tang:2020,dong2020time,martin2021restest, haq2021automatic}, and types of software testing~\cite{li2007regression, silva2017systematic, walcott2006timeaware}. Most researchers focus on (1) search algorithms: Tonella~\cite{TonellaEvo2004} proposed to iterate to generate one test case for each branch. Fraser et al.~\cite{FraserWhole} proposed to generate a test suite for all branches. Panichella et al.~\cite{PanichellaMOSA, PanichellaDynaMOSA, grano2019testing} introduced many-objective optimization algorithms; (2) fitness gradients recovery: Lin et al.~\cite{lin2020recovering} proposed an approach to address the inter-procedural flag problem. Lin et al.~\cite{LinGraph} proposed a test seed synthesis approach to create complex test inputs. Arcuri et al.~\cite{arcuri2021enhancing} integrated testability transformations into API tests. Braione et al.~\cite{braione2017combining} combined symbolic execution and SBST for programs with complex inputs; (3) readability of generated tests: Daka et al.~\cite{daka2017generating} proposed to assign names for tests by summarizing covered coverage goals. Roy et al.~\cite{roy2020deeptc} introduced deep learning approaches to generate test names; (4) fitness function design: Xu et al.~\cite{xu2017adaptive} proposed an adaptive fitness function for improving SBST. Rojas et al.~\cite{Rojas2015CombiningMC} proposed to combine multiple criteria to satisfy users' requirements.

\noindent\textbf{Coverage Criteria Combination in SBST.}
After the work of Rojas et al.~\cite{Rojas2015CombiningMC} introducing combining multiple criteria, \mynewcontent{6}{Gay et al.~\cite{GayCombine, gay2017fitness, salahirad2019choosing} conducted experiments with various combinations of coverage criteria to evaluate the efficacy of multi-criteria suites in detecting faults from the Defects4J dataset~\cite{defects4j}. Both their work and this paper concentrate on the combination of coverage criteria. The difference lies in their investigation of the impact of various combinations on fault detection, while our work aims to mitigate the coverage decrease caused by combining multiple criteria. Meanwhile, we also examine how smart selection influences bug detection. Compared to criterion combinations in their studies, smart selection (the combination of DBC, TMC, EC, and OC with two coverage goal subsets from LC and WM) is a more fine-grained multiple-criteria strategy and detects the most bugs in our evaluation. Hence, rather than criterion combinations, using combinations of coverage goals from multiple criteria for bug detection is a potential direction in future studies.} \mynewcontent{6}{Omur et al.~\cite{SAHIN2021806} introduced the Artificial Bee Colony (ABC) algorithm as a substitute for the genetic algorithm used in WS~\cite{FraserWhole} since the ABC algorithm brings more diversity in the evolving population. Sharing the same target, i.e., improve coverage when combining multiple criteria, their method replaces the genetic algorithm, and our method reduces the optimization goals. Hence, these two methods are orthogonal and can work together.}
\section{Conclusion}\label{sec:conclu}
We propose smart selection to address the coverage decrease caused by combining multiple criteria in SBST. We compare smart selection with the original combination on $400$ Java classes. The experiment results confirm that with WS and MOSA, smart selection outperforms the original combination, especially for the Java classes with no fewer than $200$ branches. However, with DynaMOSA, the differences between the two approaches are slight \mynewcontent{6}{in terms of the coverage of eight criteria}. \mynewcontent{3}{Additionally, we conduct experiments to confirm our assumptions about coverage criteria relationships. \mynewcontent{6}{We also compare their coverage performance under different budgets and evaluate their efficacy in bug detection}, \mynewcontent{6}{confirming} the advantage of smart selection over the original combination.}
\section*{Acknowledgements}
Z. Zhou and Y. Tang are partially sponsored by Shanghai Pujiang Program (No. 21PJ1410700) and National Natural Science Foundation of China (No. 62202306). Y. Zhou is partially sponsored by National Natural Science Foundation of China (No. 62172205). C. Fang and Z. Chen are partially sponsored by Science, Technology and Innovation Commission of Shenzhen Municipality (CJGJZD20200617103001003). J. He is partially sponsored by Shanghai Sailing Program (No. 22YF1428600).
\ifCLASSOPTIONcaptionsoff
\newpage
\fi
\bibliographystyle{IEEEtran}
\bibliography{ref}

\end{document}